\documentclass[]{aa}

\usepackage{amsmath}
\usepackage{natbib}
\usepackage{graphicx}
\usepackage{txfonts}
\usepackage{multirow}
\usepackage{textcomp}
\usepackage{subfigure}
\usepackage{gensymb}
\usepackage{lscape}
\usepackage{longtable}
\usepackage[pagebackref,colorlinks,citecolor=blue,urlcolor=blue,linkcolor=blue]{hyperref}
\usepackage{tikz}
\usetikzlibrary{snakes,arrows,shapes,positioning,backgrounds}

\graphicspath{{figures/}}
\usepackage{pifont}

\usepackage{color}

\begin{document}

\title{The DICE calibration project\\
design, characterization, and first results
}

\author{
N.~Regnault\inst{1}
\and
A.~Guyonnet\inst{1} 
\and
K.~Schahman\`eche\inst{1} 
\and
L.~Le Guillou\inst{1}
\and
P.~Antilogus\inst{1}
\and
P.~Astier\inst{1}
\and 
E.~Barrelet\inst{1}
\and
M.~Betoule\inst{1}
\and
S.~Bongard\inst{1}
\and
J.-C.~Cuillandre\inst{2}
\and
C.~Juramy\inst{1}  
\and
R.~Pain\inst{1}
\and
P.-F.~Rocci\inst{1}
\and
P.~Tisserand\inst{3}
\and
F.~Villa\inst{1}
}

\institute{
LPNHE, CNRS-IN2P3 and Universit\'{e}s Paris 6 \& 7, 4 place Jussieu, F-75252 Paris Cedex 05, France
\and
Canada-France-Hawaii Telescope Corporation, Kamuela, HI 96743, USA
\and
Research  School  of  Astronomy  and  Astrophysics,  Australian  National  University,  ACT  2601,  Australia.
}

\titlerunning{The DICE Calibration Project}
\authorrunning{N.~Regnault et al.}
\date{Received Mont DD, YYYY; accepted Mont DD, YYYY}

\abstract{}%
{We describe the design, operation, and first results of a photometric 
calibration project, called DICE (Direct Illumination Calibration
Experiment), aiming at achieving precise instrumental 
calibration of optical telescopes.  
The heart of DICE is an illumination device composed of 24 narrow-spectrum, high-intensity, light-emitting diodes (LED) chosen to
  cover the ultraviolet-to-near-infrared spectral range.  It
  implements a point-like source placed at a finite distance from the
  telescope entrance pupil, yielding a flat field illumination that
  covers the entire field of view of the imager. The purpose of this 
system is 
  to perform a lightweight routine monitoring of the imager passbands
  with a precision better than 5 per-mil on the relative passband
  normalisations and about $3$~{\AA}ngstr\"oms on the filter cutoff
  positions.}%
{Prior to installation, the light source is calibrated on a
  spectrophotometric bench. 
  As our
  fundamental metrology standard, we use a photodiode calibrated at the National 
  Institute of Standards and Technology (NIST).
  The radiant intensity of each beam is
  mapped, and spectra are measured for each LED.  All measurements are
  conducted at temperatures ranging from 0\celsius\ to 25\celsius\ in
  order to study the temperature dependence of the system.  The
  photometric and spectroscopic measurements are combined into a model
  that predicts the spectral intensity of the source as a function of
  temperature.}%
{We find that the calibration beams are stable at the $10^{-4}$ level
  -- after taking the slight temperature dependence of
  the LED emission properties into account.  We show that the spectral intensity of
  the source can be characterised with a precision of $3$
  \AA{ngstr\"oms} in wavelength, depending on how accurately we are
  able to calibrate the wavelength response of the mononochromator.  
  In flux, we reach an accuracy
  of about $0.2 - 0.5\%$ depending on how we understand the
  off-diagonal terms of the error budget affecting the calibration 
  of the NIST photodiode. 
  We describe how with a
  routine $\lesssim 60$-mn calibration program, the apparatus is able to
  constrain the imager passbands at the targeted precision levels.}{}

\maketitle

\section{Introduction} 

The measurement of the dark energy equation of state with type Ia
supernovae 
(SNe~Ia; see for recent examples: \citet{2012ApJ...746...85S, 2014A&A...568A..22B})   
sets very strong constraints on the accuracy of
the flux calibration of the imagers used to infer the SN~Ia luminosity
distances.  Indeed, mapping the relative variation in SN distances
with redshift boils down to comparing fluxes measured at both ends of
the visible spectrum.  A key concern is therefore the control of the
imager throughput as a function of wavelength.  In practice, this
means, on the one hand, controlling the intercalibration of the fluxes,
measured in different passbands, and on the other hand, measuring and
monitoring the passband shapes, in particular, the position of the
filter fronts.

The best technique of passband intercalibration  today
relies on observations of stellar spectrophotometric standards, whose
absolute spectral energy distribution (SED) is known with accuracy.  The
most reliable sets of standards are probably those that were established to
calibrate the instruments mounted on-board HST \citep[][]{calspec}.  
The CALSPEC flux scale is defined by
the SED of three hot DA white dwarfs, as
predicted by a NTLE model atmosphere code, itself tuned to the Balmer line
profiles of each dwarf \citep{2007ASPC..364..315B, 2013A&A...560A.106R}. The calibration of
the HST STIS and NICMOS spectrographs obtained by observing these
fundamental standards is then propagated to a larger network of redder
and mostly fainter stars \citep{2010AJ....139.1515B, 2014arXiv1406.1707B}.

This approach is robust, since it relies on more than one fundamental
calibrator.  It is also sound, since each CALSPEC release can be traced
back to a well-identified series of HST observations and to a
particular version of a stellar atmosphere model.  However, even DA
white dwarfs are complex objects, and the modelling systematics that
affect the CALSPEC flux scale are difficult to
estimate with precision.

Most recent supernova surveys have been anchored on the CALSPEC
network.  This was often implicit because people were using the SEDs of
historical bright standards recalibrated with HST
\citep{bohlin_absolute_2004, bohlin_hubble_2004} when interpreting
their magnitudes as fluxes \citep[see e.g.][]{2010ApJ...716..712A}. A
few projects, such as SNLS and SDSS, tried to explicitly tie their
calibration to specific CALSPEC calibrators
\citep{holtzman_sloan_2008, regnault_photometric_2009}.  
The calibration error budgets published by these authors were dominated by
uncertainties introduced in the metrology chain that links the standard
star observations to the science images.  Recent work by
\citet{2013A&A...552A.124B} has shown, however, that it is now possible to transfer
the CALSPEC flux scale with uncertainties comparable to the CALSPEC
internal uncertainties ($\sigma_{g-i} \sim 3-5$ mmag).

To improve on the current cosmological measurements, future
dark energy surveys will need to intercalibrate their passbands at
a few per-mil, which is below the current estimates of the CALSPEC uncertainty budget.  Since
modern, ground-based imagers display repeatabilities of 1-3 mmag 
\citep{2007A&A...470.1137M},
they can in theory be
calibrated to this precision level.  However, it is not obvious yet that
this ambitious goal is within reach of the stellar calibration
techniques presented above.

\begin{figure}
\centering
\setlength{\unitlength}{1cm}%
\begin{tikzpicture} [align=center, rounded corners, thick,meas/.style={rectangle,draw=black,minimum height=1cm,minimum width=2cm},logo/.style={minimum height=1cm,minimum width=2cm},new/.style={stealth-stealth, dashed,above,sloped,midway,red},old/.style={stealth-stealth}, auto]
  \node  [meas] (powr) at (  -2.5, 6) {POWR (NIST)\\\citep{2006Metro..43S..31H}};
  \node  [meas] (sircus) at ( 2.5, 4.5) {SIRCUS/SCF (NIST)\\\citep{2006ApOpt..45.8218B, 2000Metro..37..579B}\\\citep{NIST.SPECIAL.PUBLICATION.250.1}};
  \node  [meas] (pd) at ( -2.5, 3) {Calibrated\\Si photodiode};
  \node  [meas] (dice) at (  2.5, 1.5) {Calibrated source\\(in the telescope enclosure)};
  \node  [meas] (tel) at (  -2.5, 0) {Imager};
  \node  [meas] (sky) at (   2.5, -1.5) {Astronomical sources};
  \draw [->,dotted] (powr) to node [swap] {} (sircus);
  \draw [->,dotted] (sircus) to node [swap] {} (pd);
  \draw [->,thick] (pd) to node {\ding{202}} (dice);
  \draw [->,thick] (dice) to node {\ding{203}} (tel);
  \draw [->,thick] (tel) to node  {\ding{204}} (sky);
  \draw [dashdotted,color=gray] (0, 8) -- (0, -2);
  \node [logo] at (-2, 7.5) {\Large\sf\textcolor{gray}{DETECTORS}};
  \node [logo] at ( 2, 7.5) {\Large\sf\textcolor{gray}{SOURCES}};  
\end{tikzpicture}
\caption{Generic metrology chain from POWR, the official
  implementation of the optical watt, which is maintained at NIST, to the
  telescope imagers. \label{fig:nist_metrology_chain}}
\end{figure}
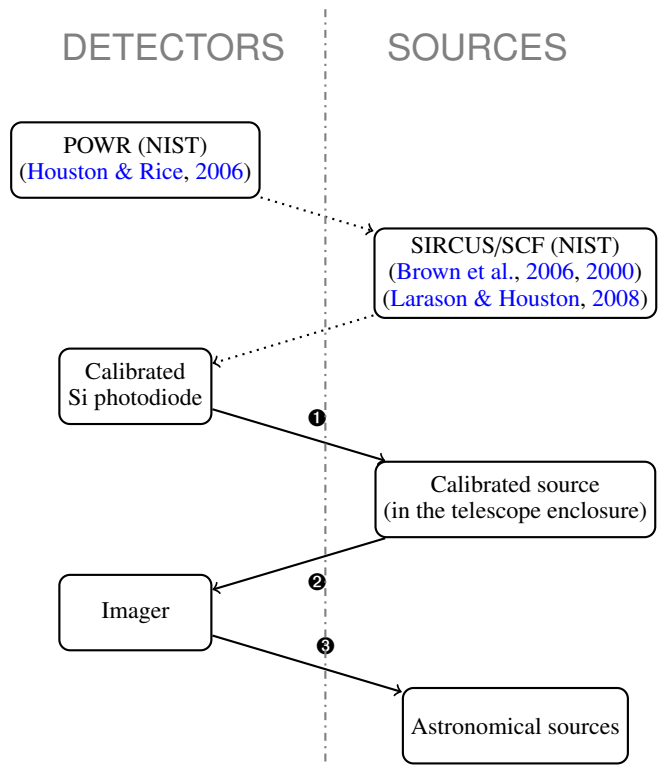

In this context, alternate calibration strategies have been
proposed by several groups.  Most have in common that they rely on laboratory flux
standards instead of stellar models.  Over the past two decades, flux
metrology has indeed experienced dramatic improvements, moving from
1--2\% source standards to detector standards accurate at the 0.1\%
level. In the visible, the {metrology} standards are generally silicon
photodiodes, whose calibration can be traced back to the Primary
Optical Watt Radiometer \citep[POWR, ][]{2006Metro..43S..31H}, an
electrical cryogenic substitution radiometer maintained by the National
Institute of Standards and Technology (NIST) as the official
implementation of the optical watt. To characterise 
the photodiodes provided to its end users, NIST maintains a sophisticated metrology chain \citep{NIST.SPECIAL.PUBLICATION.250.1} involving
intermediate light sources, notably the SIRCUS laser facility
\citep{2006ApOpt..45.8218B,2000Metro..37..579B} and the Spectral
Comparator
Facility 
(SCF). With a calibrated photodiode in hand and the goal of
calibrating another light detector, the end user has no choice but to build
an intermediate light source to transfer the NIST flux scale to his
own instrument. As a consequence, most alternatives to stellar
calibration implement variations around the generic metrology chain
presented in Fig. \ref{fig:nist_metrology_chain}.

Among those, at least two groups have proposed to  revisit the
historical measurements of \citet{1975ApJ...197..593H} using modern
flux metrology techniques. The goal is to compare the CALSPEC and NIST flux
scales directly using bright standard stars.  A first team is developing the ACCESS 
rocket-borne 40-cm telescope \citep{2010hstc.workE..10K}. The instrument 
is a spectrograph, sensitive from 350-nm to 1.7\micro m 
and calibrated directly at the SIRCUS facility. The other team, 
comprising NIST scientists, is building a spectrophotometer at the focus
of a 10-cm telescope, calibrated
with an artificial source, which will target stars 
in the $0<V<5$ magnitude range \citep[e.g.][]{2012SPIE.8450E..1SM}.
Both projects will rely on HST
observations to bridge the gap between their magnitude range and the
CALSPEC magnitude range. They will permit a direct comparison between
the CALSPEC and NIST flux scales, which is an invaluable piece of
information.  This does not, however, relieve future surveys from
monitoring their own instrument, in particular, from measuring 
their passbands {\em \emph{in situ}} and checking for a possible evolution in the long term.

Today, the required accuracy on the positioning of the passband
cut-offs is as low as a fraction of a nanometre. As an example, for SNLS, 
decreasing the uncertainty on $r$-band positioning from 3-nm to 
2\AA\ decreases the uncertainty on $w$ as much as adding 200 SNe~Ia in the 
Hubble diagram.  Passband
models are usually built using pre-installation test-bench
measurements of the imager optical components. This is not entirely
satisfactory because filters may evolve over time. For this reason, nearly
every modern survey has plans to build and operate a dedicated
calibration source, able to follow the shape and
normalisation of the imager passbands in real time.  A precursor in this domain is
the apparatus described in \citet{Doi2010}.  It consists in a movable
light source, designed to inject quasi-monochromatic light into the
optics of the SDSS 2.5-m telescope, and measure its
effective passbands {\em \emph{in situ}}.  This system permitted us to unveil significant
variations in the blue part of the imager $u$ band.  

{\citet{stubbs_toward_2006} push the concept one step further by
describing how an in-situ narrow-band illumination system can be used to transfer
the calibration of a NIST photodiode to an astronomical imager. 
Implementations of the procedure sketched in their paper are the} 
systems developed for the Mosaic imager on the CTIO 4-m Blanco telescope 
\citep{stubbs_preliminary_2007} and for the PanSTARRS GigaPixel
imager \citep{Stubbs2010}. Another notable example, {built along similar design principles,}
is the DECal system \citep{2013arXiv1302.5720M} now installed 
in the dome of the Blanco Telescope to monitor and calibrate 
the DECam imager \citep{2008SPIE.7014E..0ED}. All these systems consist
in a large diffusive screen placed in front of the telescope pupil and
illuminated with monochromatic light, generated either by a lamp
coupled to monochromators or by a tunable laser.  So far, these
projects have mostly focussed on in-situ measurements of the passband
shapes.  \citet{Stubbs2010} was able to perform a comparison between a
stellar and an instrumental calibration, reporting a qualitative
agreement of $\sim 5\%$ between both.

DICE\footnote{Direct Illumination Calibration Experiment}  was
conceived by members of the SNLS collaboration, building upon the
lessons learned working with the MegaCam wide field imager. 
It shares
most of the goals of the projects described above, but differs in
several points of its design.
{The primary goal of the DICE project is to implement the
  metrology chain sketched in Fig. \ref{fig:nist_metrology_chain},
  i.e. to transport the calibration carried by a NIST photodiode to
  the MegaCam imager and compare this calibration to CALSPEC. This
  includes performing a spectrophotometric calibration of the light
  source using the NIST photodiode as a metrology standard (Step
  \ding{202} in Fig. \ref{fig:nist_metrology_chain}), measuring the
  relative normalisations of the MegaCam passbands from series of
  calibration exposures taken with DICE (Step \ding{203}), and applying
  this calibration to observations of CALSPEC standards (Step
  \ding{204}). }

{This paper is the first in a series of three papers that will describe  Steps 
\ding{202}, \ding{203}, and \ding{204}. 
For now, we focus on Step \ding{202}}. We present the design and implementation of the DICE
calibration sources (\S\ref{sec:system_description}).  We then
concentrate on the spectrophotometric
characterisation of the DICE sources in \S\ref{sec:spectrophotometric_test_bench},
\S\ref{sec:photometric_calibration}, and
\S\ref{sec:spectroscopic_calibration}, on a dedicated test bench equipped with
a photodiode calibrated at NIST.
Spectrophotometry is notoriously difficult, and we had to develop
specific methods to remain immune to the measurement systematics. 
{In \S\ref{sec:analysis}, we venture into the area that will be covered in the 
next DICE papers: we sketch briefly 
how a DICE source can be used to calibrate a broadband imager,
and we estimate
the precision of the calibration that can be obtained from DICE observations 
by propagating the Step \ding{202} error budget established in this paper.  }
We conclude in \S\ref{sec:discussion}.


\section{The DICE light source}
\label{sec:system_description}

The design of the DICE light source has been described in
\citet{juramy:tel-00592266,barrelet_direct_2008}, and \citet{2008SPIE.7014E.166J}. 
Two demonstrators have been built and installed.  A first
apparatus (SnDICE) was designed in 2007 for the MegaCam wide field
imager \citep{Boulade2003} which equips the 3.6-m Canada-France-Hawaii
Telescope (CFHT).  In 2011-2012, a second light source (SkyDICE) was
built for the SkyMapper imager \citep{2007PASA...24....1K} with an
improved design, building on the lessons learned with SnDICE. A third
source was also built and kept at LPNHE for long-duration test-bench
studies.

A key requirement that guided our design effort is that a calibration
instrument must be as simple and easy to maintain as possible. Also, it must {intrinsically}
be as stable as possible, with additional
built-in redundancies that allow checking for 
long-term drifts of the intensity delivered into the telescope pupil.
As a rule of thumb, we consider that to calibrate an instrument at the
permil level, we must be able to measure and study any
1\textperthousand\ drift of its response with a precision of
10\% or better. This sets an ambitious stability goal on the light
source of about $10^{-4}$.
In the rest of this
section, we present the most important aspects of the light source design.

\subsection{Calibration beam}

\begin{figure}
  \begin{center}
    \mbox{
      \subfigure[DICE beam]{\includegraphics[width=0.35\linewidth,angle=-10]{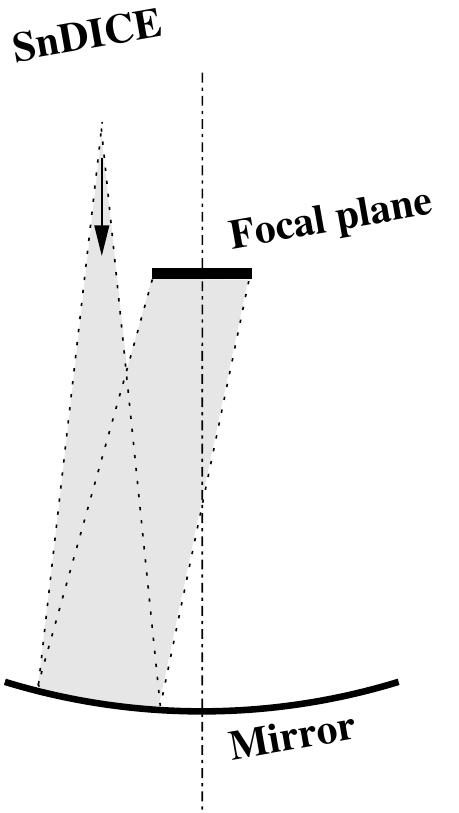}\label{fig:dice_beam}}}
      \subfigure[Science beam]{\includegraphics[width=0.35\linewidth,angle=-10]{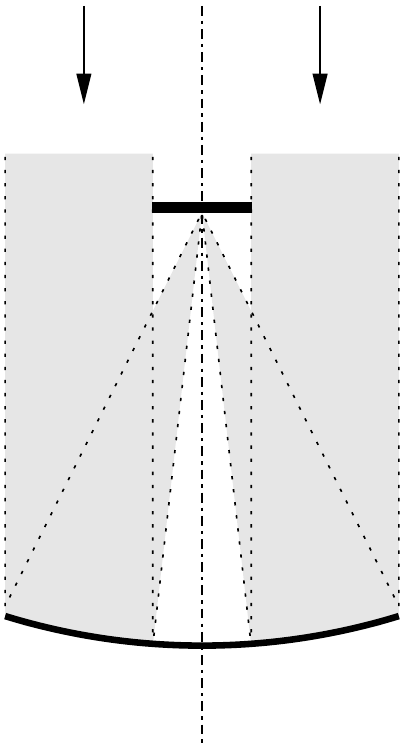}\label{fig:science_beam}}
    \caption{Left: the DICE
      calibration beam.  A DICE light source consists in a quasi-point
      source, located at a finite distance from the telescope pupil.
      Such a beam generates a quasi-uniform illumination of the focal
      plane. 
      Right: the science beam. The light from a distant source,
      such as a star, may be approximately modelled by a plane wave that fills
      the entire telescope aperture.  Such a beam is focussed
      on a single point of the focal plane.  }
    \label{fig:beams}
  \end{center}
\end{figure}

Ideally, a calibration light source should mimic the
science objects under study as much as possible. Since a supernova survey is dealing
primarily with point sources (supernovae and field stars), we should
try to generate quasi-parallel beams, covering the entirety of the
primary mirror (see figure \ref{fig:science_beam}). Such a beam
would result in a spot on the focal plane, and we could use the
photometry code used in the survey photometry pipeline to
estimate its flux, thereby avoiding the systematic errors that arise
from using different flux estimators.

Unfortunately, building a good artificial star turns out to be
difficult. We therefore deliberately opted for a different design
(figure \ref{fig:dice_beam}). SnDICE is a point source located
in the dome a few metres away from the telescope primary mirror and
close to the object plane. The source generates a conical,
quasi-Lambertian beam, of aperture $\sim 2^{\mathrm o}$, so slightly
larger than the telescope's angular acceptance. Such an illumination
results in an almost uniform focal-plane illumination.

As shown in figure \ref{fig:beams}, the calibration beam is radically
different from the science beam.  In particular, the angular
distribution of the light rays that hit the various optical surfaces
(e.g.  the interference filters) is not comparable. However, this
specific calibration beam has at least two very nice
properties. First, no intermediate elements (folding mirror, window,
screen, etc.) are present between the light emission zone and the
primary mirror.  As a consequence, the  design stays very
simple, and the system is expected to be stable in the
long term.  Second, the structure of the beam is much simpler than the
science beam, in the sense that each pixel sees photons that came
through a unique path.  In other words, there is a one-to-one
relationship between the focal plane elementary surface elements and
the calibration beam elementary solid angles. It is therefore quite
simple to predict the focal plane illumination, once one knows the
beam radiant intensity map: the former follows from the latter from
purely geometrical considerations, involving propagation of light in
free space, and through the optics.

\subsection{Light emitters}

Narrow-spectrum light emitting diodes (LED) were chosen as light
emitters.  As shown later in this paper, LEDs are extremely
stable, as long as they are fed with stable currents and operated at a
stable temperature.  It it is relatively easy today to build current
sources that are stable at a few $10^{-5}$ over a temperature range of a few
degrees, and the LED emission properties vary with temperature in a
smooth and simple manner. One of the purposes of this paper is to show
that one can build a LED-based light source, delivering beams whose
stability can be controlled at the level of a few 10$^{-4}$ over long
durations.

\begin{figure*}
\begin{center}
\subfigure[SnDICE coverage of the MegaCam passbands]{\includegraphics[width=0.48\linewidth]{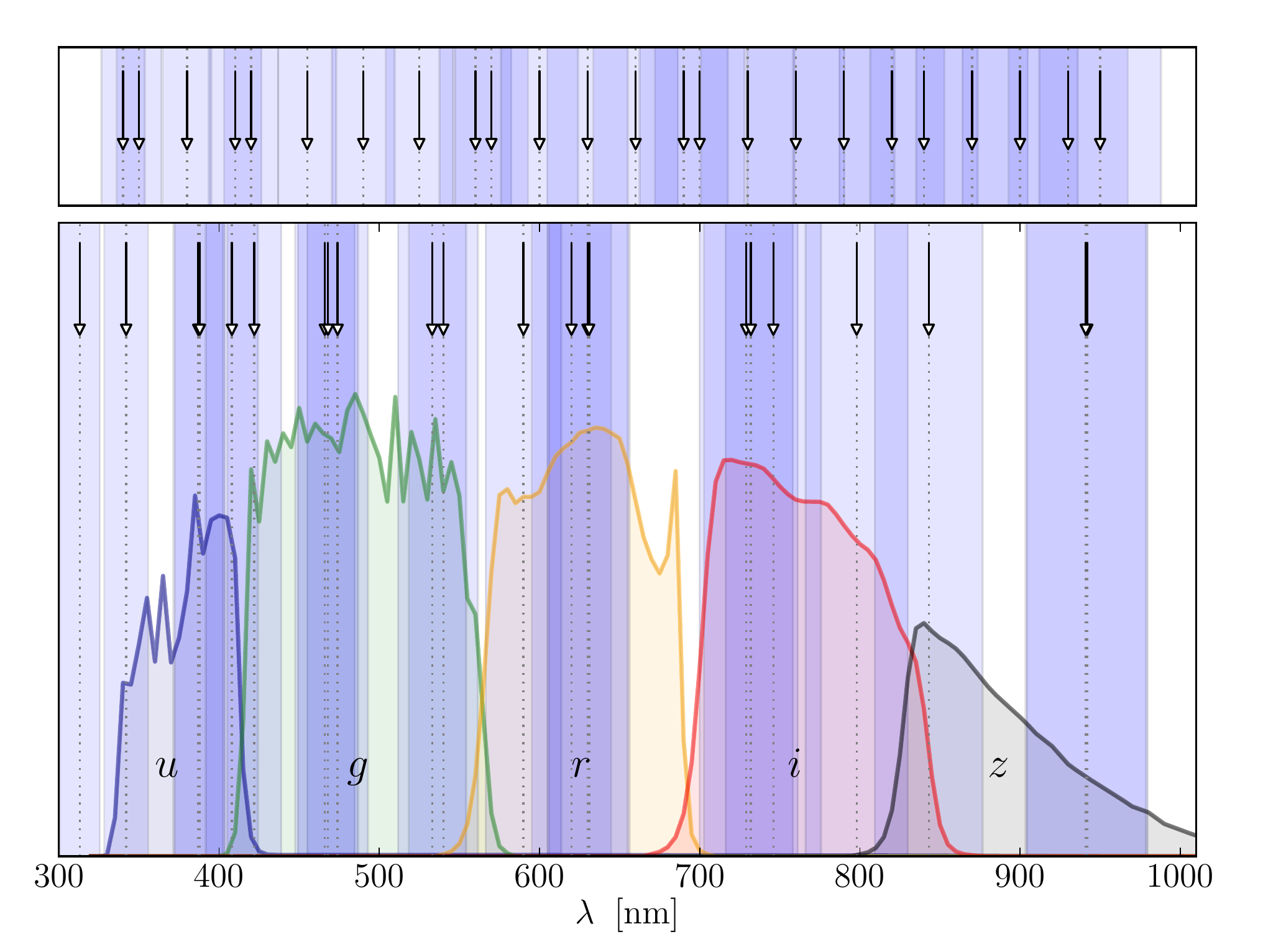}\label{fig:filter_coverage_sndice}}\quad
\subfigure[SkyDICE coverage of the SkyMapper passbands]{\includegraphics[width=0.48\linewidth]{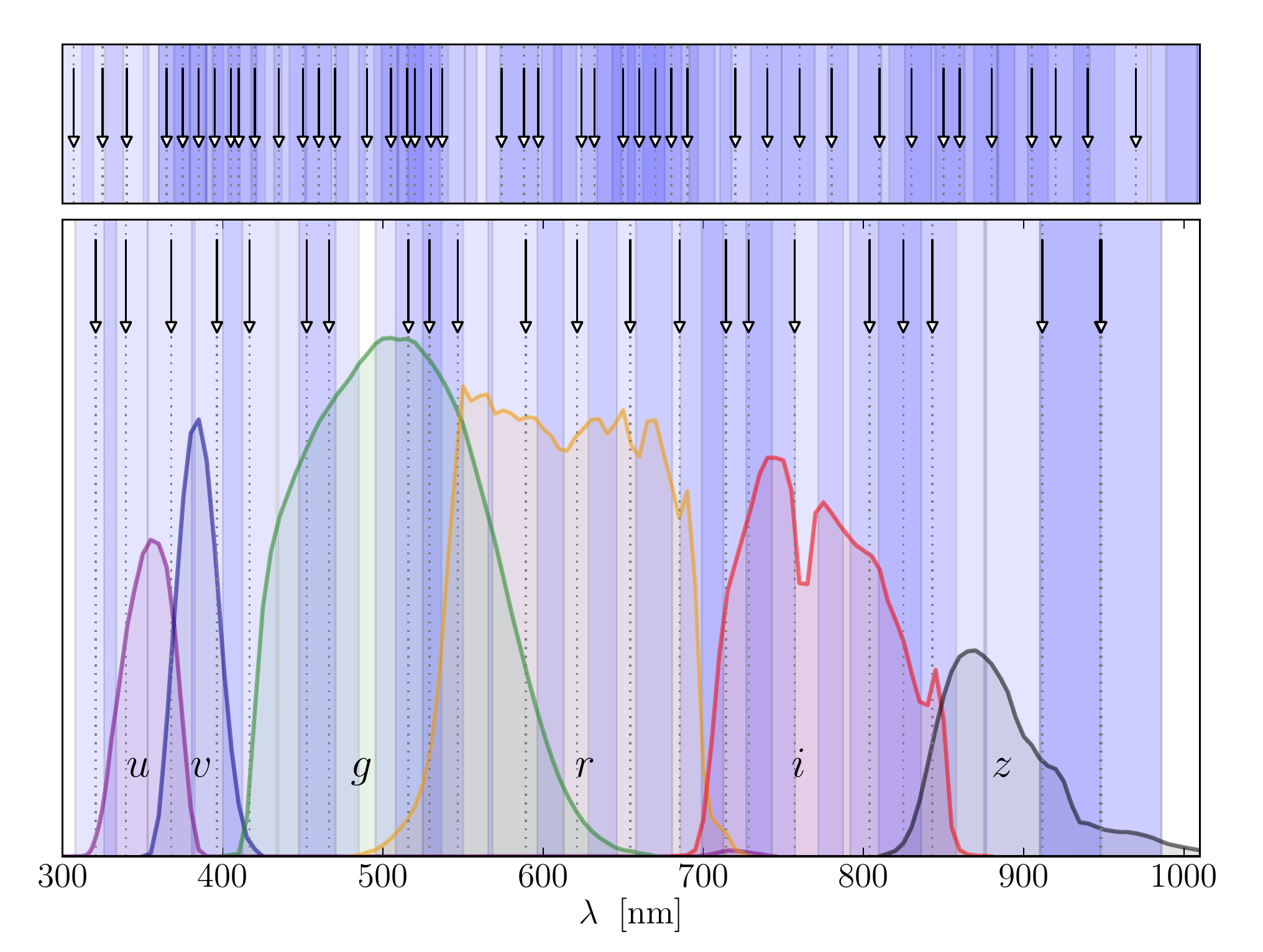}\label{fig:filter_coverage_skydice}}
\caption{{\em Lower panels:} {\em (Left)} coverage of the MegaCam filters with SnDICE (2008). The central wavelengths of all LEDs
  are displayed with arrows. The shaded regions show the wavelength
  extension of each LED SED --within 7\% of the spectrum peak.  
  In 2007, no narrow spectrum LED with sufficiently
  high power was available around 700-nm and above
  900-nm. {\em (Right)} coverage of the SkyMapper filters with SkyDICE (2010). The position 
  of about half of the LEDs were chosen so that they constrain the filter cutoffs well. 
  The other LEDs were selected to fill the gaps between filter cutoffs with about 
  one LED every 20/30-nm. 
  {\em Upper panels:} {\em (Right)} sampling from the near-UV to the near-IR achievable 
  with the LEDs listed in the OSRAM and Roithner catalogs. {\em (Left)} optimised sampling of the MegaCam passbands that 
  can be implemented today with the LEDs available on the market. 
  }
\label{fig:filter_coverage}
\end{center}
\end{figure*}

LEDs do not emit monochromatic light.  The typical FWHM of a LED
spectrum is about $\delta \lambda / \lambda \sim 5 - 7 \%$
(i.e. 20-nm to 50-nm).  This means that we need 20 to 25 LEDs to cover
the entire visible spectrum -- from $350$-nm to $1100$-nm.  In
figure \ref{fig:filter_coverage}, we show the sampling that could be
obtained with the first prototype, which was built for MegaCam
(figure \ref{fig:filter_coverage_sndice}). We also show, for comparison, what
could be achieved four years later with SkyDICE, our second prototype, 
built to calibrate SkyMapper
(figure \ref{fig:filter_coverage_skydice}).  As can be seen, the diversity of
LEDs available on the market improved very significantly in a few years.  
Today, by combining the catalogues of the
three main LED manufacturers, it is theoretically possible to cover the entire
spectral range of silicon imagers with about one LED every 10-20~nm 
(upper panel of figure \ref{fig:filter_coverage_skydice}) or 
to design some sort of ``optimal sampling'' of the passbands 
(upper panel of figure \ref{fig:filter_coverage_sndice}).

In this design, we chose to sacrifice wavelength
precision in favour of high-stability illumination.
This makes sense, since what one actually needs is a {follow-up}
more than an absolute {measurement} of the filter cutoff positions. The 
filter transmissions are measured well prior to installation, and 
our goal is instead to monitor any drift over the life time of the instrument. 
One of the aims of this paper is to estimate how precisely
one can locate a filter front with such an instrument, which delivers a coarse 
wavelength sampling. We show in section \ref{sec:analysis} that excellent precision may be attained with a small number of exposures,
as long as we can secure a precise model of the spectral intensity 
delivered by the light source.

As temperature increases, the LED emission efficiency
drops by about 0.5\%/\celsius, and the mean wavelength of the
emitted light shifts redwards by as much as 0.1 \AA/\celsius. 
We often refer to these variations as the ``cooler-brighter'' and ``cooler-bluer'' effects, respectively. As discussed in \S \ref{sec:test_bench}, these variations are
generally linear and extremely reproducible. As a consequence, once
each emitter has been characterised well, one only needs to implement
a real-time follow-up of the source temperature to account for these
effects.

{The temperature variations induced by the LED itself
  are small, if detectable.  Around 2007-2008, the typical power
  consumption of our LEDs operated at a low regime was of 50 to
  100~mW.  With the new generation of LEDs available today, it is
  closer to 10~mW.  A large fraction of this power is dissipated as
  heat, but it is easy to build a heat sink able to absorb all of it.
  We have verified that if a LED is correctly glued to a radiator with
  a heat-conducting glue, no noticeable temperature elevation of the
  radiator itself can be detected.  The only exceptions so far are
  less powerful LEDs (older models or UV LEDs), which have to be
  operated near their maximum power.  For those emitters, we found
  typical radiator temperature elevations of about 0.1\degree C, with
  a maximum of 0.2\degree C for a UV LED, after an hour of
  operation). }

\subsection{Light source design}

\begin{figure}
\centering
\includegraphics[width=\linewidth]{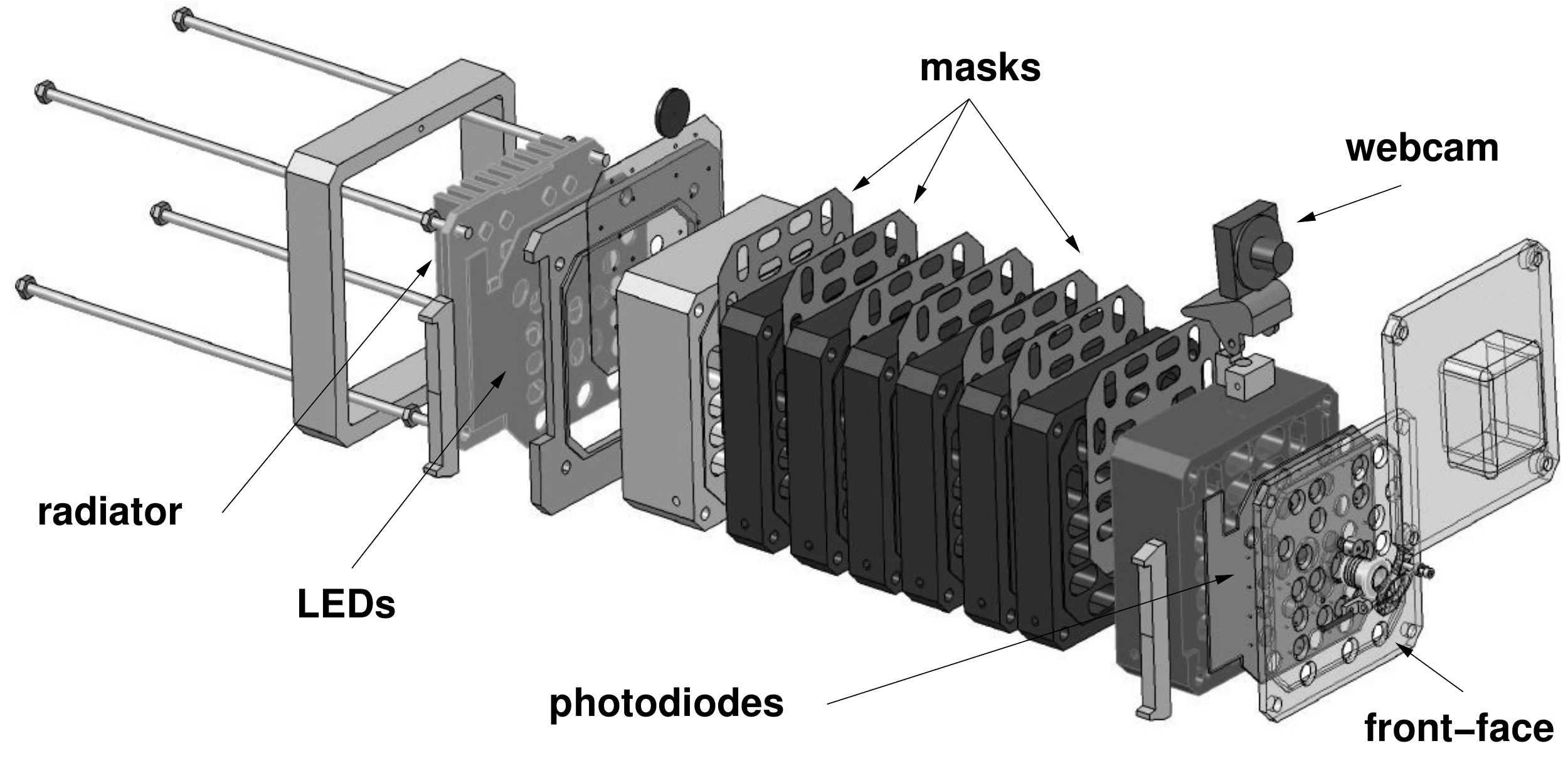}
\caption{Design of the DICE LED head. The LED head is made of 8
  aluminium blocks pierced with 24 holes corresponding to each of
  the 24 LED channels. Black
  masks are placed between each block to shape the beam
  and minimise the stray light. Each channel 
  is equipped with an off-axis control photodiode that
  monitors the light delivered by the LED in real time. The photodiodes 
  are placed on a board just behind the front face of the device. 
  The LEDs are in thermal contact with an aluminium radiator, in order to dissipate 
  waste heat and monitor the LED temperature.
  An additional central 25th channel 
  delivers a pencil beam used to control the relative
   orientation of the device with respect to the telescope. 
  \label{fig:head_design}}
\end{figure}

The mechanical design of the light source is illustrated in figure
\ref{fig:head_design}.  SnDICE and SkyDICE have very similar
designs.  SnDICE is a $\mathrm{150\ mm \times 150\ mm \times 300\ mm}$
modular box, made of eight almost identical anodised aluminium blocks, each
pierced with 25 apertures to let the light through. The LEDs are
located on the back of the device, about 260-mm from the front face.
The calibration beams exit through \diameter 9-mm apertures located on
the front face of the device.  This design permits  conical
2\degree wide beams to be generated.  SkyDICE, the second prototype, is shorter
($\mathrm{150\ mm} \times \mathrm{150\ mm} \times \mathrm{227\ mm}$)
because it needs to generate wider (3\degree) beams to cover the
larger $(5.7 \mathrm{deg}^2)$ field of view of the imager. The
modularity of the design allows us to adapt the aperture of the beam easily as a function 
of the imager field of view. 

The light source implements 24 calibration channels, each generating a
conical Lambertian beam, in order to cover the
$350\ \mathrm{nm} < \lambda < 1100\ \mathrm{nm}$ spectral range as evenly as possible. The
LED currents are chosen so that a beam generates about 1000
photoelectrons per second in each pixel. As the typical solid angle
subtended by a pixel is about $10^{-12}\ \mathrm{sr}$, this means
that the radiant intensity of the LED should generate of the order of 
$10^{15}\ \gamma/\mathrm{s/sr}$, which corresponds to about 0.5 mW/sr 
at 500-nm. 

The source is attached to the dome of the telescope. It can rotate
around an altitude and an azimuth axis, in order to control the
alignment of the LED beams with the telescope axis.  By moving the source and the telescope
simultaneously, it is possible to
illuminate nearly every region of the primary mirror.

\begin{figure}
\begin{center}
\includegraphics[width=\linewidth]{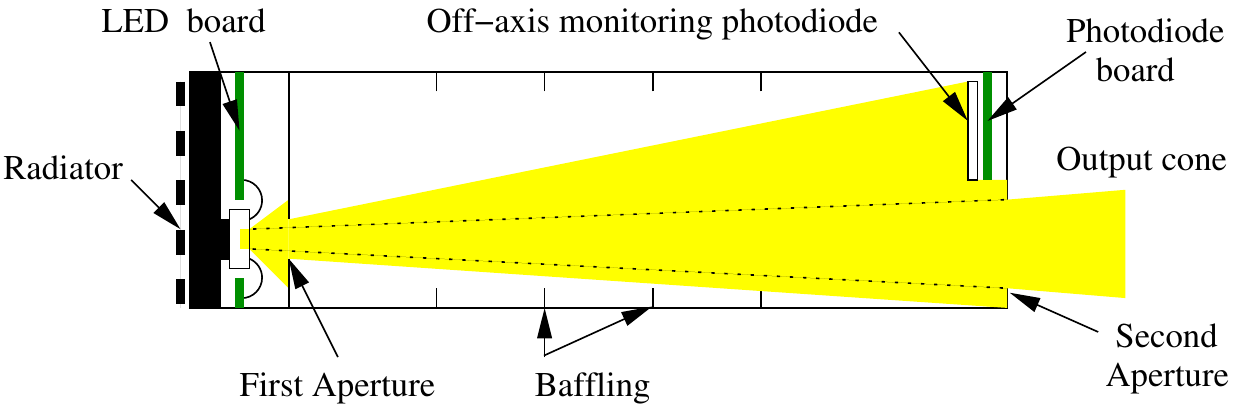}
\caption{Sketch of a calibration channel showing the emitting LED and the off-axis monitoring photodiode.
\label{fig:LED_channel}}
\end{center}
\end{figure}

A special LED channel, called the {\em \emph{artificial planet}} is used to
control the relative alignment of the device and the telescope.
The planet light is generated by a wide spectrum LED covering all the
filter passbands. The planet channel is equipped with a small
\diameter 10-mm convergent lens that transforms the isotropic LED
beam into a quasi-parallel beam.  Planet exposures produce a spot on
the imager focal plane, along with ghosts, owing to internal reflexions
between the optical components.  The position of the planet spot is a
direct measurement of the angle between the telescope optical axis
and the planet beam. It is measured with a
precision of about 3 \arcsec (the FWHM of the spot itself being of about 20 \arcsec). 
This precision is slightly better than the precision of
the motor axis encoders (0.0025 degree, 9 \arcsec).

\subsection{Generation of the LED currents}

The intensity of the light emitted by each LED is a function of the
current injected into it. The LED currents are generated (and
monitored in real time) by a custom-made backend board, connected to
the illumination system with an analogue link.  We attempted to
build a current source that achieves an electrical stability better than 
$10^{-5}$ over a temperature range of a few degrees. It is 
classically implemented as a transistor current source driven by a
programmable voltage level and stabilised with a negative feedback
loop \citep[see e.g.][p. 181]{1989arel.book.....H}. For redundancy, the feedback voltage level is also sampled at a
rate of a few kHz by a LTC1608 16-bit ADC and logged for offline
checks.

\begin{figure}
  \begin{center}
    \includegraphics[height=0.9\linewidth,angle=-90, trim=25mm 0mm 30mm 0mm, clip]{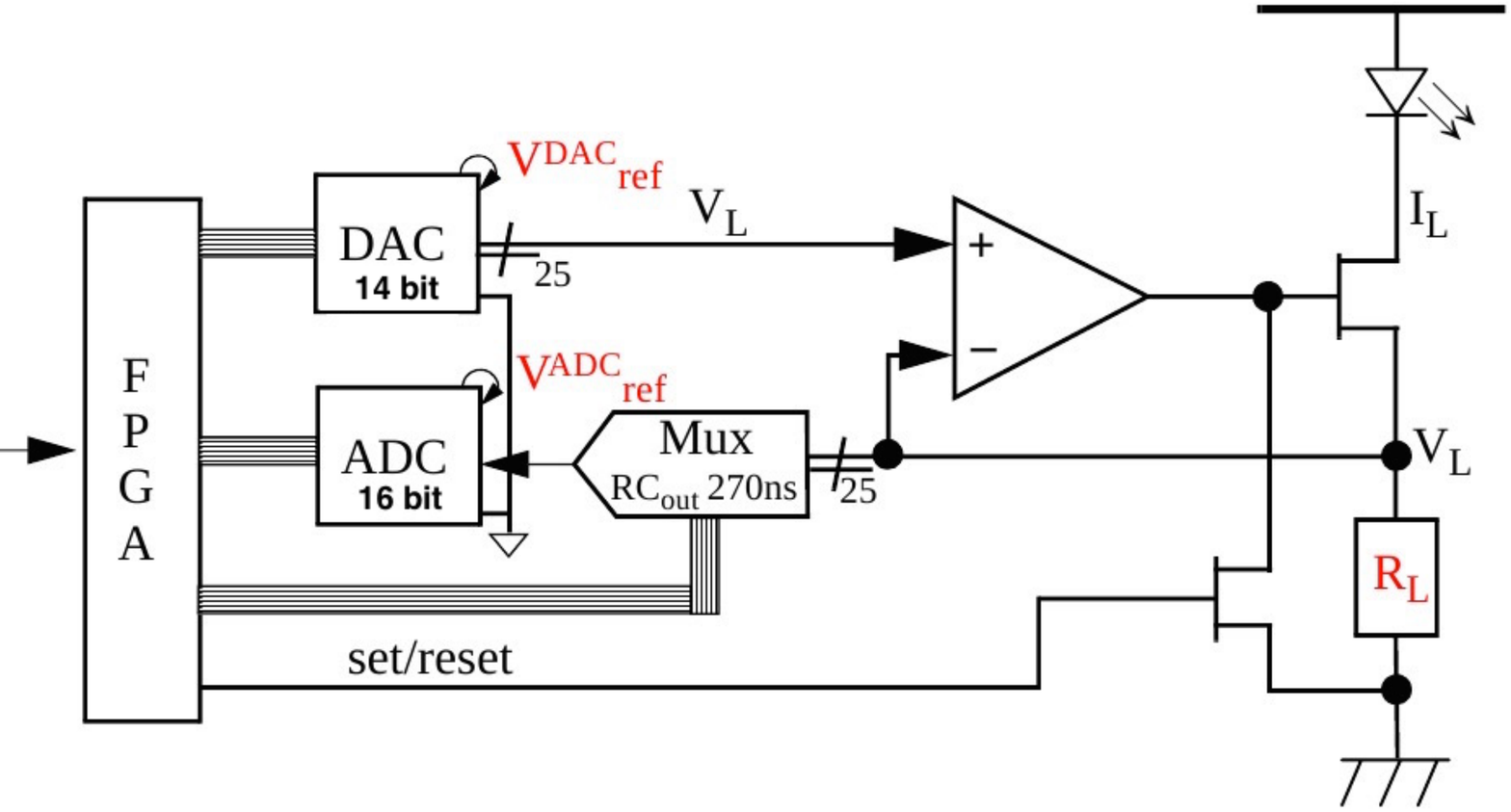}
    \caption{Schematics of the current source. }
  \end{center}
\end{figure}

\begin{figure}[t]
  \begin{center}
    \includegraphics[width=\linewidth]{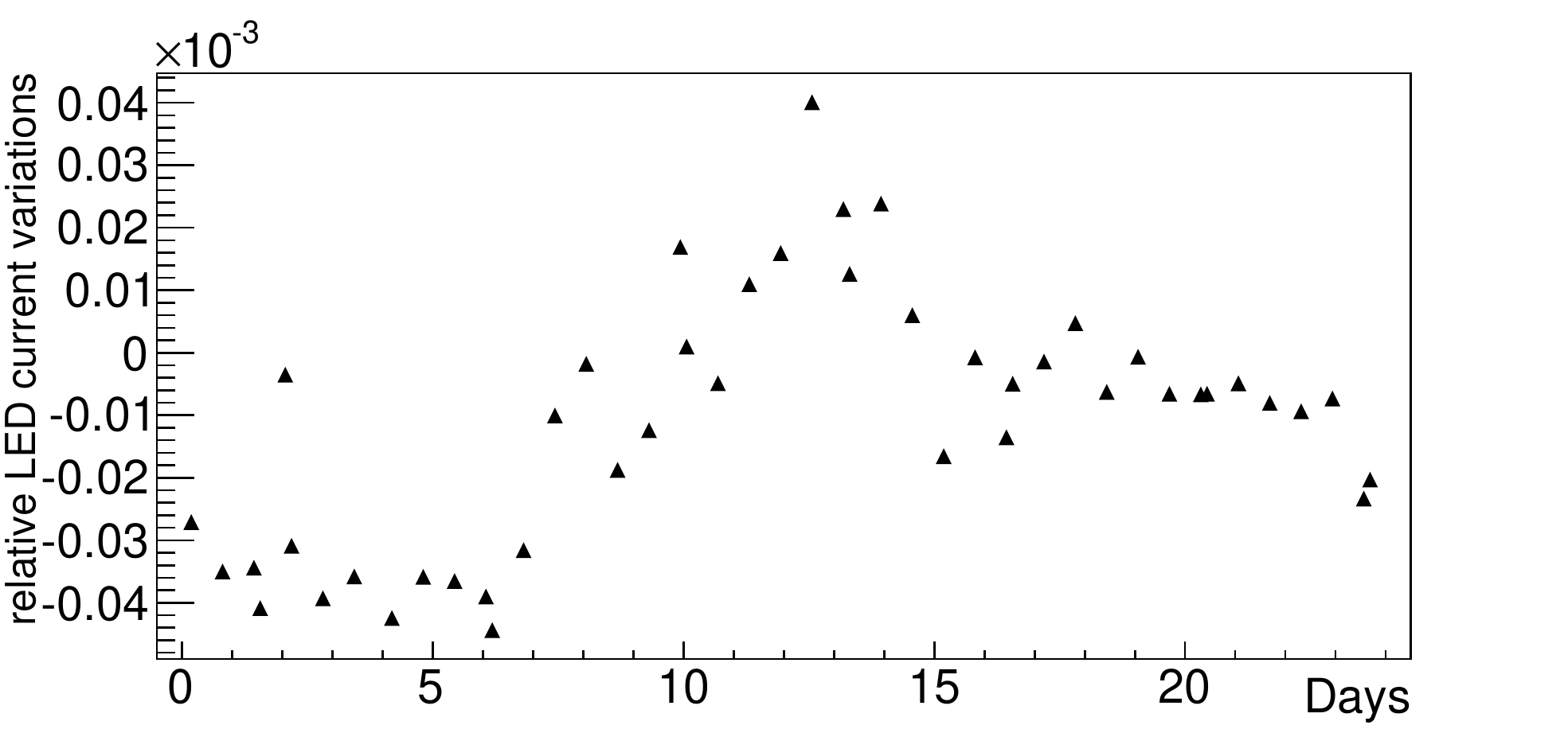}
    \caption{Relative variations in the LED current, measured as the voltage 
      drop across a resistor in series with the LED and monitored over a
      duration of 3 weeks. During this period, the LED head was operated without
      interruption on our spectrophotometric test bench. During this
      run the temperature varied by about 5\celsius. As can be seen,
      the current varies by a few $10^{-5}$ at most.}
    \label{fig:led_current_stability}
  \end{center}
\end{figure}

The accuracy and stability of this current source depends entirely on
a few high-quality commercial components, used in the board, namely the
voltage references of the ADC and DAC, the voltage division chains of
the ADC and DAC, and the serial resistors.  Extensive, long-duration
tests of the generated currents show that they are stable at the
$10^{-5}$-level (figure \ref{fig:led_current_stability}). As the
characteristics of the components may have a small temperature
dependence, the temperature of the backend board is monitored in real
time during data-taking.

\subsection{Redundancies}

Redundant checks are an essential part of the design.  No matter how
stable the source is on a test bench, we need to prove that the light
actually delivered during the calibration runs is not affected by
unexpected fluctuations or long-term drifts.  For this reason,
several critical quantities are monitored in real time during
operations.

The temperature of the LEDs is the most important of all of these.
What is measured in practice is the temperature of the radiator on
which all LEDs are glued with a thermally conductive
glue. For this purpose, we use a PT1000 thermistor glued to
the radiator.  What we monitor with this probe does not strictly correspond
to the temperature of the LED junctions.  It is, however, a surprisingly good proxy,
which allows us to empirically standardise each LED on a test bench
and use these standardisation relations later during operations.

Another critical point is the stability of the current source
itself. We characterise it with two observables. First, the temperature
of the backend board, which may have an impact on the current
generator, is monitored using a DS600 temperature sensor, mounted on
the board itself.  Second, the current actually delivered to the LED
(in practice, the voltage drop across a resistor
mounted in series with the LED) is also logged for offline checks.

Finally, we directly characterise the light actually delivered by the
LED by placing an off-axis monitoring photodiode in each LED channel,
close to the exit hole (see figure \ref{fig:LED_channel}).  We use
$\mathrm{5.8\ mm \times 5.8\ mm}$ Centronic\footnote{\url{http://www.centronic.co.uk}} 
OSD35-7 photodiodes, covering the full
wavelength range of the illumination system with good efficiency in
the UV. These control photodiodes are mounted on a board located just
behind the front panel of the instrument (see figure \ref{fig:head_design}).

All these quantities are digitised on the backend board at a
frequency that can be tuned (from 1kHz to 32 kHz), so that one can
study the possible sources of noise over a large band.  The digital
samples are stored in a 16MB buffer that is read out on demand by the
DAQ system (typically after each calibration frame).

Another miniaturised system, comprising a cooled, large area
photodiode coupled to a ultra-low noise current amplifier has been
built to monitor the light as it goes through the telescope optical
path. It has not been used extensively in our analysis. We describe it
in Appendix \ref{sec:clap}.

\subsection{Data acquisition}

The DAQ and control system of these light sources run on a dedicated 
industrial PC
(PC104), which can be installed a few metres away from the source. It
is connected to the backend board with a USB link and to the telescope
control system (TCS) with an ethernet link.  In its current
implementation, it hosts a lightweight server, which interacts with the
TCS through a variant of the {XML-RPC} protocol. The server communicates
with the backend electronics and relays the orders sent by the TCS. In
particular, it controls the LED current and the LED head orientation
motors, and it retrieves on demand the monitoring data stored on the
backend board and sends them to the TCS so that they can be stored
along with the calibration frames.

\subsection{Operations}

All calibration frames are taken during daytime in order not to
interfere with the telescope observing schedule.  This requires the
dome to be dark, or requires monitoring the ambiant luminosity with the imager
itself, complemented with an external device, such as one of the
modules described in Appendix \ref{sec:clap}.  In practice, we have
found that in the years 2008-2010, the dome of CFHT was very dark with
a contamination lower than $0.05$ ADU/s/pixel. For a 10 s exposure,
with a typical level of $\sim 5~000$ to $10~000$ ADU, this yields a
relative contamination by ambiant light of $5\ 10^{-5}$ to $10^{-4}$
at most.  The enclosure of SkyMapper is slightly less light-tight, but
the ambiant luminosity is also lower than $0.1$ ADU/s, giving relative
contaminations of a few $10^{-4}$.  {With such low contamination
  levels, we did not implement any correction at the pixel level.
  During a later series of calibration runs performed at the CFHT from
  January to August 2014, the dome was found to be significantly brighter,
  following the installation of venting apertures (several
  ADU/s/pixel).  The ambient luminosity was then monitored by
  interlacing the calibration frames with dark dome exposures taken
  in the same position with the LED turned off. This nearly doubled
  the duration of a typical calibration run from
  $\sim$ 50' to 1 h 40' (see Table
  \ref{tab:typical_dice_runs} of \S\ref{sec:calibration_runs}).}

A calibration run goes as follows.  The telescope and dome motion are
decoupled, and the telescope points inside the dome in the direction
of the illumination device.  The optical axes of both instruments are
then aligned using the planet beam (see section
\ref{sec:system_description}), as a guide.  Once the alignment of both
instruments is known, within an acceptable range, series of
calibration exposures may be taken using the main LEDs.

Several types of calibration exposures may be taken.  The stability of
the readout electronics is studied with repeated exposures of the same
LED, taking advantage of the $10^{-4}$ stability of the illumination
system.  The linearity of the imager is checked using illumination
ramps (i.e. exposures of the same LED of longer and longer exposure
time).  Finally, one measures the instrument passbands with series of
calibration frames taken with all the LEDs matching the passband under
study.

As we see in section \ref{sec:analysis}, calibrating one single
filter requires taking about four to eight exposures (each with a different
LED), of one to ten seconds each. The length of a calibration sequence is
therefore dominated by the imager readout time (40~s for MegaCam) and
never exceeds ten minutes.  As a result, it is possible to check the five
to six filters that equip MegaCam or SkyMapper in less than one hour.


\section{Spectrophotometric test bench: Overview}
\label{sec:spectrophotometric_test_bench}

\begin{figure*}[t]
  \begin{center}
    \includegraphics[width=\linewidth]{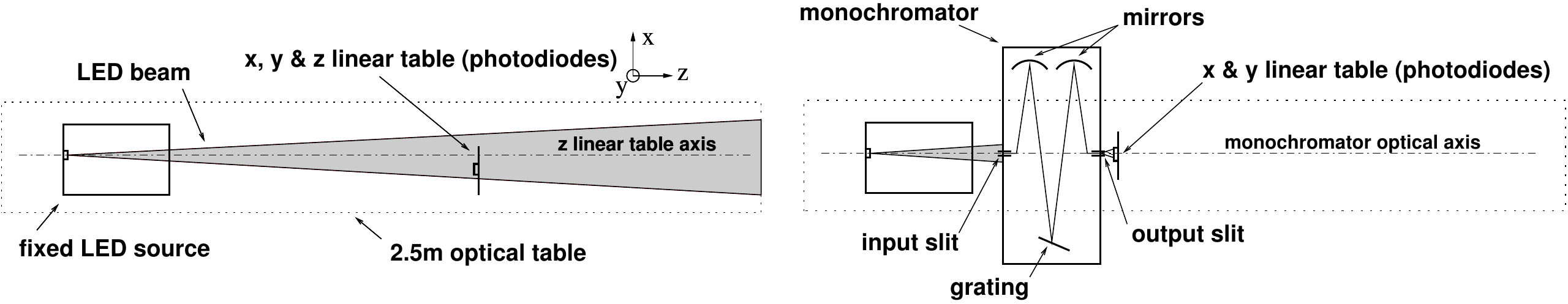}
    \caption{Spectro-photometric test bench. {\em Left panel:} photometric
      bench. The photodiode is placed on a linear table and moved in
      planes parallel to the z-axis to map each LED beam.  {\em Right
      panel:} a monochromator is inserted between the source and the
      photodiode.}
    \label{fig:calibration_test_bench}
  \end{center}
\end{figure*}

\subsection{Goal}

The primary goal of the test bench studies is to transfer the
calibration carried by the NIST photodiode to the light source (arrow
labelled \ding{202} in figure \ref{fig:nist_metrology_chain}).  In
practice this means that we want to characterise the emissivity of
each LED using the NIST photodiode as our primary standard.

The physical quantity which describes the emissivity of a point source
is its {\em \textit{\emph{spectral intensity}}}, ${\cal S}(\lambda, \vec{u})$. It is
defined as the power emitted by the source, per unit wavelength and
per unit solid angle (at a given wavelength $\lambda$ and in a given
direction $\vec{u}$).  Since the properties of many types of LEDs are
sensitive to temperature, we expect ${\cal S}$ to also be a
function of $T$: ${\cal S}(\lambda, T, \vec{u})$.  

Our goal is to build a smooth model $\hat{S}(\lambda, T, \vec{u})$ of
the LED's true spectral intensity ${\cal S}(\lambda, T, \vec{u})$, which is
valid in the temperature range $0\celsius < T <
+25\celsius$ and which is typical of what is measured in most telescope
enclosures.  Along with this model, we would like to build an
uncertainty model to 
account for
the finite precision of the test-bench measurements and the finite
stability of the source. These two ingredients will then allow us to
predict, with a known accuracy, the amount of calibration light
delivered by the source for any real calibration exposure.

To build such a model, we need to accumulate enough measurements at
temperatures typical of what is measured in a telescope enclosure.
We also need to
characterise the short-term and long-term stability of the source. In
the next sections, we describe the test-bench configurations, the
measurements that are taken, and the way we combine them to build a smooth
estimate of the LED spectral intensities.

\subsection{Test-bench setup}
\label{sec:test_bench}

A sketch of the calibration test bench is shown in figure
\ref{fig:calibration_test_bench}. 
The light source is placed on one end of a 2.5-m long optical table and mounted on an X-Y
Kinetic Systems KVP-100 linear table, with a total stroke length of
150-mm. 

A 1 cm$^2$ Hamamatsu S2281 Si photodiode was purchased from NIST to be
used as our primary standard.  Following the NIST procedures, it is
operated unbiased at room temperature\footnote{{ This is
    sometimes called ``photovoltaic mode'' as opposed to the
    photoconductive mode where the photodiode is reverse-biased.}}.
Depending on the bench setup (see below), the photocurrents generated
by the photodiode vary from a few nano-amperes (in photometric mode)
down to a few pico-amperes (in spectroscopic mode).  They are measured
with a Keithley 6514 feedback picoammeter.

The calibrated photodiode is mounted on a KVP-100 X-Y linear table
that allows for a stroke length of about 300-mm.  A third motorisation
permits moving it by about 1500-mm along the Z-axis.  The
repeatability of the KVP-100 tables is of 1 \micro m, allowing us to
control the relative positioning of the source and the
detector very precisely.

The test bench is placed in a $\mathrm{2\ m \times 2\ m \times
    3.5\ m}$ dark enclosure.  The enclosure walls are insulated, and
using a powerful air conditioning system, we manage to cool it down
to a temperature of $\sim$ 0\celsius. The bench is not strictly speaking thermalised, since its
temperature is neither regulated nor perfectly uniform within the whole
volume. We compensate for this by monitoring the temperature of the
key parts of the test bench 
using PT1000 thermistances or type-K thermocouples.

\subsection{Assumptions}
\label{sec:assumptions}
${\cal S}(\lambda, T, \vec{u})$ depends on four scalar parameters.  This
means that to pave the full parameter space, we need to accumulate a
large number of spectra.  
{However, we have verified on six LED models, chosen so as to
  cover the full spectral range and the full range of available LED
  technologies, that the spectra are essentially independent of the
  direction of emission $\vec{u}$ (see Appendix
  \ref{sec:led_illumination_isotropy} for a full discussion).  }
As a consequence, we assume that the spectral intensity of all the
LEDs that equip the DICE light sources can be written as
\begin{equation}
{\cal S}(\lambda, T) \times {\cal B}(\vec{u})
\end{equation}
where ${\cal S}$ is the spectral intensity of the LED in a
(arbitrary) reference direction, while ${\cal B}(\vec{u})$ is a
dimensionless function, which accounts for the variations in the beam
intensity as a function of the angle of emission. Here, ${\cal B}(\vec{u})$
is normalised to one in the reference direction.

In what follows, we refer to ${\cal S}(\lambda, T)$ as the LED
``spectrum'', keeping in mind that it is actually a spectral
intensity.  
The dimensionless quantity ${\cal B}(\vec{u})$ is called the ``beam
map''.

\subsection{Photometric and spectroscopic measurements}
\label{sec:photometric_and_spectroscopic_measurements}
The beam maps ${\cal B}$ can be measured simply by intercepting the
beam with a calibrated photodiode placed at a known distance from the
source and moved with respect to the source, in order to sample the
whole beam.  The measurement of ${\cal S}$ is a little more
complex. We need to perform spectroscopic measurements by inserting a
monochromator between the source and the calibrated photodiode.

The calibration of the light source is therefore performed in two
distinct steps.  First, we simply map the radiant intensity of each
calibration beam by moving the standard photodiode in a series of
planes orthogonal to the Z-axis. These calibration sequences, called
hereafter ``photometric calibration sequences'', are performed at about
10 to 15 different temperatures, ranging between $\sim 0\celsius$ and
$\sim 25\celsius$.  They allow us to study how the intensity delivered
by each LED varies with temperature and how the intensity varies with
the direction of emission (beam map).  They also permit us to assess the
stability of the source. The analysis of this dataset is discussed in
section \ref{sec:photometric_calibration}.

In a second step, we insert a {\em Digikr\"om DK240} Czerny-Turner
monochromator between the light source and the calibrated photodiode.
The LEDs are positioned in turn in front of the monochromator
entrance slit, while the photodiode senses the intensity that comes out of the exit slit.  These
measurements are performed at about ten distinct temperatures between
0\celsius\ and room temperature.  This spectroscopic dataset, combined
with the photometric measurements described above, allows us to derive
smooth models of the LED spectral intensity.  The spectroscopic
measurements are described in section
\ref{sec:spectroscopic_calibration} with additional details given in
appendix \ref{sec:spectroscopic_calibration_details}.


\section{Photometric calibration of the light source}
\label{sec:photometric_calibration}

The goal of the photometric calibration studies is to determine the
absolute normalisation of the LED spectral intensities ${\cal
  S}(\lambda, T)$ and also to measure the beam maps ${\cal
  B}(\vec{u})$ (see section \ref{sec:assumptions}).  In this
configuration of the bench, the calibrated photodiode samples directly
the beam light (left panel of figure
\ref{fig:calibration_test_bench}). The photocurrent registered with
the Keithley~6514 picoammeter can then be written as the product of
two simple quantities:
\begin{equation}
  I_{|\mathrm{phot}} = {\cal B}(\vec{u}) \times {\cal J}(T) 
\end{equation}
where ${\cal B}(\vec{u})$ is the beam map defined in section \ref{sec:assumptions}, $T$ is the LED temperature 
(or at least a proxy for it), and ${\cal
  J}(T)$  the photocurrent generated when the photodiode is placed
at a specific reference position $\vec{r_0}$ with respect to the
source.  At this reference position, ${\cal B}$ is conventionally set to 1,
the photodiode subtends a solid angle $\delta\Omega_0$, and we have
\begin{equation}
{\cal J}(T) = \delta\Omega_0 \times \int \eta(\lambda)\ {\cal S}(\lambda, T)\ d\lambda,
\end{equation}
{where $\eta(\lambda)$ is the photodiode efficiency reported by NIST}. The main difficulty is to control the relative
positions and orientations of the photodiode with respect to the
source. 

Since acquiring detailed beam maps is time consuming ($\sim$ 30 to 60 minutes per map),
we optimise the photometric measurements as follows.  In a
first series of measurements, we concentrate on a few specific beam
locations, keeping the photodiode fixed, while varying the bench
temperature.  These sequences, called ``minimaps'' are much faster to acquire (about 260 seconds)
and yield about ten independent measurements of each selected beam location 
(in particular, the central region of the beam, taken as a reference). 
They are targeted at
measuring ${\cal J}(T)$ for each LED, i.e. the relative variations in
the LED emission with temperature. The analysis of the minimaps is
presented in section \ref{sec:minimaps} below.

We also realise fine-grained maps of the calibration beams at two or
three temperatures, keeping the temperature of the bench as constant
as possible during data taking.  From this data, we obtain the 
${\cal B}(\vec{u})$ maps described above, and we verify that these maps are stable with
temperature (see section \ref{sec:beam_maps}).  Such measurements are
taken at several distances to the source, in order to verify the
projectivity of the beam against scattered light and to check that we control the LED-source
geometry well.

\subsection{Minimaps: LED emission versus temperature}
\label{sec:minimaps}

\begin{figure}
  \begin{center}
    \includegraphics[width=\linewidth]{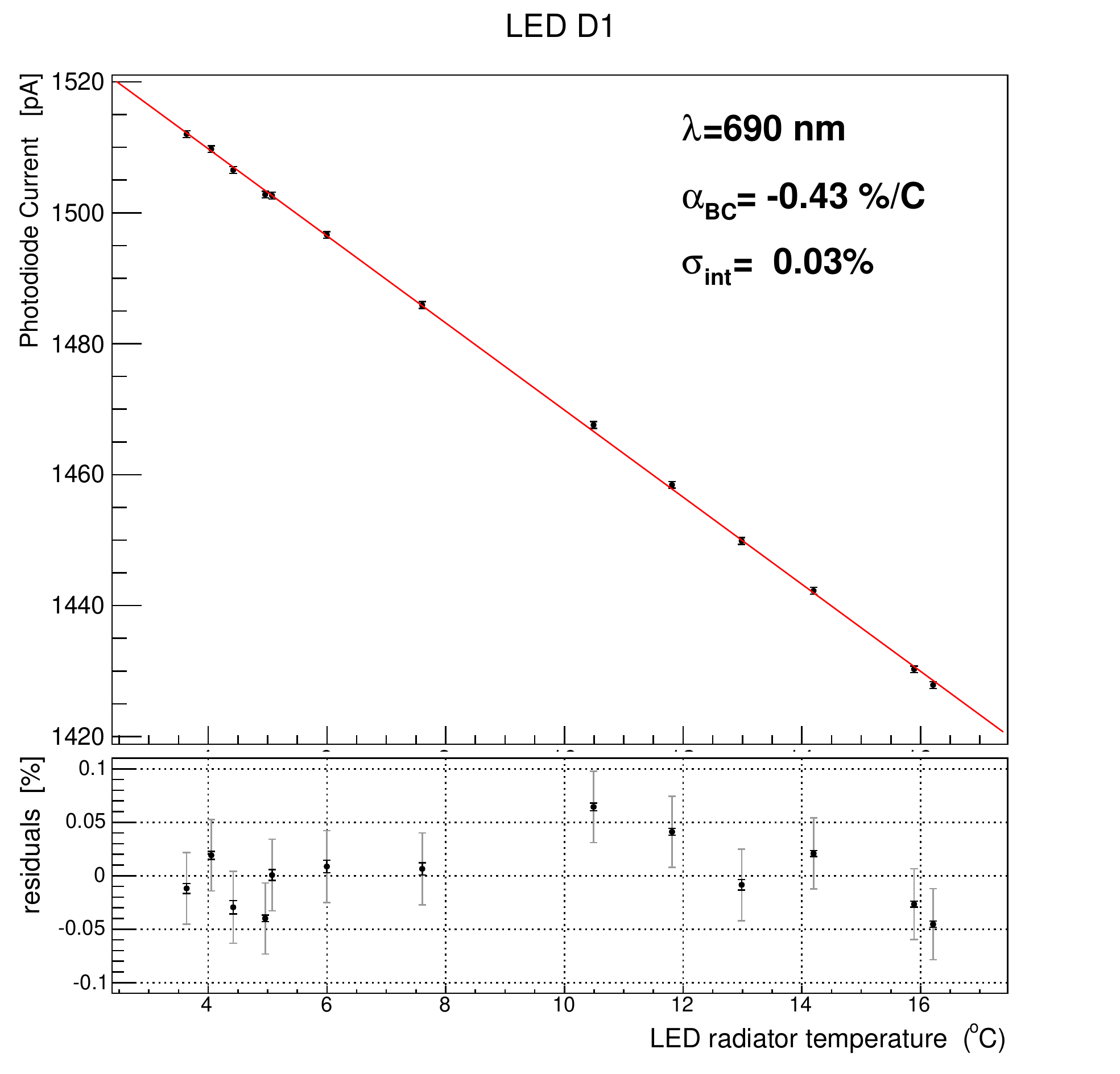}
    \caption{Top panel: Cooler-brighter relation measured from 13
      minimaps for a Roithner-Lasertechnik APG2C1-690 emitter,
      mounted on SkyDICE.  Bottom panel: residuals. The small black error
      bars display the uncertainties on the minimap measurements
      (combined).  The larger grey error bars take the
      measurement repeatability between maps into account (i.e. over time scales of
      one hour).}
    \label{fig:minimaps_iphot_vs_T}
  \end{center}
\end{figure}

In most cases, the LED emissivity decreases with temperature.  
We refer to this as the ``cooler-brighter'' relation already mentioned above. 
It generally obeys a linear law:
\begin{equation}
  {\cal J}(T) = \alpha_{BC} \times (T_{\mathrm{LED}}-T_0) + {\cal J}_0
\end{equation}
where ${\cal J}(T)$ is the photodiode
current defined above,$T$  a proxy for the LED
temperature, typically the temperature of the LED radiator, and $T_0$ is
an arbitrary temperature pivot.

\begin{figure}
  \begin{center}
    \includegraphics[width=\linewidth]{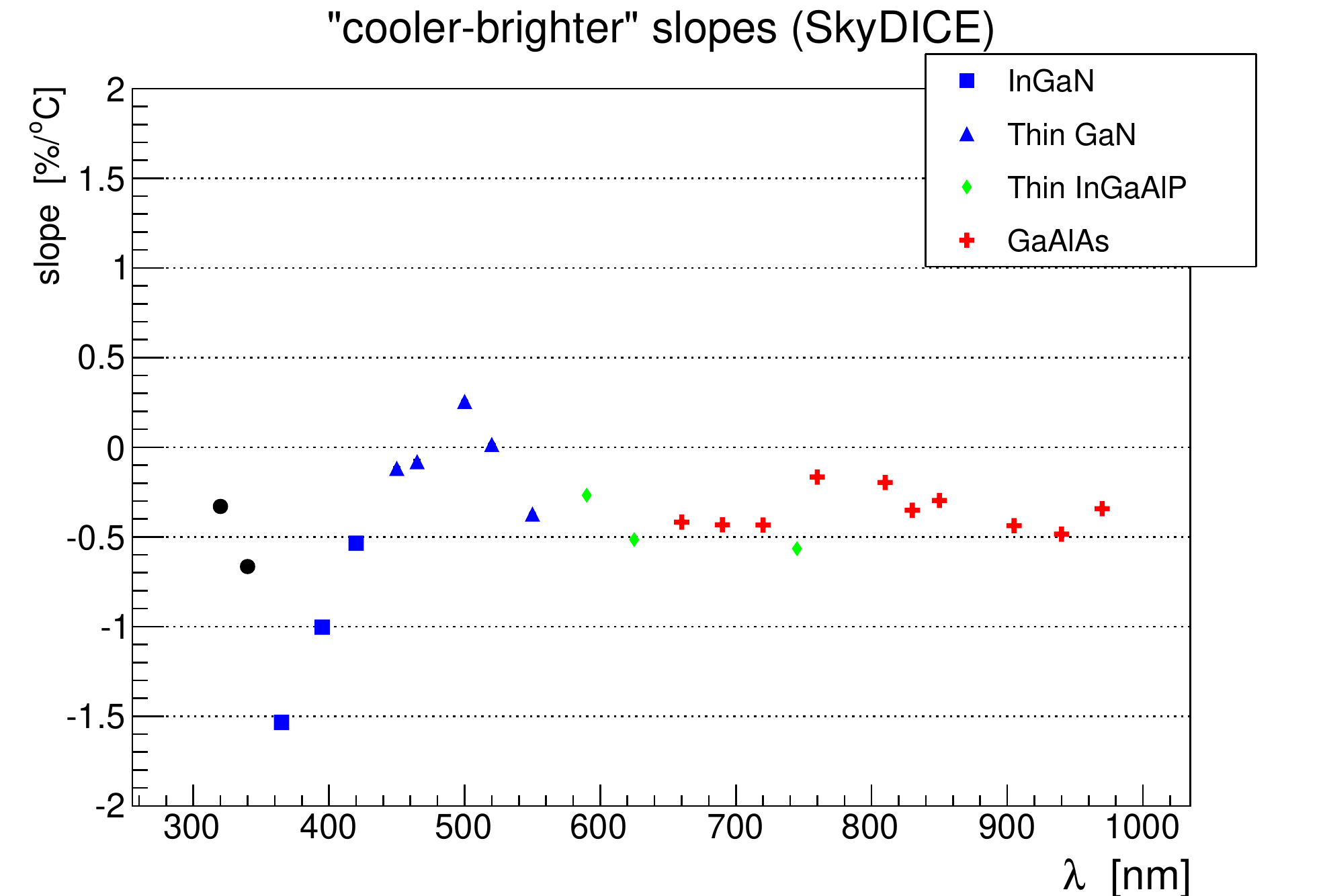}
    \caption{Cooler-brighter slopes for the SkyDICE LEDs.  The various
      LED technologies are coded with different symbols (see Appendix
      \ref{sec:led_selection} for details). The amplitude of the effect
      is similar for all LEDs at about $0.5\%/\celsius$.  For two LEDs, which are operated at very low currents, we
      actually observe a ``brighter-warmer'' relation.}
    \label{fig:brighter_cooler_slopes}
  \end{center}
\end{figure}

Ten to fifteen minimaps are taken for each LED at temperatures ranging
from 0\celsius\ to 25\celsius.  The cooler-brighter law 
is fitted on the measurements of the central region of the beam.
An example of the fit, along with the residuals to the linear law, is shown in
figure \ref{fig:minimaps_iphot_vs_T} for one of the LEDs that equips
SkyDICE. 

Our determinations of the LED radiant intensities at 25\celsius\ and
of the cooler-brighter slopes are reported in Tables
\ref{tab:led_properties} and \ref{tab:skydice_leds} in the appendix for
SnDICE and SkyDICE, respectively. Whenever possible, the SkyDICE LEDs
have been tuned to deliver of the order of 0.1 mW/sr. On the test
bench, placing the 1~cm$^2$ NIST photodiode about 2 m away from the
source, this translates into typical photodiode currents of a few
$10^3$ pA, as shown in figure \ref{fig:minimaps_iphot_vs_T}. Such a
current level is easily measured with a picoammeter.
In figure \ref{fig:brighter_cooler_slopes} we summarise the intensity of the
cooler-brighter effect for all the LEDs mounted on SkyDICE.  On
average, it is slightly lower than 0.5\%/\celsius\ for all LEDs (regardless 
of the LED technology), 
except for two blue InGaN emitters that can reach over 1\%/\celsius.
We notice that two LEDs emitting around 500-nm actually exhibit a
{\em \emph{warmer}}-brighter relation.  
It may be because the nominal currents that were chosen for these
LEDs are below the normal range recommended by the vendor.  What
matters for our application is that the effect is well measured and reproducible.

Only one of the LEDs mounted on SkyDICE clearly displays small
deviations from a linear cooler-brighter law. It is a Golden
Dragon\textsuperscript\textregistered\ LED of type LD W5AM, emitting
at $\sim 450$-nm.  The intensity of the effect is small ($\lesssim
0.2\%$ peak-to-peak over the full temperature range), and we model it
with a second-order polynomial.  For all the other LEDs, it is not
possible with the data in hand to distinguish between small deviations
from linearity from the bench or source instabilities.

Special care has been taken into evaluating the
measurement repeatability empirically on various time scales.  The black error
bars in the lower panel of figure \ref{fig:minimaps_iphot_vs_T}
display the empirical variability of the central beam region, measured
on each single minimap (i.e.  over about $260\ \mathrm{seconds}$).  It
is about $0.1\ \mathrm{pA}$ (about $5\ \mathrm{nW/sr}$) and depends
slightly on whether the cooling system is on or off.  When combining
the ten minimap measurements together, it represents a negligible
contribution to the flux uncertainty: about 1.5 nW/sr, for a nominal
flux of about 0.1 mW/sr, i.e. a few $10^{-5}$.  The grey error bars in
the same figure represent the map-to-map variability.  This
contribution depends on the LEDs, so we attribute it to some
longer term-variability of the source, and to a minor extent, to
some additional variability of the bench. We model it with a noise
pedestal $\sigma_{\mathrm{int}}$, adjusted iteratively to obtain a
reduced $\chi^2$ of unity.  For most LEDs, it represents a little less
than $5\ 10^{-4}$ of the LED nominal flux.  For the two faintest UV
LEDs, it is slightly larger ($2\ 10^{-3}$).

\subsection{Long-term stability of the DICE illumination device}

\begin{figure}[t]
  \begin{center}
    \includegraphics[width=\linewidth]{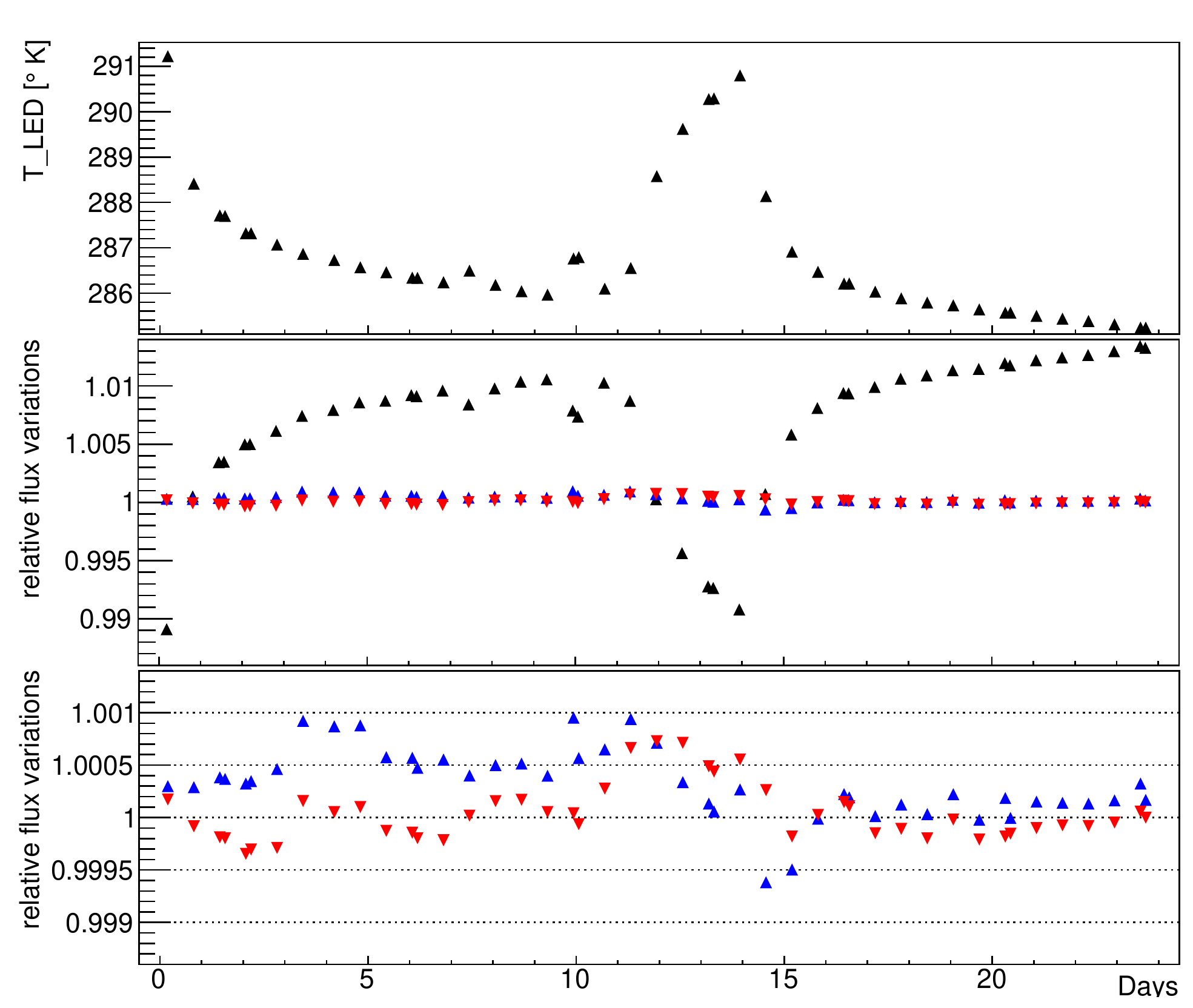}
    \caption{Upper panel: variations in the source temperature over a
      3-week run designed to test the system stability. Middle panel:
      black triangles: relative variations of the raw flux delivered
      by the source. Blue lower triangles: relative variations in the
      flux after correction for the cooler-brighter effect. Red upper
      triangles: relative variations in the flux after correction by
      the control photodiode measurements.  Lower panel: zoom on the
      corrected flux baselines.}
    \label{fig:led_flux_baseline}
  \end{center}
\end{figure}

In the previous section, we have shown, from the study of the
residuals to the brighter-cooler law, that the source is stable at the
level of a few $10^{-4}$~{(rms)} over the duration of a calibration run (i.e.
24 hours).  Since the stability of the light source is a critical
aspect of the design, it has been checked over longer durations (i.e.
weeks). 
For these tests, the bench was configured in
photometric mode, and repeated minimaps of each calibration channel
were acquired.  The system cycles through all LEDs over and over again
for a little more than three weeks.
The tests were performed at room temperature, each LED being fed with
its nominal current.  All the auxiliary quantities (LED temperature,
backend board temperature, control photodiode current, etc.) were
logged, along with the NIST photodiode measurements.

Figure \ref{fig:led_flux_baseline} shows the flux baseline that was
acquired for a typical LED, along with the LED temperatures.
Significant, anti-correlated variations of the temperature ($\sim 5
\celsius$) and the LED flux ($\sim$ 2\%) are noticeable over the
course of the run. This is the consequence of the cooler-brighter
relation studied in the previous section.

\begin{figure}[t]
  \begin{center}
    \includegraphics[width=\linewidth]{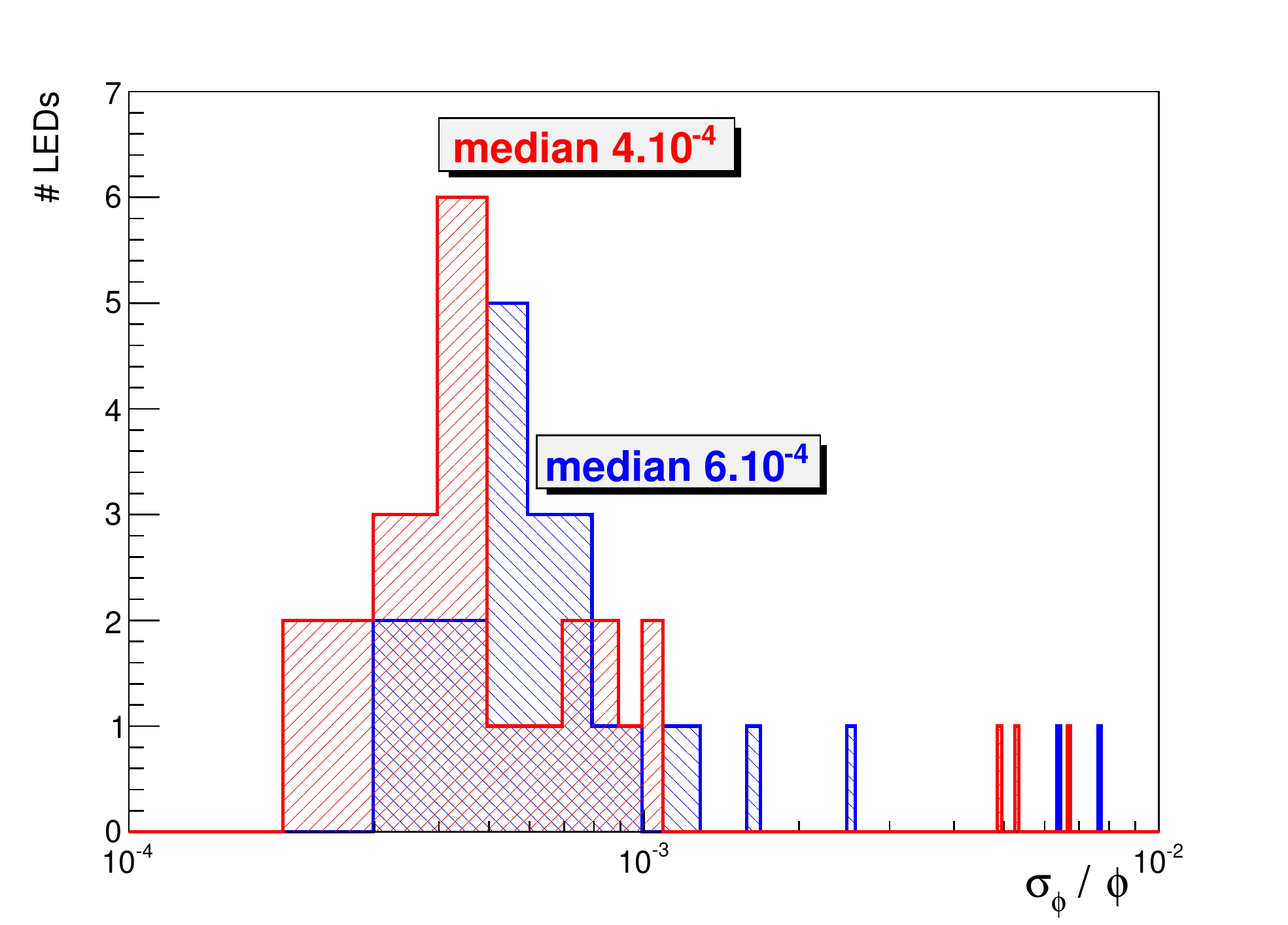}
    \caption{Red histogram: stability of the LED flux (i.e. the NIST photodiode current) after
      correction for the cooler-brighter effect (and for the backend
      board temperature variations). Blue histogram: stability of the
      LED channels after correction for the control photodiode
      measurements. The two outliers above $5.  10^{-3}$ are LEDs in
      the blue histogram that correspond to channels that are
      intrinsically unstable.  The outliers in the red histogram
      correspond to the same LEDs, plus one LED whose control
      photodiode is faulty. }
    \label{fig:stability_after_temperature_standardization}
  \end{center}
\end{figure}

We can take advantage of the cooler-brighter effect to predict the
variations in the LED flux, since we monitor the temperature of the
LED radiator $T_{\mathrm{LED}}$.  We also measure the temperature of
the backend board that generates the LED currents, $T_{\mathrm{bb}}$,
which has been found to correlate slightly with the LED flux for some
channels.  We use these two variables to predict the variations in the
LED flux over the long run:
\begin{equation}
  \Delta\tilde\varphi_{LED} = a_{\mathrm{LED}} \times \delta T_{\mathrm{LED}} + a_{\mathrm{bb}} \times \delta T_{\mathrm{bb}}
.\end{equation}
We find that $T_{\mathrm{LED}}$ and $T_{\mathrm{bb}}$ permit
parametrizing most of the variations in the LED flux over time as shown
in figure \ref{fig:led_flux_baseline}. We show the residual
dispersions attained over three weeks of continuous monitoring in
figure \ref{fig:stability_after_temperature_standardization}.  It is in the range $2. 10^{-4} < \sigma_{\tilde\varphi_{LED}} <
10^{-3}$~{(rms)} with a median value of $6\ 10^{-4}$~{(rms)}.  Only two LEDs are less
stable than $5\ 10^{-3}$. One is a UV LED (340-nm) with a large
threshhold voltage (about 5V).  Such voltages are slightly higher than
what our current source was designed for, and it seems to become less
stable in this regime. Fortunately, the modern UV LEDs on the market
require lower threshhold voltages, so we should not have to deal
with this problem in the future. The other unstable LED, a red APG2C1-760
distributed by Roithner-Lasertechnik, has displayed a true instability
during the run, its flux jumping by almost 2\% about two weeks after
the beginning of the operations.  This behaviour has not been explained yet
and has not been seen with any other LED. It has no counterpart on the
monitoring data, besides the LED control photodiode placed close to
the exit hole (see Appendix \ref{sec:on_led_9}).

We have developed an alternate and redundant standardisation method that relies 
on the direct measurements of the LED fluxes performed with the off-axis control
photodiodes.  Since the photodiode currents are digitised on the
backend board, we also include the backend temperature in the
standardisation relation:
\begin{equation}
  \Delta\hat\varphi_{LED} = a_f \times \delta \phi_{pd} + a_{\mathrm{bb}}' \times \delta T_{\mathrm{bb}}
.\end{equation}
Again, we observe very low residual dispersions of the standardised
flux ($2\  10^{-4} < \sigma_{\hat\varphi_{LED}} < 10^{-3}$) {(rms)} with a median
value of $4\ 10^{-4}$~{(rms)}, as shown in figure
\ref{fig:stability_after_temperature_standardization}.  Again, there
are a few outliers: the two LEDs discussed above, as well as a third
LED whose control photodiode seems to be malfunctioning.  These channels still display repeatabilities that are better than 1\%
over  weeks.

We plan to extend the duration of these intensive stability tests to
longer durations of the order of a few months. In a near future, we
will also re-calibrate the sources that are currently in operation
(SnDICE at CFHT and SkyDICE at SSO).  This will allow us to check the
stability of their calibration over durations of a few years.

As a conclusion, we have built a calibrated source that
is intrinsically stable at the $10^{-2}$-level. The fluctuations of the
light delivered by each LED can be parametrised as a function of 
simple auxiliary observables, notably the temperature of the LED radiator, $T_{\mathrm{LED}}$. Using
two independent methods (one based on monitoring $T_{\mathrm{LED}}$ and 
another based on the measurements performed with the off-axis control
photodiodes), we have shown that we can correct for the variations of the
source at the level of a few $10^{-4}$. {During operations, we generally choose to 
use the temperature method as our primary standardisation technique and to use
the control photodiode for redundancy checks. The photodiode and temperature probes are read
before and after each calibration exposure, therefore either of the two techniques 
can be used for each exposure. Since both methods 
give similar results, they can be interchanged in the future. } 

Because of its stability, our light source 
qualifies as a ``standard light source''. Once calibrated, it may be used to
disseminate a flux calibration to remote locations.

\subsection{Beam maps: LED emission versus direction}
\label{sec:beam_maps}

\begin{figure}[h]
  \begin{center}
    \includegraphics[width=\linewidth]{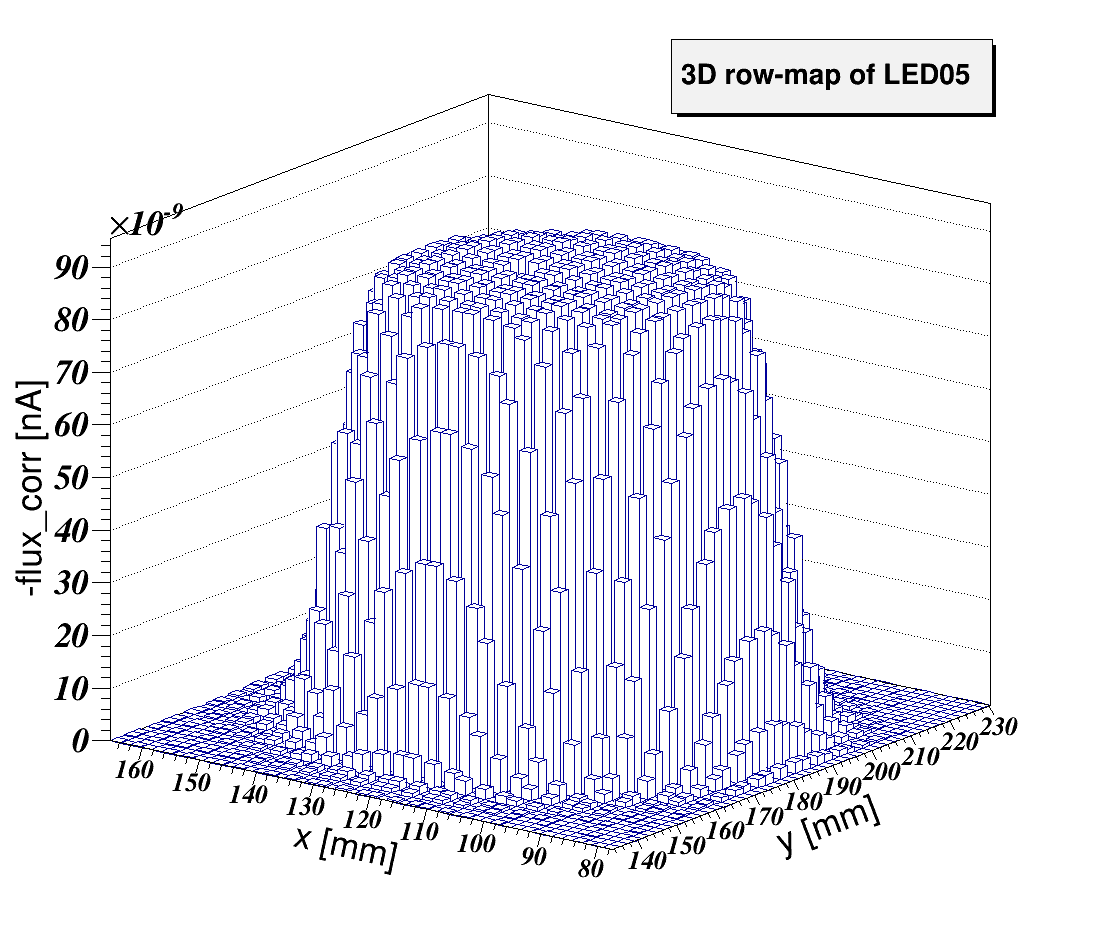}
    \caption{Beam map of LED {LVW5AM} (OSRAM, Golden
      Dragon\textregistered\ series) mounted on the SkyDICE device.}
    \label{fig:led_beam_map}
  \end{center}
\end{figure}

We now describe how we derive fine-grained beam maps (${\cal B}$) from
the detailed photometric scans realised at nearly constant temperature 
(figure \ref{fig:led_beam_map}).
We have found that the relative variations in the LED emissivity with
direction are small (1\% at most) and smooth.  They sometimes display a
complicated shape, which does not suggest an analytic parametrisation. 
We therefore develop them on the basis of splines:
\begin{equation}
{\cal B}(\vec{u}) = \sum_p \beta_p B_p(\vec{u})
;\end{equation}
and we fit for the $\beta$ parameters. The main source of systematics
here is the determination of $\delta\Omega$, i.e. the control of the
test-bench geometry. 

The stability of the beam maps themselves has been studied in detail
over long durations. The temperature fluctuations may indeed dilate
the metal, hence slightly change the relative positioning of the LED
with respect to the masks and pieces of baffles that shape the
beam. We have found that the fluctuations of the map profiles are
always lower than 5 $10^{-4}$.


\section{Spectroscopic calibration of the light source}
\label{sec:spectroscopic_calibration}

We now turn to the determination of the LED spectra.  This requires
additional measurements, which are performed on our spectroscopic test
bench (see right panel of figure \ref{fig:calibration_test_bench}).  This bench is similar to the photometric test bench described
above, except for the presence of a Czerny-Turner type monochromator,
inserted between the LED source and the calibrated photodiode.  In
this configuration, the photocurrent generated by the photodiode is
then given by
\begin{equation}
  I_{|\mathrm{spec}} = \left[\eta(\lambda) \cdot T_m(\lambda) \cdot {\cal S}(\lambda, T)\right] \otimes W_m(\lambda) 
  \label{eqn:i_spec}
\end{equation}
where $\eta(\lambda)$ is the photodiode efficiency reported by NIST,
$T_m(\lambda)$ and $W_m(\lambda)$ are the transmission
and the spectral response of the monochromator, respectively.

The key point here is the control of the monochromator
calibration. From the equation above, we see that we need to check (1)
the wavelength calibration of the monochromator, (2) its transmission
$T_m(\lambda),$ and (3) its spectral response, $W_m(\lambda)$.  This
work is described in detail in Appendix
\ref{sec:spectroscopic_calibration_details} and briefly summarised in
section \ref{sec:monochromator_systematics} below.

\subsection{Modelling the LED spectra}
\label{sec:modeling_led_spectra}

Our goal is to build a smooth model of the LED spectral intensities
as a function of wavelength and temperature. We choose to develop 
this model on the basis of two-dimensional B-splines:
\begin{equation}
  \hat{S}(\lambda, T) = \sum_p \theta_p B_p(\lambda,T)
\end{equation}
where the $B_p(\lambda, T)$ functions are two-dimensional splines of order 
3 to describe the wavelength
variations, and order 2 to model the temperature
variations.
We use three nodes placed every 10\celsius\ 
to describe the temperature and O(100) nodes, placed every 2-nm to capture the spectral shapes.

Because the uncertainty on the absolute normalisation of the monochromator
transmission is difficult to assess, we decided to fit for the
(unknown) normalisation $f_s$ of each spectrum measurement $s$.  This
means that the only piece of information we extract from the spectroscopic
measurements is related to the spectrum {\em \emph{shapes}} and not the spectrum
normalisations.  Our model of the photodiode current (equation
\ref{eqn:i_spec} above) becomes
\begin{equation}
  I_{|\mathrm{spec}} = f_s \times  \sum_p \theta_p\ \left[\eta(\lambda) \cdot T_m(\lambda) \cdot {B}_p(\lambda, T)\right]  \otimes W_m (\lambda) 
  \label{eqn:i_spec_developed}
.\end{equation}
By construction, there is a perfect degeneracy between the $\theta_p's$ and the $f_s's$. 
We break it by
requiring that the absolute normalisation of the $\hat{S}$ models is
determined by the photometric measurements (performed without a
monochromator).  In practice, this is done by adding the following
term to the fit $\chi^2$:
\begin{equation}
  \sum_s w_s \cdot \left({\cal J}\left(T_s\right) - \int \eta(\lambda) \cdot \hat{S}(\lambda, T_s)\ d\lambda\right)^2
  \label{eqn:constraints_on_the_spectrum_normalization}
\end{equation}
where ${\cal J}(T)$ is the photocurrent measured in a reference
position, as defined in section \ref{sec:photometric_calibration}.
With this approach, the uncertainty affecting the monochromator
transmission is entirely absorbed, and the absolute normalisation of
the LED spectral intensity models is set exclusively by the
photometric measurements. 

We estimate the LED spectral intensities by fitting the model above on
the photodiode current measurements and marginalizing on the
$f_s$ nuisance parameters.

\subsection{Results}

\begin{figure}
  \begin{center}
    \includegraphics[width=\linewidth]{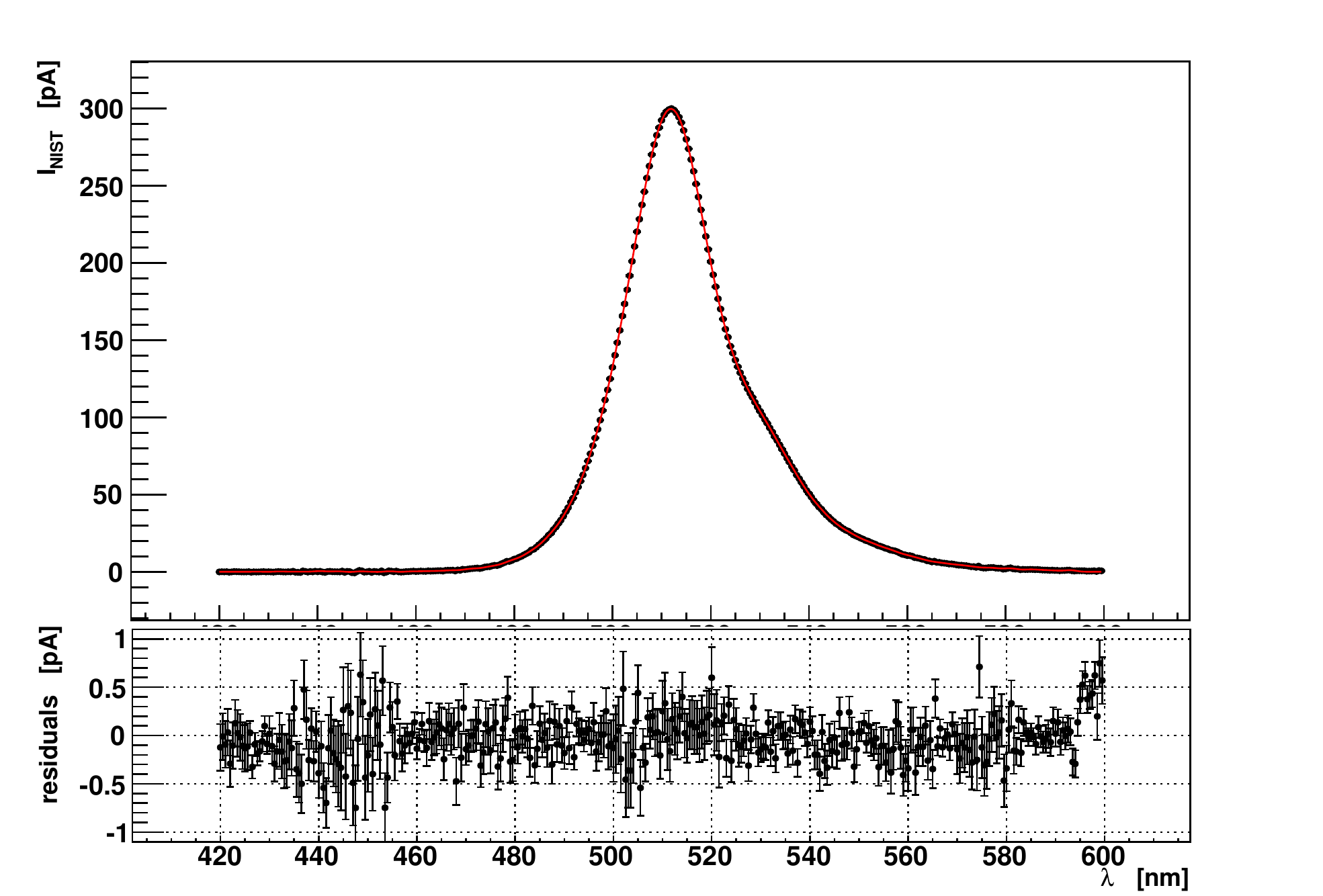}
    \caption{Upper panel: Comparison between a typical LED spectrum
      (SkyDICE Golden Dragon\textsuperscript\textregistered\ LV W5AM)
      and the fitted model (red line). Bottom panel: residuals.}
    \label{fig:spectral_model_residuals}
  \end{center}
\end{figure}

\begin{figure*}
  \begin{center}
    \mbox{
      \subfigure[Golden Dragon\textsuperscript\textregistered\ LT W5SM]{\includegraphics[width=0.5\linewidth]{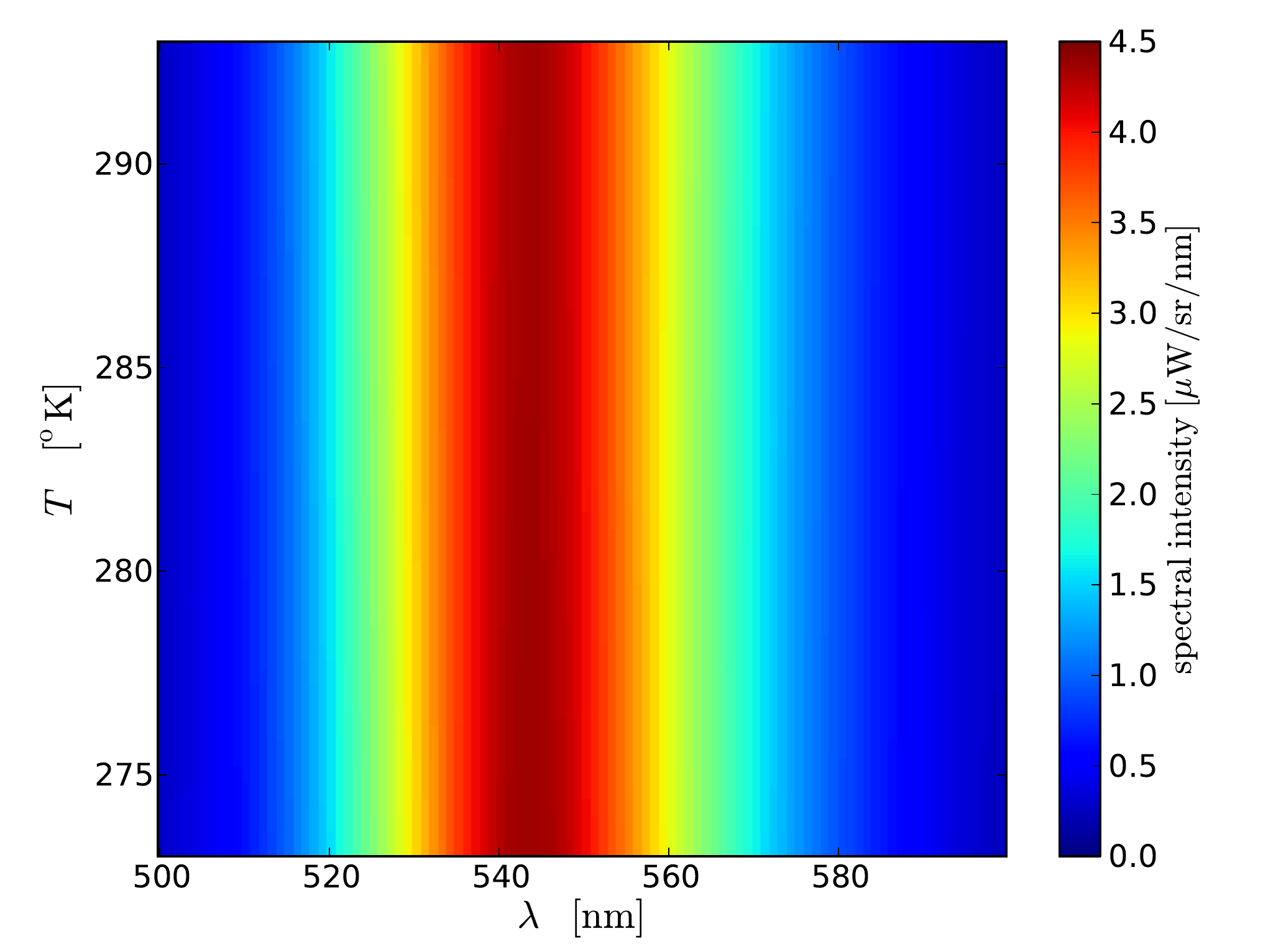}}
      \subfigure[APG2C1-850 (Roithner LaserTechnik)]{\includegraphics[width=0.5\linewidth]{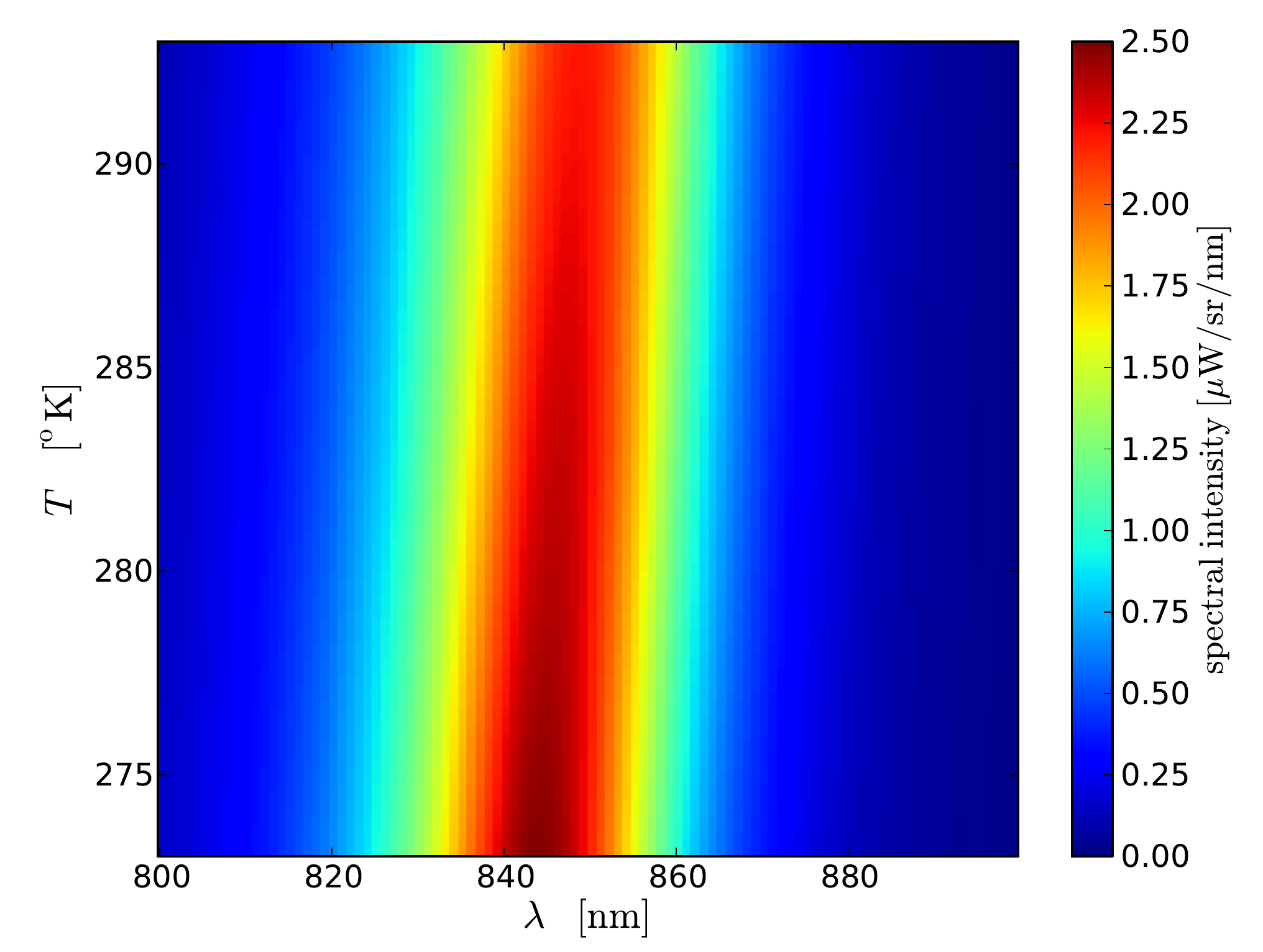}}}
    \caption{Spectral intensity surfaces $\hat{S}(\lambda, T)$ for two
      SkyDICE LEDs.  The model incorporates the cooler-brighter and
      cooler-bluer effects. \label{fig:spectral_intensity_surfaces}}
  \end{center}
\end{figure*}

The spectral intensity of the SnDICE and SkyDICE LEDs, ${\cal
  S}(\lambda)$, have been reconstructed using the method described
above, combining the spectroscopic and photometric data.  On average,
about 15 spectra and a similar number of minimaps have been taken
for each LED at temperatures ranging from $2\celsius$ to $25\celsius$.
In figure \ref{fig:spectral_model_residuals} we show a comparison
between the model and a typical spectrum.  Figure
\ref{fig:spectral_intensity_surfaces} shows the spectral intensity
surfaces obtained for typical blue and red LEDs.  These models
summarise the behaviour of the source.  Their normalisation is set by
the photometric dataset. By construction  they incorporate the
cooler-brighter and cooler-bluer effects.  Once we know the
operating temperature and the distance between the source and the
telescope aperture, we can predict the spectral intensity delivered on
the primary mirror.

\begin{figure}
  \begin{center}
    \includegraphics[width=\linewidth]{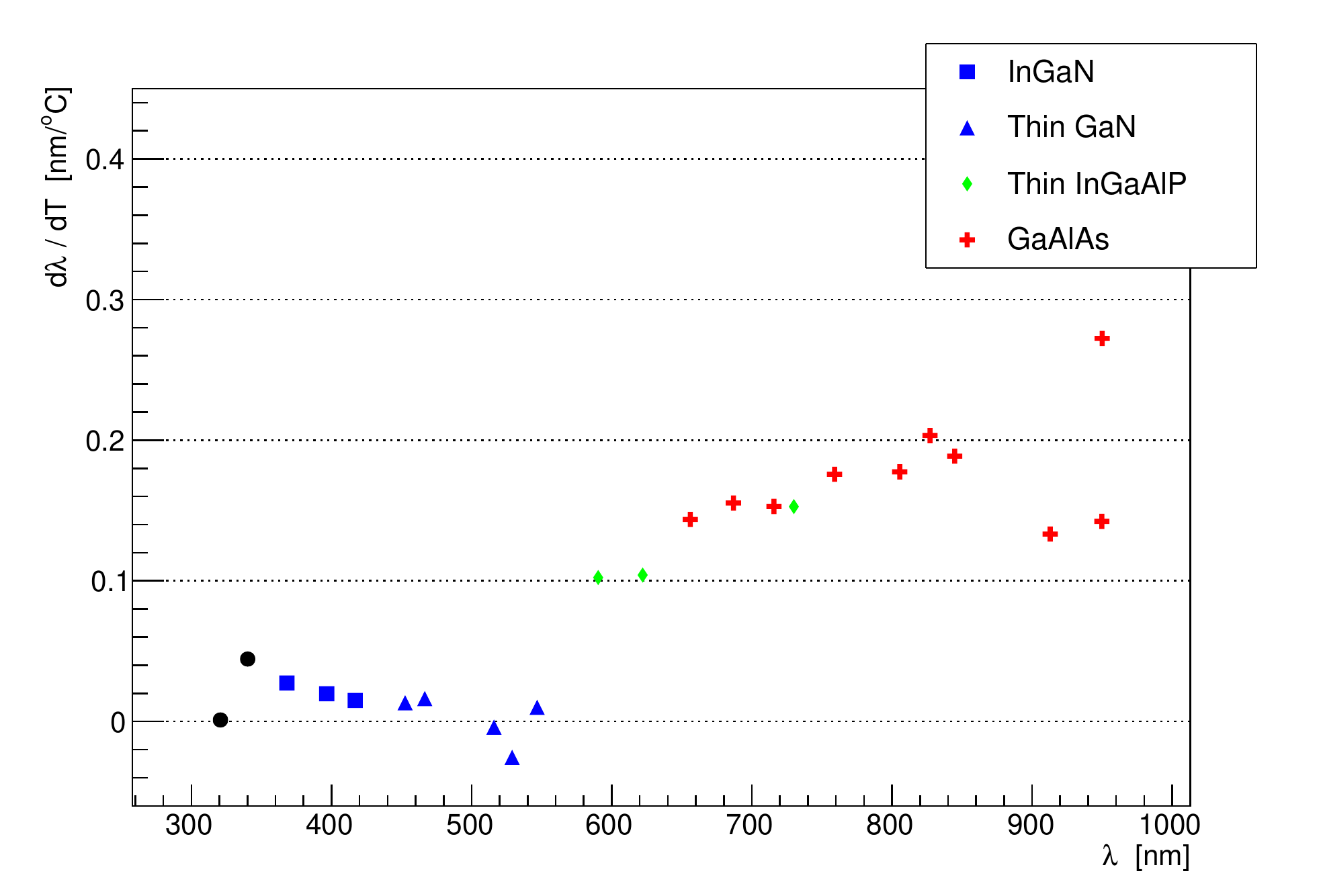}
    \caption{Cooler-bluer slopes measured on the SkyDICE spectra.}
    \label{fig:skydice_cooler_bluer}
  \end{center}
\end{figure}

As can be seen in figure \ref{fig:spectral_intensity_surfaces}, the
cooler-bluer effect does not have the same intensity for all
LEDs. In figure \ref{fig:skydice_cooler_bluer} we report the
variations in the mean spectrum wavelength as a function of
temperature (estimated from the model, for all the LEDs that equip
SkyDICE).  On average, the effect is about 1\AA/\celsius. A trend
is clearly visible, indicating that the red LEDs are more sensitive to
the effect than the blue LEDs. The two LEDs that have been 
found to exhibit a {\em \emph{warmer-brighter}} behaviour (section \ref{sec:minimaps})
also follow a {\em \emph{warmer-bluer}} relation. 

The spectral intensity models are affected by uncertainties owing to the
finite number of spectra and flux measurements.  These uncertainties
are statistical in nature, but behave as systematics in any subsequent
analysis relying on these spectral intensity models.  For this
reason, we are careful to propagate them in an exact way
(i.e. including the off-diagonal terms). They are derived from the
covariance matrix of the spectral intensity fit presented in the
previous section.  We find that with the datasets in hand, the
relative uncertainties on the LED broadband fluxes predicted with the
spectral intensity model are below 0.05\%.

\subsection{Test-bench systematics}
\label{sec:test_bench_systematics}

\begin{table*}
  \begin{center}
    \caption{Summary of the test bench systematics.}
    \renewcommand{\arraystretch}{1.2}
    \begin{tabular}{l|c|c}
      \hline
      \hline
                                             & Uncertainty     & Comment \\
                                             & $(1 \sigma)$    &         \\
      \hline
      \multicolumn{3}{c}{Monochromator} \\
      \hline
      Wavelength calibration                 &  0.3 nm  &         \\
      $\alpha_{\mathrm{Ebert}}$                &  1\degree       &                                                      \\
      $\epsilon_{\mathrm{blaze}}$ (grating \#1)  &  0.17\degree    & $\left<\lambda_{LED}\right> < 450\ \mathrm{nm}$       \\
      $\epsilon_{\mathrm{blaze}}$ (grating \#2)  &  0.28\degree    & $450\ \mathrm{nm} < \left<\lambda_{LED}\right> < 750\ \mathrm{nm}$             \\
      $\epsilon_{\mathrm{blaze}}$ (grating \#3)  &  0.43\degree    & $\left<\lambda_{LED}\right> > 750\ \mathrm{nm}$       \\
      \hline
      \multicolumn{3}{c}{Hamamatsu S2281 calibrated @ NIST} \\
      \hline
                                             &                  &  \multirow{3}{*}{\begin{minipage}{3.5cm}{\begin{center}From the calibration uncertainties\\provided by NIST.\end{center}}\end{minipage}}             \\
      $\alpha_{NIST}$                         &  $2.9\ 10^{-5}$ nm$^{-1}$  & \\
      $\beta_{NIST}$                          &  $0.002$         &        \\
      \hline
    \end{tabular}
    \label{tab:test_bench_systematics}
  \end{center}
\end{table*}

We now examine the systematic uncertainties that have an impact on
our spectral intensity models, $\hat{S}(\lambda, T)$.  We identify
three main sources of systematics: the wavelength calibration of the
monochromator, the transmission of the same monochromator, and the
accuracy of the NIST photodiode efficiency $\eta(\lambda)$.

\subsubsection{Monochromator}
\label{sec:monochromator_systematics}

\paragraph{Wavelength calibration} The wavelength calibration of the
device is checked (at several temperatures), using a series of
calibration lamps, notably a sodium lamp and a polymetallic lamp. From this dataset  we
obtain a correction to the calibration given by the
manufacturer.  This correction does not exceed 0.1-nm in amplitude.
We find a small linear dependence of the wavelength calibration with
temperature (see Table \ref{tab:absolute_wavelength_calibration}) of
about 0.025~nm/\celsius, which is taken into account.  The dominant
systematics affecting the wavelength calibration is related to the
positioning uncertainty of the source with respect to the
monochromator entrance slit.  We found that this contribution never
exceeds 0.1-nm. Given the limited precision of the alignment procedure
and the small number of calibration runs, we conservatively set the
wavelength calibration systematics to 0.3-nm (see Table \ref{tab:test_bench_systematics}).

\paragraph{Transmission} 

The transmission of the monochromator, $T_m(\lambda)$, is measured at
several discrete wavelengths.  For Czerny-Turner designs, it is
relatively easy to compute the global shape of $T_m(\lambda)$, as a
function of two specific angles: the so-called ``Ebert angle'' $\alpha_E$,
which depends on the optical design of the device, and the ``blaze angle''
$\epsilon_{\mathrm{blaze}|i}$ of each grating $i$ being used.  
We use these continuous models (one for each grating) fitted on the
discrete transmission measurements as our estimates of the
monochromator transmission (see figure
\ref{fig:monochromator_transmission}).  At first order, $T_m(\lambda)$
only depends  on the $\epsilon_{\mathrm{blaze}}$ parameter.  The
uncertainty on the monochromator transmission is estimated by
propagating the uncertainties on our estimates of the $\epsilon_{\mathrm{blaze}}$ angles. They
are reported in Table \ref{tab:test_bench_systematics}.

{The method used to extract the LED spectrophotometric
  models (equations \ref{eqn:i_spec_developed} and
  \ref{eqn:constraints_on_the_spectrum_normalization}) ensures that
  the model normalisation only depends on the photometric measurements
  (without the monochromator). This means that the uncertainties on
  the monochromator transmission have no impact on the LED-to-LED
  calibration and may only affect the spectrum shapes.  To fix ideas,
  we find that varying the Ebert angle by one degree has an impact on
  the spectrum mean wavelength ($\left<\lambda\right> = \int \lambda
  S(\lambda) d\lambda / \int S(\lambda) d\lambda$), which is lower than
  $1\AA$.  We also find that altering the monochromator transmission
  model around 550-nm, as suggested by the residuals of figure
  \ref{fig:monochromator_transmission}, results in a shift in
  wavelength that is also lower than 1\AA.}

\subsubsection{NIST}
\label{sec:nist_systematics}

\begin{figure}
  \begin{center}
    \includegraphics[width=\linewidth]{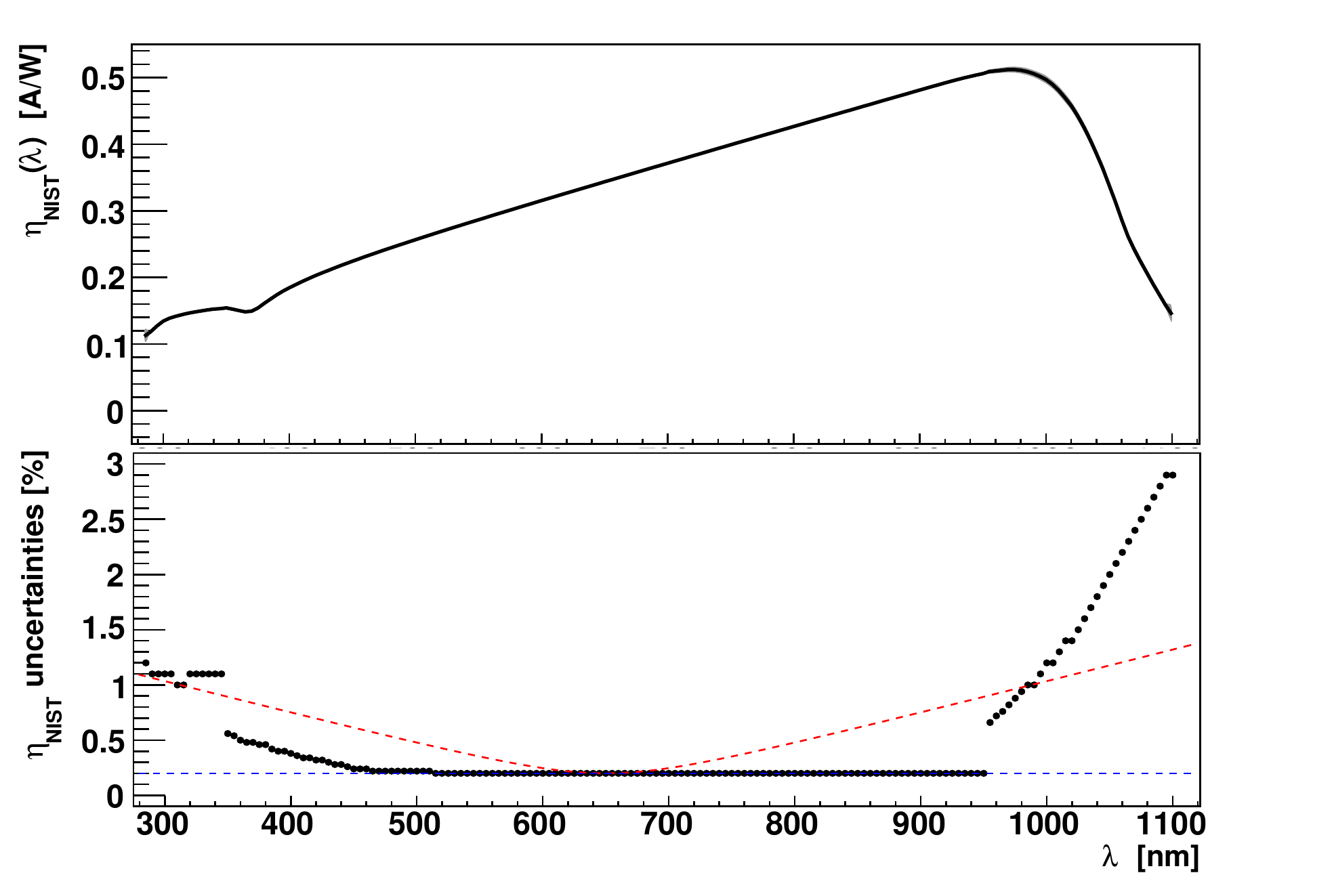}
    \caption{Upper panel: efficiency (in A/W) of the Hamamatsu S2281
      photodiode calibrated at NIST, purchased for this project.
      Bottom panel: black points: uncertainties reported by NIST (in
      percentage); dashed red line: diagonal elements of the ``worst
      case'' interpretation of the NIST error budget, as discussed in
      section \ref{sec:nist_systematics}; dashed blue line: diagonal
      elements of the ``best case'' interpretation of the NIST ERROR
      budget discussed in the same section.}
    \label{fig:nist_efficiency}
  \end{center}
\end{figure}

The calibration of the NIST photodiode is itself
uncertain. NIST provides its clients with an error budget 
and reports uncertainties of about $\sim 0.2\%$ between 400-nm and 950-nm 
and up to 1\% in the UV and near-IR (figure \ref{fig:nist_efficiency}).
We expect 
a fraction of the uncertainties affecting measurements at different wavelengths to be correlated. 
And since we are primarily interested in the {\em \emph{relative}} calibration of 
our passbands, it is essential for us to account for these off-diagonal terms.
As of today, we have not been able to obtain this information from NIST.
We therefore built two different error models, depending on how we choose to interpret
the NIST uncertainties. 

In the ``best case scenario'', we assume that what is uncertain in the
photodiode efficiency is primarily a grey scale. This means
that the NIST uncertainties are positively fully correlated.  Since our goal 
is to measure the {\em \emph{relative}} normalisation of the imager passbands, 
this would be
an ideal situation, because the uncertainty on the global flux scale
cancels out when comparing two different passbands.

In the ``worst case scenario'', we assume that there is a ``colour
uncertainty'' affecting the efficiency reported by NIST.  This can be
modelled with two random variables $\alpha_{\mathrm{NIST}}$ and
$\beta_{\mathrm{NIST}}$, of variance $\sigma_\alpha$ and
$\sigma_\beta$, and such that
\begin{equation}
  \eta_{\mathrm{NIST}}(\lambda) = \eta_{\mathrm{true}}(\lambda) \times \left(\alpha_{\mathrm{NIST}}\ (\lambda-\bar{\lambda}) + \beta_{\mathrm{NIST}} + 1\right)
,\end{equation}
where $\sigma_\alpha$ and $\sigma_\beta$ are chosen so that the
uncertainties computed from the equation above stay compatible with
the error budget reported by NIST, and $\bar{\lambda}$ is a pivot value
(chosen to be $\bar{\lambda} = 650\ \mathrm{nm}$).  Their values are listed in
Table \ref{tab:test_bench_systematics}.

This model is of course a little extreme because it assumes a correlation
length that spans the entire visible range. This would mean that the
uncertainty reported by NIST lies with POWR, the primary flux standard.  This is not
the case, since for the first steps of its metrology chain, NIST reports
uncertainties of a few $10^{-4}$, about one order of magnitude lower.
It is more likely that the main fraction of the uncertainty budget is introduced in
the last step at the SCF facility, and it is difficult to see how SCF
could introduce error with an infinite correlation length in
wavelength.  However, in the present state of our knowledge, exploring
these two extreme cases is the best we can do, because they bracket the 
true uncertainties.

\subsubsection{Propagating the bench systematics}

The test bench systematics affect how we reconstruct the LED spectral
intensities ($\hat{S}$). They may, for example, shift the $\hat{S}(\lambda)$ 
models in wavelength or
distort their shape, and this must be accounted for.
Regardless of the propagation method we
choose, we need to quantify how the bench errors change the
reconstructed LED spectral intensities, $\hat{S}(\lambda, T)$.  To
do this, we compute the derivatives of the $\hat{S}(\lambda, T)$
models with respect to all the identified systematics.  This is done
by shifting each term of the systematics vector in turn
and
re-determining the model parameters 
as described in section
\ref{sec:modeling_led_spectra} above.  
They are used to propagate the test-bench systematics, 
(see section
\ref{sec:analysis} and Appendix \ref{sec:passbandcal} for details). 

\subsection{Conclusion}

At this stage, we have modelled the emissivity of each LED as a smooth
function that predicts the spectral intensity of the beam in any
given direction and at any temperature.  The bench statistical
uncertainties were encoded into the covariance matrices of the
$\vec{\theta}$ parameters.  The main systematics were quantified.
They come primarily from our calibration of the monochromator and from
the error budget reported at NIST. How they will be propagated in the
final analysis is discussed in section
\ref{sec:analysis} and Appendix \ref{sec:passbandcal}.

We now have all we need to measure the transmission of an imager from
series of calibration frames taken with these light sources. This is the subject of
the next section.


\section{Discussion: calibrating broadband observations}
\label{sec:analysis}

We have built a very stable ($0.01\%$) light source and characterised
it on a test bench with an accuracy of $\sim$ 0.3 nm\ in wavelength
and $\sim 0.1\%$ in flux.  Now the question is how we use it to
calibrate a real imager.  And first of all: what do we need to
measure in order to calibrate an imager?

\subsection{Transmissions}

Passbands are known long before the
first star light hits the focal plane.  The transmissions of all
optical components and the quantum efficiency curves of the detectors are
measured before assembly.  
These measurements are combined to build a synthetic passband model:
\begin{equation} 
  T(\lambda) = g \times {\cal A} \times R_{\mathrm{mirror}}(\lambda) \times
  T_{\mathrm{optics}}(\lambda) \times T_{\mathrm{filter}}(\lambda) \times
  \varepsilon(\lambda)
  \label{eqn:synthetic_passband_model}
\end{equation}
where $\varepsilon(\lambda)$ is the quantum efficiency of the CCD, $g$
is the gain of its readout chain, ${\cal A}$ is the area of the mirror,
and the other terms are the various transmissions and reflectivities of
the optical elements.  Note that ${T(\lambda)}$ is a dimensioned quantity:
here it has units of ADU/$\gamma$/m$^2$.

The absolute normalisation of $T(\lambda)$ varies with time and must be
monitored: the gain $g$ of the readout electronics may fluctuate by a
few per-mil on timescales of a few hours; also, alterations of the
optical surfaces (dust deposits, ageing of the coatings, etc.) slowly
degrade the transmission of the instrument by as much as
5-10\% per year.  The resulting attenuation of the telescope
transmission is slightly wavelength dependent. Therefore, the {\em
  \emph{relative}} normalisation of the passbands with respect to each other
may itself vary by a few percentage points per year. The main purpose of
calibration is therefore to monitor these variations. In most
applications, what we need to measure is the evolution of the relative
normalisation of the passbands with respect to each other.

The {\em \emph{shape}} of the passbands is not expected to vary very
significantly over time.  This is a design requirement.
However, several studies have reported evidence of slow evolution of
the passband shape.  For example, \citet{Doi2010} report a
30\% decrease in the short-wavelength side of the SDSS 2.5-m $u$-band channels,
probably due to ageing of the CCD anti-reflective coatings.  Another
example can be found in \citet{2013A&A...552A.124B}, who show hints that the
red fronts of the $r$-band and $i$-bands that equips MegaCam are about 8-nm
off with respect to the scans provided by the manufacturer (resulting in 
a disagreement of 4 nm on the filter mean wavelength). In this
case, there is indirect evidence that this evolution of the passbands took
place in an early phase of the life of the instrument. There
is therefore a strong incentive for future surveys to monitor the
shape of their effective passbands.

\subsection{Constraining passbands}
\label{sec:calibration_fit}

Instead of measuring the filter transmissions again with a monochromatic
beam, the DICE strategy consists in taking the (known) synthetic
passbands as a starting point and in constraining small alterations
to these passbands from series of measurements.

\begin{figure}
  \begin{center}
    \includegraphics[width=\linewidth]{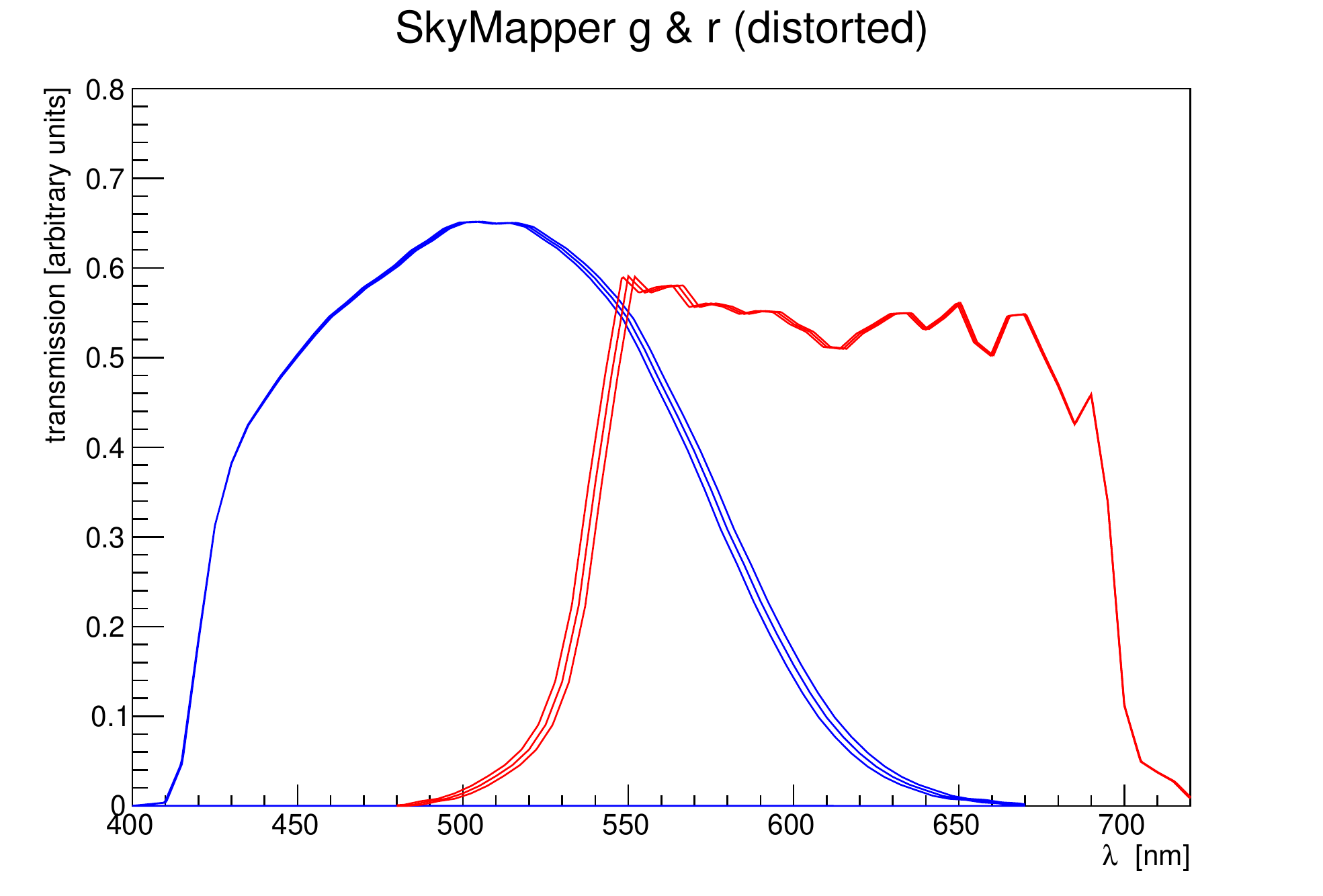}
    \caption{Distortion of  the red and blue fronts of the
      SkyMapper $g$ and $r$ passbands, respectively.}
    \label{fig:passband_distorsions}
  \end{center}
\end{figure}

The broadband flux of an astrophysical object is primarily sensitive
to (1) the normalisation of the passband 
and (2) the position (in wavelength) of the blue and red filter
cutoffs. 
These three quantities are therefore
what should be monitored in the long run.  To do this, we alter the
synthetic passband model ${T(\lambda)}$ presented in equation
\ref{eqn:synthetic_passband_model} to allow for a different
normalisation and for small (potential) variations in the filter
cutoffs.  The latter is done by composing ${T(\lambda)}$ with a linear
function that shifts and stretches (or dilates) $\lambda$ around the
filter mean wavelength:
\begin{equation}
  \lambda \mapsto \lambda' = \alpha (\lambda-\bar{\lambda}) + \beta
.\end{equation}
In practice, we reparametrise the function above, so that it
depends directly on the filter cutoff displacements, denoted
$\delta\lambda_{\mathrm{blue}}$ and $\delta\lambda_{\mathrm{red}}$
hereafter. This way, one can shift each front easily, and they are essentially
independent of the other, as shown in figure
\ref{fig:passband_distorsions}.

By doing so, we reduce our problem to fitting only three parameters
per filter: a normalisation, $\cal N$, and two filter front
displacements. 
For an imager equipped with $N$ (typically 5) filters, we end up with
$3 \times N$ calibration parameters, which we group in a single
vector, noted $\vec{\vartheta}_t$:
\begin{equation}
  \vec{\vartheta}_t = \begin{pmatrix}
    \mathrm{{\cal N}}_{u}\\
    \delta\lambda_{\mathrm{blue}|u}\\
    \delta\lambda_{\mathrm{red}|u}\\
    \vdots\\
    \mathrm{{\cal N}}_{z}\\
    \delta\lambda_{\mathrm{blue}|z}\\
    \delta\lambda_{\mathrm{red}|z}\\
  \end{pmatrix}
.\end{equation}
To constrain these unknowns, we build a model that predicts the flux
registered on the focal plane. These predictions are built from the
LED spectral intensity estimates $\hat{S}_l(\lambda, T)$, derived in
section \ref{sec:spectroscopic_calibration}, and the re-parametrised
passbands that we denote as $T_b(\lambda, \vec{\vartheta}_t)$. The model
simply writes as
\begin{equation}
  {\varphi}_{bl} = \delta\omega\ \ \int \hat{S}_l(\lambda,T)\ {{T}_b}(\lambda, \vec{\vartheta}_t)\ d\lambda
  \label{eqn:illu_model}
\end{equation}
where $\delta\omega$ is the solid angle covered by a focal plane pixel
or superpixel (about $10^{-12}$ sr for a single pixel, and $10^{-8}$
sr for a typical $128\times 128$ superpixel). Comparing measurements
of the calibration flux performed with the imager with the
predicted flux $\varphi_{bl}$ yields constraints on
$\vec{\vartheta}_t$.  

The details of the method are described in Appendix
\ref{sec:passbandcal}.  Classically, it consists in minimising a
$\chi^2$ function built from the model above and the flux
measurements.  We are careful to propagate the effect of the
identified uncertainties to the final estimates.  The statistical
uncertainties affecting the LED spectral intensity models are
accounted for.  The systematics, those related to the test bench
measurements, as well as those affecting the flux measurements on
site, are incorporated into the fit as nuisance parameters and
marginalised over.

\subsection{Lightweight calibration runs}
\label{sec:calibration_runs}

The system has been designed to be run in routine mode every day or
so. During a typical calibration run, each passband is sampled with
the corresponding LEDs.  Figure \ref{fig:filter_coverage} shows that
the typical $u(v)griz$ passbands are covered by three to nine LEDs, depending
on their extension in wavelength, with the filter cut-ons and cut-offs sampled by one or two LEDs\footnote{Except for the early version, 
SnDICE, which presented an under-sampled region around
  700-nm, precisely at the location of the red front of the
  $r$-filter, and the blue front of the $i$-filter}.  In Table
\ref{tab:typical_dice_runs}, we estimate the number of exposures that
have to be taken during typical calibration runs for various 
designs and the time requested to complete these runs.
Conservatively  taking about one minute per exposure, plus 20 minutes for the
overheads (filter changes, alignment of the source, and the telescope,
etc.), we find that a typical run requires a little less than one hour of
daytime.  This calibration program is short enough to be run every
day or so just after and/or before telescope operations.

Significant gains in precision can be obtained by combining several
calibration runs. Indeed, if the normalisation of the passbands does
change slightly from one night to the next, the position of the filter
cut-offs is not expected to move in a measurable way overnight.  As a
consequence, we will typically fit for one normalisation per filter
and per run, but only one position parameter for each filter cut-off.
This represents $\sim N_{\mathrm{bands}} \times N_{\mathrm{runs}} + 2
\times N_{\mathrm{bands}}$ (e.g.  60 parameters for 10 MegaCam
calibration runs and 135 parameters for 25 runs).  As shown
below, this allows us to improve the precision on all
calibration parameters sizeably, at the price of a slightly more complex
procedure.

\subsection{Simulated datasets}

To conclude, we assess the quality of the constraints that can be
obtained from series of calibration runs of the type
described above.  The main goal here is to show that the problem is
well constrained and to give estimates of the uncertainty budget.
These estimates are derived from the quantitative analysis  of the accuracy of the source
test bench calibration (section \ref{sec:spectroscopic_calibration}).  
These are lower bounds, because we do not
discuss the uncertainties associated with the imager
measurements of the calibration light in detail.  Based on preliminary analyses
of the calibration frames taken with MegaCam and SkyMapper, we assume
that they are of about 0.5\% (uncorrelated).

\begin{table}[t]
  \begin{center}
  \caption{Typical MegaCam and SkyMapper calibration runs.}
  \renewcommand{\arraystretch}{1.2}
  \begin{tabular}{l|cccccc|c}
    \hline
    \hline
              & $u$ & $v$ & $g$ & $r$ & $i$ & $z$ & Duration \\
              &     &     &     &     &     &     & (\# frames / {\bf mn}) \\
    \hline
    \begin{minipage}[c]{0.2\linewidth}\begin{center}MegaCam\\{\tiny (SnDICE)}\end{center}\end{minipage}   
              &  4   &  -- &  7   &  5   &  7   & 4    & 27 / {\bf 47} \\
    \hline 
    \begin{minipage}[c]{0.2\linewidth}\begin{center}SkyMapper\\{\tiny (SkyDICE)}\end{center}\end{minipage} 
              &  4    &  4  &  8   &  6    &  5    &   6   &  33 / {\bf 53} \\
    \hline
  \end{tabular}
  \label{tab:typical_dice_runs}
  \end{center}
\end{table}

We generate synthetic flux measurements, using MegaCam/SkyMapper
passband models as follows.  We first simulate realistic test bench
datasets, by modelling the ``true'' LED spectral intensities
$S_{\mathrm{true}}(\lambda, T)$ with Gaussians of width $\sigma \sim
15$-nm, normalised to a radiant intensity of 0.1~mW/sr, and emulating
the true SnDICE and SkyDICE coverage. We estimate the LED spectral
intensities with the methods described in sections
\ref{sec:photometric_calibration} and
\ref{sec:spectroscopic_calibration}.  We use the ``true'' (Gaussian)
spectra to generate the broadband fluxes and the reconstructed
spectral intensity models $\hat{S}(\lambda, T)$ to estimate the
calibration parameters.
The calibration parameters
$\vec{\vartheta}_t$ are estimated following the method described in
section \ref{sec:calibration_fit} and Appendix \ref{sec:passbandcal}
(using the reconstructed LED spectral intensity models).  The method
yields the full stat+syst covariance matrix of the calibration
parameters, and we report the diagonal elements as a function of the
number of calibration runs.

The quality of the constrains on $\vec{\vartheta}_t$ is a function of
the quality of the flux measurements performed on site, but also of
how we manage to sample each filter.  We therefore study not only what
may be obtained with the sources already in place, but also what could
be done with optimised designs, such as those discussed in Appendix
\ref{sec:led_selection}. All designs involve the same number of
LEDs (24), so that the duration of the corresponding calibration runs is
similar to what is reported in Table \ref{tab:typical_dice_runs}.

\paragraph{Filter cutoffs} The precision on the filter cut-off
positions only slightly depends on the number of runs that are
combined into the calibration fit. In five to ten runs, depending on
the quality of the cut-off coverage, it hits a systematics floor of
about 0.3-nm, where the dominant systematics here is the precision of the
monochromator wavelength calibration.  It affects all the filter
cut-off estimates performed with a given source.  

\begin{figure}
\begin{center}
  \includegraphics[width=\linewidth]{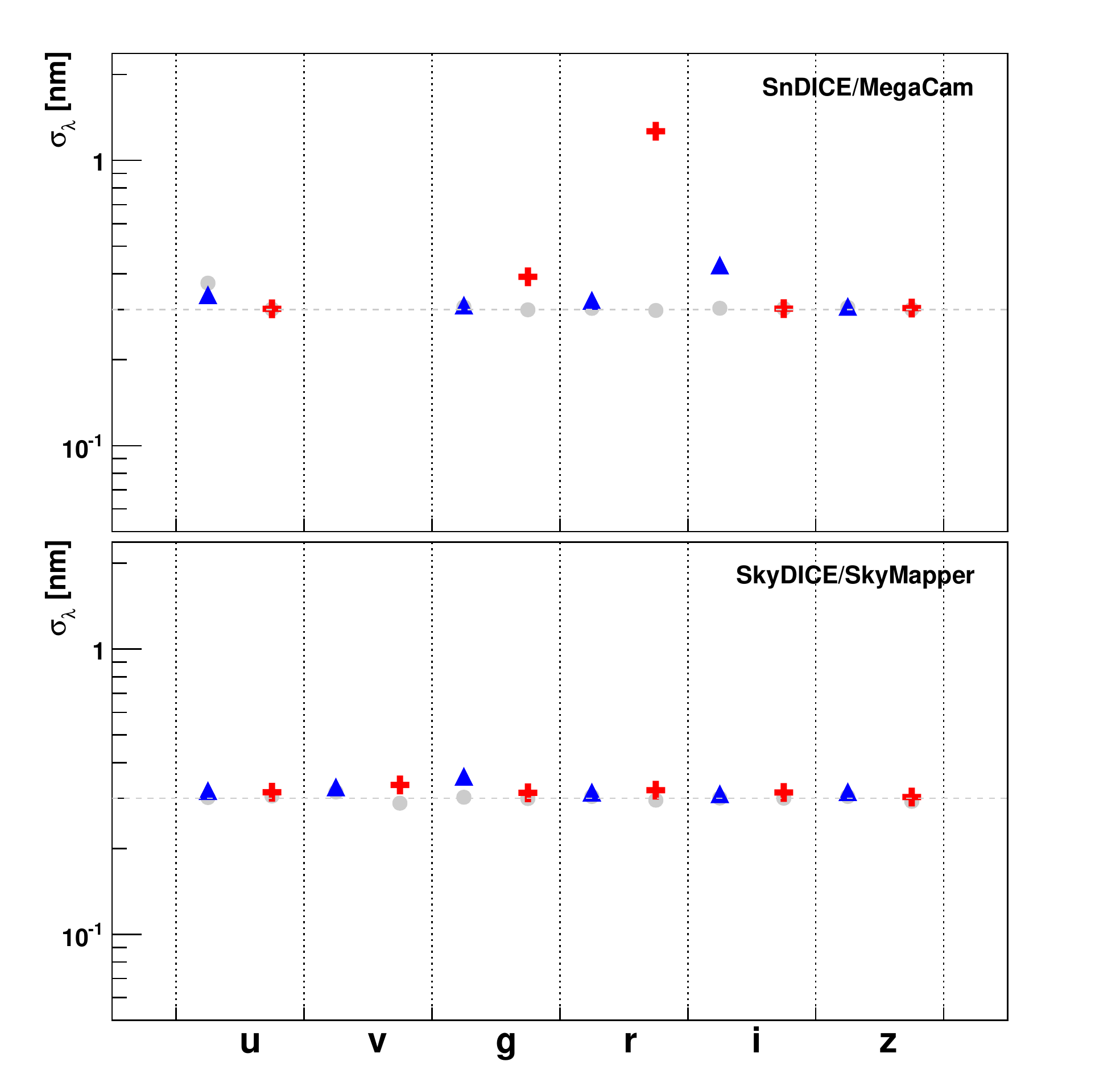}
  \caption{Upper panel: Uncertainties in nanometres on the blue
    (blue triangles) and red (red crosses) cut-offs of the MegaCam
    passbands, after 20 calibration runs. Grey points: the same,
    for an optimised design such as the one described in Appendix
    \ref{sec:led_selection}.  Lower panel: the same, for
    SkyMapper.}
  \label{fig:sigma_filter_front_summary}
\end{center}
\end{figure}

In figure \ref{fig:sigma_filter_front_summary} we show, for SnDICE
(MegaCam), SkyDICE (SkyMapper), and their associated optimised
designs, the uncertainties on the filter cut-off positions obtained by
combining about 20 calibration runs. SkyDICE is very close to an
optimal design, and it can constrain the SkyMapper filter cut-offs
extremely well. On the other hand, SnDICE suffers from its identified
lack of coverage around 700-nm. In particular, it yields only marginal
constraints on the red cutoff of MegaCam $r$-band.

\begin{figure}
  \begin{center}
    \includegraphics[width=\linewidth]{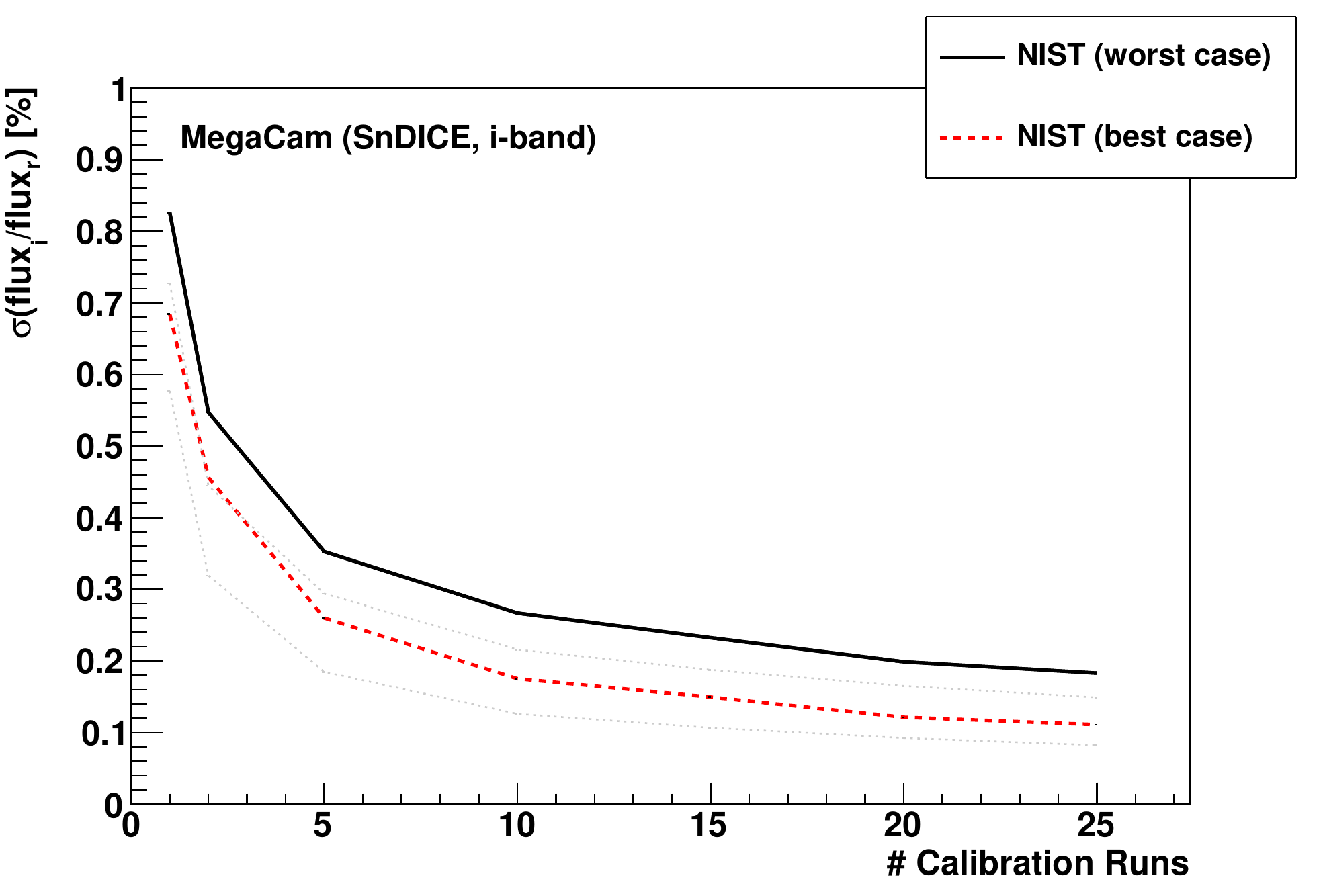}
    \caption{Uncertainty on the MegaCam $i$-band normalisations,
      relative to the $r$-band, as a function of the number of
      calibration runs.  The uncertainty level actually depends on how
      we interpret the NIST error budget colour uncertainty (black
      line) or gray scale uncertainty (dashed red line) -- see section
      \ref{sec:test_bench_systematics} for details.  The grey lines
      show the calibration uncertainties obtained with an optimised
      filter sampling (see Appendix
      \ref{sec:led_selection}). \label{fig:filter_normalization_vs_nb_runs}}
  \end{center}
\end{figure}

\paragraph{Passband normalisations} The uncertainties on the
normalisation of the MegaCam and SkyMapper passbands (relative to $r$)
improve sizeably with the number of runs, as shown in figure
\ref{fig:filter_normalization_vs_nb_runs}. In one run, we are able to
reach a sub-percent accuracy.  After about ten calibration runs, the
uncertainty is divided by two, below 0.5\% in all bands.

The level of the systematics floor depends on how we interpret the
NIST uncertainties.  In the ``best case scenario'' discussed in
section \ref{sec:nist_systematics}, in which the NIST uncertainties
are interpreted as a grey-scale uncertainty, the impact of the NIST
error budget is small, and we are able to reach uncertainties of 0.3\%
or below.  In the other scenario, where the NIST errors are negatively
correlated, the error floor can be significantly higher. 

\begin{figure}[t]
\begin{center}
  \includegraphics[width=\linewidth]{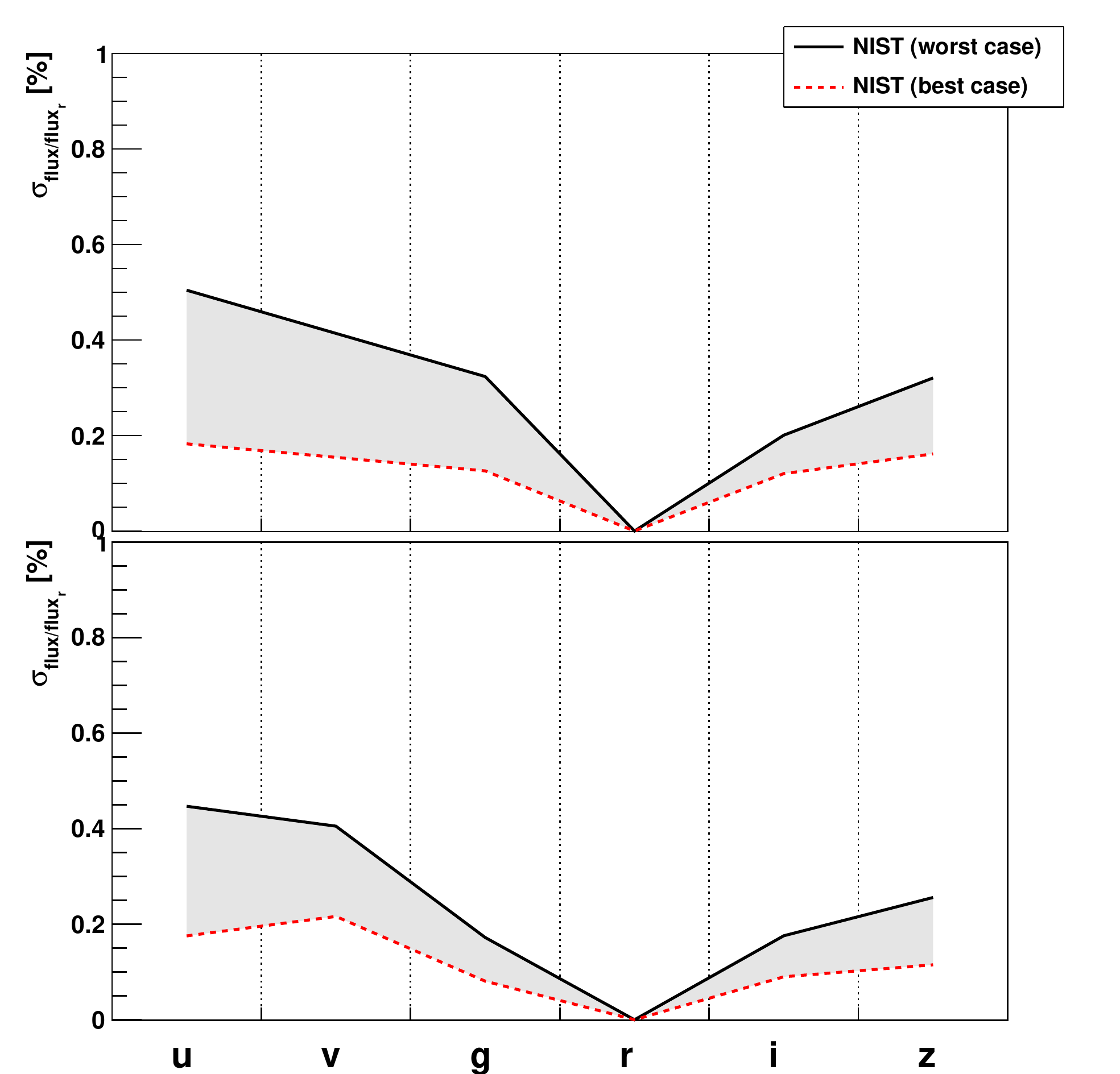}
  \caption{Upper panel: uncertainty on the MegaCam passband
    normalisation (relative to $r$) obtained when combining 20 SnDICE
    runs. Lower panel: the same, for SkyMapper.\label{fig:sigma_norm_summary}}
\end{center}
\end{figure}

Figure \ref{fig:sigma_norm_summary} summarises the normalisation
uncertainties for SnDICE (MegaCam) and SkyDICE. The true uncertainties
are somewhere in the grey band, probably closer to the ``best-case
scenario'' line.  This shows that neither our design choices nor the
finite precision of the test bench measurements may limit the
precision of the calibration that we will obtain from such measurements.
All essentially depends on how accurately we can measure the
calibration flux delivered on the focal plane.


\section{Conclusion}
\label{sec:discussion}

We have presented the concept and performances of a stable LED-based 
calibrated light source,
designed for the photometric calibration of wide field imagers. The
design has been deliberately kept as simple as possible. The light is
generated by a series of 24 narrow spectrum LEDs chosen to cover the
imager spectral range. Each LED implements a point source, generating
a conical beam, which yields a uniform illumination of the focal
plane.  No intermediate optical surface is present between the LED
emission zone and the telescope primary mirror.  This makes the
hardware very stable and easy to maintain in the long run. 

In this paper, we have described the first step of our
calibration program  in detail (labelled \ding{202} on figure \ref{fig:nist_metrology_chain}):
the transfer of the NIST flux scale to the light
source itself.  The emissivity of each LED is characterised by its
spectral intensity, i.e.  the emitted power per unit solid angle and
per unit wavelength.  The LED spectral intensities are measured on a
spectrophotometric test bench, using a monochromator and a Hamamatsu
S2281 photodiode calibrated at NIST.  The test bench data are combined
into a spectral intensity model that captures the variations in the
LED emissivity as a function of temperature (the so-called
``brighter-cooler'' and ``bluer-cooler'' effects).  This model is then
used on site to predict the light actually delivered by the source at
a given temperature.

The source is intrinsically stable at the level of $\lesssim 1\%$.  It
is equipped with several monitoring systems to control its
stability during operations.  We have shown that if we monitor either
the temperature of the LEDs or directly the light emitted by the LED,
we can control the relative variations in the calibration light with
a precision that ranges from a few $10^{-4}$ for the best channels to
about $10^{-3}$.  This means that we have built a source that is as
stable as the silicon photodiodes used to disseminate the NIST flux
scale.  This source can then be used to calibrate other light
detectors (e.g.  astronomical imagers).

In the last section of this introductory paper, we sketched the
second step in our calibration program (labelled \ding{203} in figure
\ref{fig:nist_metrology_chain}) {in order to give a complete
  picture of the system and its main use case.}
We described how we extract simultaneously the relative normalisation 
of the telescope passbands and the position of the passband
cutoffs from a relatively small number of non-monochromatic 
calibration exposures (4 to 9, depending on the passband width).  
All the uncertainties, either those related to the test bench measurements 
or those affecting the flux measurements performed on site, can be 
propagated to the final result (including the non-diagonal terms).

As an illustration, we quantified the impact of the test bench
systematics on the calibration parameters.  Regarding the filter
normalisations (relative to the $r$-band), we showed that the
systematics floor is well below 0.5\%.  {\em \emph{The exact value actually
  depends on how we interpret the NIST error budget. This is an
  important point that will have to be clarified in the future}.}
Regarding the passband cutoffs, we have shown that, although the
source is not monochromatic, the constraints we obtain are essentially
limited by the accuracy on the wavelength calibration of our
monochromator.

{As of today, more than 30 calibrations runs have been
  accumulated with MegaCam, about ten of them, in conjunction with
  direct observations of CALSPEC standards.  The analysis of this
  dataset is ongoing.  It constitutes the next step in our
  calibration programme, and it will be described in upcoming papers,
  currently in preparation.
  
  Paper 2 will be devoted to analysing the
  calibration exposures, in particular determining the
  passband normalisations and cutoffs, as described in section
  \ref{sec:analysis} (Step \ding{203} of Fig.
  \ref{fig:nist_metrology_chain}).  A specific analysis pipeline is
  being developed for this purpose.  The pre-processing stage (e.g.
  bias subtractions, gain equalization, etc.) is extremely similar to
  what is implemented in a classical image processing framework.  The
  downstream operations are more specific: the goal is to measure the irradiance delivered on the focal plane by an extended beam,
  and one has to deal with non-standard foreground effects, such as (1)
  the diffraction patterns caused by dust and optical defects in the
  light path, (2) the angular distribution of the beam on the MegaCam
  interference filters, and (3) stray light (or ghosts) coming from
  reflections within the telescope optics.}  Our experience with the
MegaCam imager has shown that ghost contamination can be as high as
20\% in some regions of the focal plan.  {This effect is
  chromatic and can therefore affect the relative calibration of
  bands.  How we deal with it has been the central topic of this paper. }

{Paper 3 will describe Step \ding{204}
  of our calibration programme, i.e.  the comparison between the NIST
  and CALSPEC flux scales, using the observations of the CALSPEC
  objects, taken in conjunction with the SnDICE2 calibration
  runs. There are two main difficulties: one is the control of the
  atmospheric transmission at the time of the observations. We deal
  with this by carefully scheduling the observations, along
  with using a semi-empirical model of the atmosphere, trained on
  observations performed at the UH 2.2-m with the SNIFS instrument.
  (Some of these observations were taken in conjunction with the
  MegaCam observations of the CALSPEC standards.)  The other
  difficulty resides in assessing the systematic differences between
  calibrations obtained from flat field exposures, on one hand, and
  point source measurements, on the other.  Low-angle scattering
  of light within the telescope optics is the main effect that has to
  be understood. }

{For now, we have fulfilled the first step of our programme, which is to
demonstrate that it is possible to build LED-based sources, to
characterise them spectrophotometrically with an accuracy of a few
$10^{-4}$ in flux and 3\AA\ in wavelength, and to control their
behaviour in the long run. The DICE sources can be used to disseminate
a NIST-based calibration to a telescope.}



\begin{acknowledgements}
  This work was supported by CNRS/IN2P3, the CFHT Corporation, the
  ``Programme National Cosmologie et Galaxies'' (PNCG), and the Universit{\'e}
  Pierre et Marie Curie.  The authors acknowledge funding from the
  {\'E}mergence-UPMC-2009 research programme.  PFR received a PhD grant from
  {\'E}mergence-UPMC-2009. We are indebted to Philippe~Bailly, Julien~Coridian, 
  Colette~Goffin, Herv{\'e}~Lebbolo, 
  Philippe~Repain, Alain~Vallereau, and Daniel~Vincent from the LPNHE staff, who helped
  build the DICE light sources and the associated test bench. 

  The authors wish to thank the CFHT Corporation and its staff
  for the precious advice and continuous support.  We are particularly
  grateful to Gregory Barrick, Tom Benedict, 
  Kevin Ho, Derrick Salmon, Jim Thomas and Christian Veillet for their
  help before, during and after the SnDICE installation.

  The SkyDICE project was supported by the Research School of
  Astronomy and Astrophysics (RSAA) at Australian National University.
  We are thankful to Daniel Bayliss, Gabe Bloxham, Peter Conroy, Bill Roberts, Brian
  Schmidt, Denise Sturgess, Annino
  Vaccarella, Peter Verwayen, and Colin Vest, for their support during
  the design and installation effort.
\end{acknowledgements}



\begin{thebibliography}{33}
\expandafter\ifx\csname natexlab\endcsname\relax\def\natexlab#1{#1}\fi

\bibitem[{{Amanullah} {et~al.}(2010){Amanullah}, {Lidman}, {Rubin}, {Aldering},
  {Astier}, {Barbary}, {Burns}, {Conley}, {Dawson}, {Deustua}, {Doi}, {Fabbro},
  {Faccioli}, {Fakhouri}, {Folatelli}, {Fruchter}, {Furusawa}, {Garavini},
  {Goldhaber}, {Goobar}, {Groom}, {Hook}, {Howell}, {Kashikawa}, {Kim}, {Knop},
  {Kowalski}, {Linder}, {Meyers}, {Morokuma}, {Nobili}, {Nordin}, {Nugent},
  {{\"O}stman}, {Pain}, {Panagia}, {Perlmutter}, {Raux}, {Ruiz-Lapuente},
  {Spadafora}, {Strovink}, {Suzuki}, {Wang}, {Wood-Vasey}, {Yasuda}, \&
  {Supernova Cosmology Project}}]{2010ApJ...716..712A}
{Amanullah}, R., {Lidman}, C., {Rubin}, D., {et~al.} 2010, \apj, 716, 712

\bibitem[{Barrelet \& Juramy(2008)}]{barrelet_direct_2008}
Barrelet, E. \& Juramy, C. 2008, Nuclear Instruments and Methods in Physics
  Research A, 585, 93

\bibitem[{{Betoule} {et~al.}(2014){Betoule}, {Kessler}, {Guy}, {Mosher},
  {Hardin}, {Biswas}, {Astier}, {El-Hage}, {Konig}, {Kuhlmann}, {Marriner},
  {Pain}, {Regnault}, {Balland}, {Bassett}, {Brown}, {Campbell}, {Carlberg},
  {Cellier-Holzem}, {Cinabro}, {Conley}, {D'Andrea}, {DePoy}, {Doi}, {Ellis},
  {Fabbro}, {Filippenko}, {Foley}, {Frieman}, {Fouchez}, {Galbany}, {Goobar},
  {Gupta}, {Hill}, {Hlozek}, {Hogan}, {Hook}, {Howell}, {Jha}, {Le Guillou},
  {Leloudas}, {Lidman}, {Marshall}, {M{\"o}ller}, {Mour{\~a}o}, {Neveu},
  {Nichol}, {Olmstead}, {Palanque-Delabrouille}, {Perlmutter}, {Prieto},
  {Pritchet}, {Richmond}, {Riess}, {Ruhlmann-Kleider}, {Sako}, {Schahmaneche},
  {Schneider}, {Smith}, {Sollerman}, {Sullivan}, {Walton}, \&
  {Wheeler}}]{2014A&A...568A..22B}
{Betoule}, M., {Kessler}, R., {Guy}, J., {et~al.} 2014, \aap, 568, A22

\bibitem[{{Betoule} {et~al.}(2013){Betoule}, {Marriner}, {Regnault},
  {Cuillandre}, {Astier}, {Guy}, {Balland}, {El Hage}, {Hardin}, {Kessler}, {Le
  Guillou}, {Mosher}, {Pain}, {Rocci}, {Sako}, \&
  {Schahmaneche}}]{2013A&A...552A.124B}
{Betoule}, M., {Marriner}, J., {Regnault}, N., {et~al.} 2013, \aap, 552, A124

\bibitem[{{Bohlin}(2007)}]{2007ASPC..364..315B}
{Bohlin}, R.~C. 2007, in Astronomical Society of the Pacific Conference Series,
  Vol. 364, The Future of Photometric, Spectrophotometric and Polarimetric
  Standardization, ed. C.~{Sterken}, 315

\bibitem[{{Bohlin}(2010)}]{2010AJ....139.1515B}
{Bohlin}, R.~C. 2010, \aj, 139, 1515

\bibitem[{Bohlin \& Gilliland(2004{\natexlab{a}})}]{bohlin_absolute_2004}
Bohlin, R.~C. \& Gilliland, R.~L. 2004{\natexlab{a}}, Astronomical Journal,
  128, 3053

\bibitem[{Bohlin \& Gilliland(2004{\natexlab{b}})}]{bohlin_hubble_2004}
Bohlin, R.~C. \& Gilliland, R.~L. 2004{\natexlab{b}}, Astronomical Journal,
  127, 3508

\bibitem[{{Bohlin} {et~al.}(2014){Bohlin}, {Gordon}, \&
  {Tremblay}}]{2014arXiv1406.1707B}
{Bohlin}, R.~C., {Gordon}, K.~D., \& {Tremblay}, P.-E. 2014, ArXiv e-prints

\bibitem[{{Boulade} {et~al.}(2003){Boulade}, {Charlot}, {Abbon}, {Aune},
  {Borgeaud}, {Carton}, {Carty}, {Da Costa}, {Deschamps}, {Desforge},
  {Eppell{\'e}}, {Gallais}, {Gosset}, {Granelli}, {Gros}, {de Kat}, {Loiseau},
  {Ritou}, {Rouss{\'e}}, {Starzynski}, {Vignal}, \& {Vigroux}}]{Boulade2003}
{Boulade}, O., {Charlot}, X., {Abbon}, P., {et~al.} 2003, in Society of
  Photo-Optical Instrumentation Engineers (SPIE) Conference Series, Vol. 4841,
  Society of Photo-Optical Instrumentation Engineers (SPIE) Conference Series,
  ed. {M.~Iye \& A.~F.~M.~Moorwood}, 72--81

\bibitem[{{Brown} {et~al.}(2000){Brown}, {Eppeldauer}, \&
  {Lykke}}]{2000Metro..37..579B}
{Brown}, S.~W., {Eppeldauer}, G.~P., \& {Lykke}, K.~R. 2000, Metrologia, 37,
  579

\bibitem[{{Brown} {et~al.}(2006){Brown}, {Eppeldauer}, \&
  {Lykke}}]{2006ApOpt..45.8218B}
{Brown}, S.~W., {Eppeldauer}, G.~P., \& {Lykke}, K.~R. 2006, \ao, 45, 8218

\bibitem[{CALSPEC(2000)}]{calspec}
CALSPEC. 2000, Calspec, http://www.stsci.edu/hst/observatory/cdbs/calspec.html

\bibitem[{{DePoy} {et~al.}(2008){DePoy}, {Abbott}, {Annis}, {Antonik},
  {Barcel{\'o}}, {Bernstein}, {Bigelow}, {Brooks}, {Buckley-Geer}, {Campa},
  {Cardiel}, {Castander}, {Castilla}, {Cease}, {Chappa}, {Dede}, {Derylo},
  {Diehl}, {Doel}, {DeVicente}, {Estrada}, {Finley}, {Flaugher}, {Gaztanaga},
  {Gerdes}, {Gladders}, {Guarino}, {Gutierrez}, {Hamilton}, {Haney}, {Holland},
  {Honscheid}, {Huffman}, {Karliner}, {Kau}, {Kent}, {Kozlovsky}, {Kubik},
  {Kuehn}, {Kuhlmann}, {Kuk}, {Leger}, {Lin}, {Martinez}, {Martinez},
  {Merritt}, {Mohr}, {Moore}, {Moore}, {Nord}, {Ogando}, {Olsen}, {Onal},
  {Peoples}, {Qian}, {Roe}, {Sanchez}, {Scarpine}, {Schmidt}, {Schmitt},
  {Schubnell}, {Schultz}, {Selen}, {Shaw}, {Simaitis}, {Slaughter}, {Smith},
  {Spinka}, {Stefanik}, {Stuermer}, {Talaga}, {Tarle}, {Thaler}, {Tucker},
  {Walker}, {Worswick}, \& {Zhao}}]{2008SPIE.7014E..0ED}
{DePoy}, D.~L., {Abbott}, T., {Annis}, J., {et~al.} 2008, in Society of
  Photo-Optical Instrumentation Engineers (SPIE) Conference Series, Vol. 7014,
  Society of Photo-Optical Instrumentation Engineers (SPIE) Conference Series,
  0

\bibitem[{{Doi} {et~al.}(2010){Doi}, {Tanaka}, {Fukugita}, {Gunn}, {Yasuda},
  {Ivezi{\'c}}, {Brinkmann}, {de Haars}, {Kleinman}, {Krzesinski}, \& {French
  Leger}}]{Doi2010}
{Doi}, M., {Tanaka}, M., {Fukugita}, M., {et~al.} 2010, \aj, 139, 1628

\bibitem[{{Hayes} \& {Latham}(1975)}]{1975ApJ...197..593H}
{Hayes}, D.~S. \& {Latham}, D.~W. 1975, \apj, 197, 593

\bibitem[{Holtzman {et~al.}(2008)Holtzman, Marriner, Kessler, Sako, Dilday,
  Frieman, Schneider, Bassett, Becker, Cinabro, {DeJongh}, Depoy, Doi,
  Garnavich, Hogan, Jha, Konishi, Lampeitl, Marshall, {McGinnis}, Miknaitis,
  Nichol, Prieto, Riess, Richmond, Romani, Smith, Takanashi, Tokita, van~der
  Heyden, Yasuda, \& Zheng}]{holtzman_sloan_2008}
Holtzman, J.~A., Marriner, J., Kessler, R., {et~al.} 2008, Astronomical
  Journal, 136, 2306

\bibitem[{{Horowitz} \& {Hill}(1989)}]{1989arel.book.....H}
{Horowitz}, P. \& {Hill}, W. 1989, {The Art of Electronics - 2nd Edition}

\bibitem[{{Houston} \& {Rice}(2006)}]{2006Metro..43S..31H}
{Houston}, J.~M. \& {Rice}, J.~P. 2006, Metrologia, 43, 31

\bibitem[{Juramy(2006)}]{juramy:tel-00592266}
Juramy, C. 2006, Theses, {Universit{\'e} Pierre et Marie Curie - Paris VI}

\bibitem[{{Juramy} {et~al.}(2008){Juramy}, {Barrelet}, {Schahmaneche},
  {Bailly}, {Bertoli}, {Evrard}, {Ghislain}, {Guimard}, {Huppert}, {Imbault},
  {Laporte}, {Lebbolo}, {Repain}, {Sefri}, {Vallereau}, {Vincent}, {Antilogus},
  {Astier}, {Guy}, {Pain}, {Regnault}, {Attapatu}, {Benedict}, {Barrick},
  {Cuillandre}, {Gajadhar}, {Ho}, \& {Salmon}}]{2008SPIE.7014E.166J}
{Juramy}, C., {Barrelet}, E., {Schahmaneche}, K., {et~al.} 2008, in Society of
  Photo-Optical Instrumentation Engineers (SPIE) Conference Series, Vol. 7014,
  Society of Photo-Optical Instrumentation Engineers (SPIE) Conference Series

\bibitem[{{Kaiser} {et~al.}(2010){Kaiser}, {McCandliss}, {Pelton}, {Sahnow},
  {Dixon}, {Feldman}, {Gaither}, {Lazear}, {Moos}, {Riess}, {Rauscher}, {Kruk},
  {Kimble}, {Benford}, {Foltz}, {Gardner}, {Hill}, {Kahle}, {Malumuth}, {Mott},
  {Waczynski}, {Wen}, {Woodgate}, {Bohlin}, {Deustua}, {Kurucz}, {Lampton},
  {Perlmutter}, \& {Wright}}]{2010hstc.workE..10K}
{Kaiser}, M.~E., {McCandliss}, S.~R., {Pelton}, R., {et~al.} 2010, in Hubble
  after SM4. Preparing JWST

\bibitem[{{Keller} {et~al.}(2007){Keller}, {Schmidt}, {Bessell}, {Conroy},
  {Francis}, {Granlund}, {Kowald}, {Oates}, {Martin-Jones}, {Preston},
  {Tisserand}, {Vaccarella}, \& {Waterson}}]{2007PASA...24....1K}
{Keller}, S.~C., {Schmidt}, B.~P., {Bessell}, M.~S., {et~al.} 2007, \pasa, 24,
  1

\bibitem[{{Larason} \& {Houston}(2008)}]{NIST.SPECIAL.PUBLICATION.250.1}
{Larason}, T.~C. \& {Houston}, J.~M. 2008

\bibitem[{{Marshall} {et~al.}(2013){Marshall}, {Rheault}, {DePoy}, {Prochaska},
  {Allen}, {Behm}, {Martin}, {Veal}, {Villanueva}, {Williams}, \&
  {Wise}}]{2013arXiv1302.5720M}
{Marshall}, J.~L., {Rheault}, J.-P., {DePoy}, D.~L., {et~al.} 2013, ArXiv
  e-prints

\bibitem[{{McGraw} {et~al.}(2012){McGraw}, {Zimmer}, {Zirzow}, {Woodward},
  {Lykke}, {Cramer}, {Deustua}, \& {Hines}}]{2012SPIE.8450E..1SM}
{McGraw}, J.~T., {Zimmer}, P.~C., {Zirzow}, D.~C., {et~al.} 2012, in Society of
  Photo-Optical Instrumentation Engineers (SPIE) Conference Series, Vol. 8450,
  Society of Photo-Optical Instrumentation Engineers (SPIE) Conference Series

\bibitem[{{Montalto} {et~al.}(2007){Montalto}, {Piotto}, {Desidera}, {de
  Marchi}, {Bruntt}, {Stetson}, {Arellano Ferro}, {Momany}, {Gratton},
  {Poretti}, {Aparicio}, {Barbieri}, {Claudi}, {Grundahl}, \&
  {Rosenberg}}]{2007A&A...470.1137M}
{Montalto}, M., {Piotto}, G., {Desidera}, S., {et~al.} 2007, \aap, 470, 1137

\bibitem[{{Rauch} {et~al.}(2013){Rauch}, {Werner}, {Bohlin}, \&
  {Kruk}}]{2013A&A...560A.106R}
{Rauch}, T., {Werner}, K., {Bohlin}, R., \& {Kruk}, J.~W. 2013, \aap, 560, A106

\bibitem[{Regnault {et~al.}(2009)Regnault, Conley, Guy, Sullivan, Cuillandre,
  Astier, Balland, Basa, Carlberg, Fouchez, Hardin, Hook, Howell, Pain,
  Perrett, \& Pritchet}]{regnault_photometric_2009}
Regnault, N., Conley, A., Guy, J., {et~al.} 2009, Astronomy and Astrophysics,
  506, 999

\bibitem[{{Stubbs} {et~al.}(2010){Stubbs}, {Doherty}, {Cramer}, {Narayan},
  {Brown}, {Lykke}, {Woodward}, \& {Tonry}}]{Stubbs2010}
{Stubbs}, C.~W., {Doherty}, P., {Cramer}, C., {et~al.} 2010, \apjs, 191, 376

\bibitem[{{Stubbs} {et~al.}(2007){Stubbs}, {Slater}, {Brown}, {Sherman},
  {Smith}, {Tonry}, {Suntzeff}, {Saha}, {Masiero}, \&
  {Rodney}}]{stubbs_preliminary_2007}
{Stubbs}, C.~W., {Slater}, S.~K., {Brown}, Y.~J., {et~al.} 2007, in
  Astronomical Society of the Pacific Conference Series, Vol. 364, The Future
  of Photometric, Spectrophotometric and Polarimetric Standardization, ed.
  {C.~Sterken}, 373--+

\bibitem[{Stubbs \& Tonry(2006)}]{stubbs_toward_2006}
Stubbs, C.~W. \& Tonry, J.~L. 2006, Astrophysical Journal, 646, 1436

\bibitem[{{Suzuki} {et~al.}(2012){Suzuki}, {Rubin}, {Lidman}, {Aldering},
  {Amanullah}, {Barbary}, {Barrientos}, {Botyanszki}, {Brodwin}, {Connolly},
  {Dawson}, {Dey}, {Doi}, {Donahue}, {Deustua}, {Eisenhardt}, {Ellingson},
  {Faccioli}, {Fadeyev}, {Fakhouri}, {Fruchter}, {Gilbank}, {Gladders},
  {Goldhaber}, {Gonzalez}, {Goobar}, {Gude}, {Hattori}, {Hoekstra}, {Hsiao},
  {Huang}, {Ihara}, {Jee}, {Johnston}, {Kashikawa}, {Koester}, {Konishi},
  {Kowalski}, {Linder}, {Lubin}, {Melbourne}, {Meyers}, {Morokuma}, {Munshi},
  {Mullis}, {Oda}, {Panagia}, {Perlmutter}, {Postman}, {Pritchard}, {Rhodes},
  {Ripoche}, {Rosati}, {Schlegel}, {Spadafora}, {Stanford}, {Stanishev},
  {Stern}, {Strovink}, {Takanashi}, {Tokita}, {Wagner}, {Wang}, {Yasuda},
  {Yee}, \& {Supernova Cosmology Project}}]{2012ApJ...746...85S}
{Suzuki}, N., {Rubin}, D., {Lidman}, C., {et~al.} 2012, \apj, 746, 85

\end{thebibliography}

\appendix

\section{LED selection}
\label{sec:led_selection}

\paragraph{SnDICE LEDs}\begin{figure}
\centering
\includegraphics[width=\linewidth]{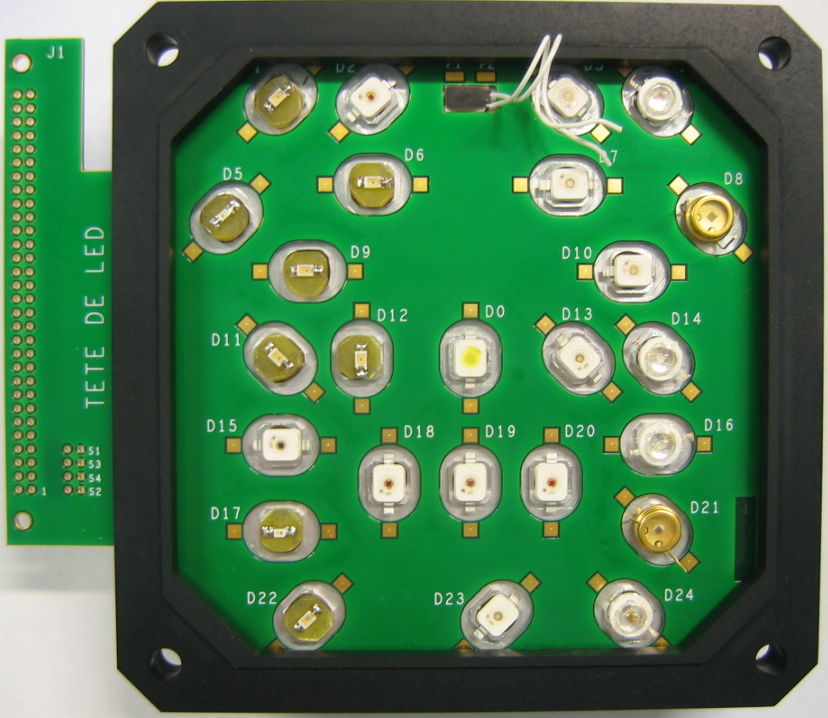}
\caption{SnDICE LEDs mounted on their front-end board.  The
  characteristics of all LEDs are summarised in Table
  \ref{tab:led_properties}.  The central LED in slot D0 is used to
  generate the planet beam.  The flat-top LEDs mounted in a white
  packaging are the high-intensity OSRAM Golden
  Dragon\textsuperscript\textregistered.  The remaining slots host
  lower intensity LEDs purchased from Marubeni and Seoul
  Semiconductors. They cover the redder wavelength range.}
\label{fig:sndice_led_board}
\end{figure}

We now give additional information about the calibration LEDs used for
SnDICE.  In the wavelength range $400\ \mathrm{nm} < \lambda < 630 \
\mathrm{nm}$, we use a series of high-intensity LEDs, manufactured by
OSRAM Semiconductors GmbH and marketed under the name Golden
Dragon\textsuperscript\textregistered.  The Golden
Dragon\textsuperscript\textregistered\ LEDs are flat top InGaN (up to
$\sim 550\ \mathrm{nm}$) and InGaIP (beyond $550\ \sim \mathrm{nm}$)
emitters, able to deliver up to a few hundred milliwatts at maximum current. 
This is actually
far more than what is needed. At the telescope, they are therefore
operated on a low regime and emit about $\sim 0.5$ mW/sr.

At longer wavelengths, we use AlGaAs LEDs, manufactured by
Marubeni America Corporation (SMC series).  These components are much
less powerful, delivering a maximum of ten milliwatts.  They
nevertheless suit our needs perfectly, as we only need a fraction of
this illumination.

In the UV, the LEDs that were available on the market in 2007 were
unfortunately much less powerful.  We selected a series of components
manufactured by Seoul Semiconductors. They all emit a maximum of a few
$10^{-2}$ mW per steradian.  As a consequence, SnDICE requires very
long exposures (hundreds of seconds) to accumulate enough flux in the
MegaCam $u$-band.

Figure \ref{fig:sndice_led_board} shows the SnDICE LED mother board.
The Golden Dragon LEDs are the flat top components mounted in white
packaging.  The Marubeni emitters are the flat top LEDs, on the
left. Whenever possible, we tried to choose flat top devices, which
are Lambertian emitters with a very good approximation.  All the blue
LEDs procured from Seoul Semiconductors have either a flat (D8, D21)
or hemispheric (D4, D14, D16, D24) protection.  They are nevertheless
quasi-Lambertian emitters in the angular range we are considering
($\pm 1\degree$).

The wavelength coverage of the MegaCam passbands is shown in figure
\ref{fig:filter_coverage_sndice} and discussed earlier in this
paper. It is not very satisfactory, mainly because of the gap around
700-nm, which leaves the red cutoff of the $r$-band and the blue cutoff
the the $i$-band unconstrained. SnDICE was designed in 2007, and the
LED diversity was not then what it is today.  However, the SnDICE
coverage was considered perfectly adequate for tests and feasibility
studies.

The types and properties of the LEDs that equip SnDICE are summarised
in Table \ref{tab:led_properties}.  We report the coefficients that
characterise the brighter-cooler and bluer-cooler relations. As can be
seen, the mean wavelength of the LED spectra vary typically by
1\AA/\celsius\ (note however, that some LEDs seem to display an inverse
behaviour).  The importance of the flux variations as a function of
temperature varies from LED to LED, from a fraction of one percentage point per
degree up to 3\%/\celsius\ for the UV LEDs.  Again, one LED seems to
display a different behaviour, at least at the nominal current at which
it is operated. These relations are (1) well measured and (2) very
reproducible, which is all that matters for our application.

\begin{table*}[t]
\begin{center}
  \caption{Summary of the SnDICE LED characteristics.}
\label{tab:led_properties}
\begin{tabular}{ccrccccl}
\hline
\hline
Model                 & Type              & $i_{LED}/i_{max}$   &  $\left<\lambda\right>$   &  $d\lambda/dT$               & Radiant intensity       & $d\Phi / (\Phi dT)$                    & channel       \\
                     &                   &                      &    (@ 25 \celsius)  &                            & (@ 25 \celsius) &                                 &               \\
                     &                   & (mA)                &   (nm)                     & $(\mathrm{nm / \degree C})$ & ($\mathrm{mW / sr}$)    & ($\mathrm{\% / \degree C}$) &               \\
\hline                                                                                                                   
S8D31C$^\odot$        & ALGaN       & 15 / 20         &   $312.5$              & $-0.035(027)$            & $0.02181(00021)$     & $-2.2129(0053)$ & D4          \\
S8D34C$^\odot$        & ALGaN       & 15 / 20         &   $342.3$              & $-0.027(008)$             & $0.01885(00003)$    & $-2.1448(0101)$ & D24         \\
T9F34C$^\odot$        & ALGaN       & 15 / 20         &   $387.5$              & $+0.059(005)$             & $0.01072(00002)$    & $-3.0912(0166)$ & D21         \\ 
T9F31C$^\odot$        & ALGaN       & 15 / 20         &   $388.0$              & $+0.086(005)$             & $0.01539(00001)$    & $-2.1406(0050)$ & D8          \\
S8D40$^\odot$         & InGaN (?)   & 17.5 / 350      &   $408.0$              & $+0.029(001)$             & $0.49891(00018)$    & $-0.4522(0033)$ & D14         \\ 
S8D42$^\odot$         & InGaN (?)   & 17.5 / 350      &   $422.2$              & $+0.015(002)$             & $0.20007(00019)$    & $+0.1983(0025)$ & D16         \\ 
LBW5SG$^{\ddagger}$    & InGaN      & 25 / 500        &   $466.4$              &  $+0.017(001)$             & $0.23207(00012)$    & $-0.2156(0005)$ & D10        \\ 
LBW5SG$^{\ddagger}$    & InGaN      & 25 / 500        &   $468.7$              &  $+0.010(001)$             & $0.33045(00016)$    & $-0.2153(0016)$ & D23        \\ 
LBW5SM$^{\ddagger}$    & ThinGaN    & 25 / 500        &   $474.3$              &  $+0.018(000)$             & $0.65071(00055)$    & $-0.2241(0007)$ & D3         \\ 
LTW5SG$^{\ddagger}$    & InGaN      & 25 / 500        &   $533.5$              &  $-0.012(000)$             & $0.12352(00003)$    & $-0.1301(0005)$ & D13        \\ 
LTW5SM$^{\ddagger}$    & ThinGaN    & 25 / 500        &   $540.6$              &  $+0.014(001)$             & $0.31282(00023)$    & $-0.5900(0013)$ & D7         \\ 
LYW5SM$^{\ddagger}$    & Thinfilm InGaAlP & 25 / 500  &   $590.6$              &  $+0.105(003)$             & $0.05893(00006)$    & $-0.3915(0022)$ & D20        \\ 
LAW5SM$^{\ddagger}$    & Thinfilm UnGaAlP & 25 / 500  &   $620.2$              &  $+0.121(001)$             & $-$                 & $-$                 & D19         \\ 
LRW5SM$^{\ddagger}$    & Thinfilm InGaAlP & 25 / 500  &   $630.3$              &  $+0.114(001)$             & $0.38592(00029)$    & $-0.2303(0005)$  & D2        \\ 
LRW5SM$^{\ddagger}$    & Thinfilm InGaAlP & 25 / 500  &   $630.6$              &  $+0.117(001)$             & $0.36542(00017)$    & $-0.3644(0021)$ & D18        \\ 
SMC750$^\star$        & AlGaAs           & 30 / 50         &    $729.4$              & $+0.170(002)$       & $0.23181(00018)$    & $-0.8754(0013)$ & D22        \\
SMC735$^\star$        & AlGaAs           & 30 / 50         &    $731.8$              & $+0.172(003)$       & $0.23947(00028)$    & $-0.4147(0008)$ & D12        \\
SMC735$^{\star\dagger}$       & AlGaAs    & 25 / 500        &    $733.1$              &  $+0.168(003)$      & $-$                 & $-$             & D6$^\dagger$  \\
SMC750$^\star$        &  AlGaAs          & 30 / 50         &    $746.8$              & $+0.163(002)$       & $0.29190(00003)$    & $-0.2369(0009)$ & D9            \\
SMC810$^{\star\dagger}$   & AlGaAs        & 30 / 50         &    $798.4$              & $+0.192(002)$       & $ - $                & $-$             & D1            \\
SMC810$^{\star\dagger}$   & AlGaAs        & 30 / 50         &    $-$                  & $-$                 & $ - $                & $-$             & D11$^\dagger$ \\
SFH 4230$^{\ddagger}$  & GaAs (?)        & 25 / 1000       &   $843.3$              &  $+0.223(001)$       & $0.37082(00037)$     & $ -0.1613(0008)$ & D15         \\  
SFH 4203$^{\bowtie}$  & GaAs            & 75 / 100        &    $942.4$              & $+0.308(003)$        & $0.69169(00091)$     & $ -0.2485(0050)$  & D17           \\
SFH 4203$^{\bowtie}$  & GaAs            & 75 / 100        &    $941.4$              & $+0.282(003)$        & $0.74315(00118)$     & $ -0.5340(0020)$ & D5            \\
\hline
\end{tabular}
\end{center}
\begin{list}{*}{}
\item[$\odot$] Seoul Semiconductor Co., Ltd -- \url{http://www.acriche.com}
\item[$\ddagger$] Golden Dragon\textsuperscript\textregistered\ series, OSRAM Opto Semiconductors GmbH -- \url{http://www.osram-os.com}
\item[$\star$] Marubeni America Corporation -- \url{http://tech-led.com}
\item[$\bowtie$] OSRAM Opto Semiconductors GmbH -- \url{http://www.osram-os.com}
\item[$\dagger$] dead channel
\end{list}
\end{table*}

\paragraph{SkyDICE LEDs} The SkyDICE source was designed about three years later. At that time,
the LED market had literally exploded, and a much higher diversity of
narrow spectrum LEDs was then available.  Figure
\ref{fig:filter_coverage_skydice} shows the resulting sampling of the
SkyMapper filters.  As discussed earlier in this paper, this coverage is
much better, although the LED density in the blue part of the spectrum
could be improved, in order to obtain tighter constrains on the narrow
$u$- and $v$-bands.

\begin{table*}[t]
\begin{center}
\caption{Summary of the SkyDICE LED characteristics.}
\label{tab:skydice_leds}
\begin{tabular}{ccrccccl}
\hline
\hline
Model                 & Type              & $i_{LED}/i_{max}$   &  $\left<\lambda\right>$   &  $d\lambda/dT$               & Radiant intensity       & $d\Phi / (\Phi dT)$                    & channel       \\
                     &                   &                     &    (@ 25 \celsius)        &                              & (@ 25 \celsius) &                                 &               \\
                     &                   & (mA)                &   (nm)                     & $(\mathrm{nm / \degree C})$ & ($\mathrm{mW / sr}$)    & ($\mathrm{\% / \degree C}$) &               \\
\hline 
UVTOP315-FW-TO39$^\clubsuit$ & AlGaN        &  12.2 /  20 & $320.7$        & $+0.001$(014) & 0.0866(0002) &  $-0.330 (014)$ & D8   \\
UVTOP335-FW-TO39$^\clubsuit$ & AlGaN        &  12.2 /  20 & $340.4$        & $+0.044$(025) & 0.1081(0002) &  $-0.665 (009)$ & D21  \\
APG2C1-365-S    $^\clubsuit$ & InGaN        &  21.4 / 350 & $368.3$        & $+0.027$(006) & 0.6625(0006) &  $-1.534 (004)$ & D2   \\
APG2C1-385$^{\clubsuit\dagger}$ & InGaN      &   $-$ / 350 & $-$            & $-$           &  $-$         &  $-$            & D23$^\dagger$\\
APG2C1-395$^\clubsuit$       & InGaN        &  10.7 / 350 & $396.7$        & $+0.020$(019) & 0.9606(0006) &  $-1.003 (003)$ & D3   \\
APG2C1-420$^\clubsuit$       & InGaN        &   2.1 / 350 & $417.0$        & $+0.015$(016) & 0.2486(0003) &  $-0.535 (005)$ & D22  \\
LD W5AM$^\ddagger$           & Thin GaN     &   3.1 / 500 & $452.6$        & $+0.013$(009) & 1.1194(0010) &  $-0.117 (005)$ & D4   \\
LB W5SM$^\ddagger$           & Thin GaN?    &   3.1 / 500 & $466.6$        & $+0.016$(019) & 1.6429(0027)  &  $-0.080 (009)$ & D17  \\
LV W5AM$^\ddagger$           & Thin GaN?    &   3.1 / 500 & $515.9$        & $-0.004$(012) & 0.6586(0001) &  $+0.256 (001)$ & D5   \\
LT W5SM$^\ddagger$           & Thin GaN     &   3.1 / 500 & $546.8$        & $+0.010$(008) & 0.6469(0002) &  $+0.016 (001)$ & D24  \\
LT W5AM$^\ddagger$           & Thin GaN     &   3.1 / 500 & $528.8$        & $-0.025$(020) & 0.3338(0001) &  $-0.372 (001)$ & D6   \\
LY W5SM$^\ddagger$           & Thin InGaAlP &  15.3 / 500 & $590.8$        & $+0.102$(009) & 0.3544(0004) &  $-0.268 (007)$ & D20  \\
LA W5SM$^\ddagger$           & Thin InGaAlP &   3.1 / 500 & $622.6$        & $+0.104$(020) & 0.2292(0001) &  $-0.515 (003)$ & D7   \\
APG2C1-660$^\clubsuit$       & GaAlAs       &  15.3 / 500 & $656.8$        & $+0.144$(009) & 0.5411(0004) &  $-0.418 (004)$ & D18  \\
APG2C1-690$^\clubsuit$       & GaAlAs       &   3.1 / 500 & $687.6$        & $+0.155$(020) & 0.1697(0001) &  $-0.433 (002)$ & D1   \\
APG2C1-720$^\clubsuit$       & GaAlAs       &   3.1 / 500 & $716.5$        & $+0.153$(020) & 0.2319(0001) &  $-0.434 (001)$ & D16  \\
APG2C1-760$^\clubsuit$       & GaAlAs       &   3.1 / 500 & $759.9$        & $+0.176$(017) & 0.2686(0001) &  $-0.166 (001)$ & D15  \\
APG2C1-810$^\clubsuit$       & GaAlAs       &   3.1 / 500 & $806.4$        & $+0.177$(031) & 0.2785(0001) &  $-0.197 (002)$ & D10  \\
APG2C1-830$^\clubsuit$       & GaAlAs       &   3.1 / 500 & $828.2$        & $+0.203$(011) & 0.3401(0003) &  $-0.352 (004)$ & D19  \\
APG2C1-850$^\clubsuit$       & GaAlAs       &   3.1 / 500 & $845.7$        & $+0.189$(025) & 0.2519(0002) &  $-0.297 (004)$ & D12  \\
SFH421$^\ddagger$            & Thin InGaAlP &   6.1 / 100 & $730.8$        & $+0.153$(009) & 0.4493(0001) &  $-0.566 (001)$ & D13  \\
APG2C1-905$^\clubsuit$       & GaAlAs       &   3.1 / 500 & $913.6$        & $+0.133$(087) & 0.1497(0001) &  $-0.437 (004)$ & D9   \\
APG2C1-940$^\clubsuit$       & GaAlAs       &  15.3 / 500 & $950.5$        & $+0.142$(016) & 0.5216(0007) &  $-0.485 (007)$ & D14  \\
APG2C1-970$^\clubsuit$       & GaAlAs       &   3.1 / 500 & $951.4$        & $+0.272$(059) & 0.1441(0001) &  $-0.343 (003)$ & D11  \\
\hline
\hline
\end{tabular}
\end{center}
\begin{list}{*}{}
\item[$\clubsuit$] Roithner Lasertechnik -- \url{http://www.roithner-laser.com/}.
\item[$\ddagger$] Golden Dragon\textsuperscript\textregistered\ series, OSRAM Opto Semiconductors GmbH -- \url{http://www.osram-os.com/osram_os/en/}.
\item[$\dagger$] Dead channel.
\end{list}
\end{table*}

The types and properties of the SkyDICE LEDs are summarised in Table
\ref{tab:skydice_leds}.  Again, in the visible, we use InGaN and
InGaAlP Golden Dragon\textsuperscript\textregistered\ components.  In
the IR, we use a new family of high-intensity GaAlAs LEDs distributed
by Roithner Lasertechnik GmbH.  Finally, in the UV, down to $\sim
360$-nm, the situation has improved considerably in terms of emitted
power. In this region, we rely on a series of InGaN LEDs from Roithner
Lasertechnik. 

These new-generation LEDs are more powerful than their SnDICE
counterparts, and they have to be operated at even lower currents.  Again,
the parameters of the brighter-cooler and bluer-cooler relations are
reported in Table \ref{tab:skydice_leds}.  They are similar in
magnitude to what was measured with SnDICE.  As mentioned
earlier in this paper, the amplitude of these effects do strongly
depend on the LED technology and mean wavelength.

\paragraph{Optimised designs} The diversity of narrow spectrum LEDs
continues to improve at a rapid pace (see upper right panel of figure
\ref{fig:filter_coverage}).  A quick glance at the Roithner and OSRAM
catalogues shows that there are about 50 LEDs covering the
wavelength range $245\ \mathrm{nm} < \lambda < 1100\ \mathrm{nm}$,
allowing us to sample the imager wavelength range with almost one LED
every 10 to 20-nm.  One may therefore wonder whether it is
feasible to design a calibration source that would sample the imager
response in some kind of ``optimal'' way.

There are many ways to define optimality. We seek to identify the set
of LEDs that would allow us to obtain the best constraints on the
model described in section \ref{sec:analysis} after one single
calibration run.  We proceed as follows. We first identify the LED
positions that would yield the best constraints on the filter fronts:
these are the positions that maximise the derivatives of the
$\varphi_{bl}$ model as a function of the filter cutoff displacements
$\delta\lambda_b$ and $\delta\lambda_r$.  We match these positions as
precisely as possible using the LEDs available on the market.  In
general, we are able to find a LED within 10 nm of the optimal
position. It is possible to play on the diversity of LEDs of
the same model to reach optimality even better. Finally, we cover the
gaps between the filter fronts as densely as possible, with about one
LED every 25-nm.  

For MegaCam, this procedure yields the coverage shown in the upper panel (left) of figure
\ref{fig:filter_coverage}.
The dotted lines show our ``optimal'' positions.  The blue bands show
the peak emission of the spectra ($\pm 5\%$ of the peak position).
The LEDs available in the OSRAM and Roithner catalogues are indicated
with arrows. In practice, we have seen on simulations that the optimal
design (dotted lines) and its actual implementation (arrows) are
nearly indistinguishable in terms of performances.  For SkyMapper, we
have not been able to identify a design that would give constraints
significantly better than those obtained from the current source. 

In section \ref{sec:analysis}, we systematically compare the
performances of the existing DICE sources with these optimised
designs.  In particular, we show that, although SnDICE is far from
optimality, SkyDICE is nearly optimal, except for the narrow SkyMapper
$v$-band.


\section{Glitch observed on a APG2C1-760}
\label{sec:on_led_9}

\begin{figure*}
  \begin{center}
    \includegraphics[width=\linewidth]{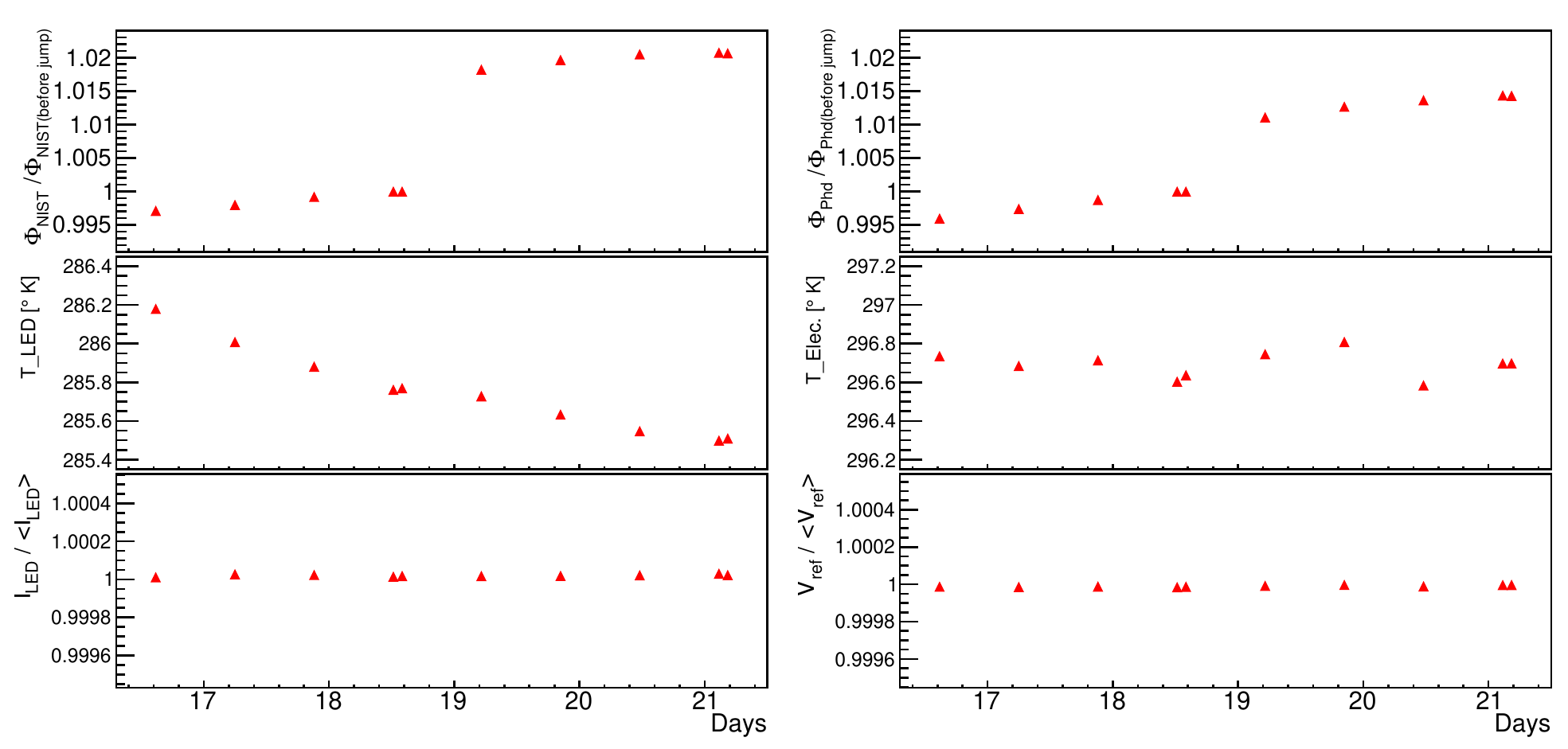}
    \caption{Time series registered during a long duration test of a a
      Roithner-Lasertechnik APG2C1-760 LED mounted on a DICE
      source. After 18 days of operations, the flux registered by the
      NIST photodiode (upper left panel) suddenly increased by almost
      2\%.  Only a fraction of this variation was detected by the
      internal off-axis control photodiode (upper right panel).  The
      other panels display the timeline of the other quantities
      monitored by the backend electronics: the LED temperature
      (middle left panel), the backend board temperature (middle right
      panel), the LED current (lower right panel), and the board
      reference tension (lower right). No corresponding glitch can be
      observed on these quantities. }
    \label{fig:glitch_apg2c1_760}
  \end{center}
\end{figure*}

In section \ref{sec:photometric_calibration}, we mentioned a 2\%
glitch observed on a Roithner-Lasertechnik APG2C1-760 LED during a
long-duration test.  In figure \ref{fig:glitch_apg2c1_760} we show the
timelines registered by the source and the bench sensors. We see in
the upper left-hand panel that after about two weeks of uninterrupted
operation, the LED flux suddenly increased by almost two percent.
This increase was detected by the off-axis control photodiode that
monitors the LED flux (upper right panel).  However, for a reason that
is not fully understood, it registered only a fraction of the
variation ($\sim 1\%$).

The other panels of figure \ref{fig:glitch_apg2c1_760} display the
timelines of the source parameters, monitored along with the LED flux:
the LED temperature, the temperature of the backend board, the LED
current, and the reference tension of the electronics.  No
corresponding glitch was registered on any of these quantities.

As of today, we have no clear explanation for this phenomenon. It may
be due to a sudden variation in the structure of the LED junction. We
have not been able to observe any similar event in any of our
subsequent long duration tests.  In any case, should such an event
occur during operations, it would be detectable by the off-axis
control photodiode, and the corresponding LED would not be used in
subsequent operations (until recalibration of the source).


\section{Spectroscopic calibration bench}
\label{sec:spectroscopic_calibration_details}

\paragraph{Test bench setup and operations}

The test bench setup is presented in figure
\ref{fig:calibration_test_bench}.  The monochromator is operated in a
slightly non-standard way, in the sense that the LEDs are not placed
just in front of the entrance slit, but at a distance of 270 mm. Since
the LED emissive zone is very small, the light that enters the
monochromator is a quasi-pencil beam, with an angle distribution less
than 0.2\degree wide.

The value of the mean angle of the incident beam is defined by the
relative positioning of the LED and the entrance slit. We
(conservatively) estimate a positioning error of the LED head with
respect to the monochromator of about 1 mm. This
translates into an uncertainty on the incident beam angle of
0.2\degree.

The output is a larger beam, convolved with the monochromator
response.  At the level of the photodiode, it produces an illumination
of width 4 mm and height 6 mm with an extended penumbra zone. It is
almost entirely contained in the S2281 NIST photodiode used to measure
it.
{Since we were
  concerned that there could be chromatic flux losses affecting the
  measurements, the beam was scanned spatially and in wavelength in
  order to check for diffuse light and parasitic reflexions. We found that the flux loss is essentially constant (0.5\%) as a
  function of wavelength in the $400\ \mathrm{nm}<\lambda< 800\ \mathrm{nm}$ range. It
  then increases up to 1\% for redder wavelengths. 
  The LED SED being narrow, the shapes of the
  measured spectra are not affected by this effect.}

The size of the entrance and output slits define the monochromator 
wavelength passband --through the dispersion relation.
The latter has been remeasured by
studying images of monochromatic sources with various slits
apertures. We find $3.2 \pm 0.11$ nm per millimetre aperture in
accordance with the manufacturer indications.  We set the slit
openings to 625 \micro m and the monochromator step to 2 nm 
for optimal sampling, during the full calibration
process. This choice is a trade-off between fineness of sampling and
S/N (which is critical for UV LEDs).

Apart from the grating and picoammeter range, which differ for each LED, the measurement
protocol is the same for all calibration channels. The temperature is
set to the desired value, and we let the bench thermalise. Each LED is positioned in front of the entrance slit, and a fast
wavelength scan is performed in order to check the SED extension. Each
spectrum is then sampled every 2 nm. At each step, we alternate 
measurements of the dark current and the LED flux. Spectra are
obtained at least at six different temperatures ranging from
0 \celsius\ to 25 \celsius.

\paragraph{Wavelength Calibration}

The monochromator does not come with a temperature-dependent
wavelength calibration.  During the tests, the monochromator
temperature has been found to vary between {5}{\celsius} and
{25}{\celsius.} (The bench is never fully thermalised, and the
monochromator is always slightly warmer than the LEDs.)  It is therefore
essential to check for potential temperature-dependent effects. The
absolute wavelength calibration of the monochromator was determined
 at {25}{\celsius}, {18}{\celsius}
and {13}{\celsius} using two calibration lamps: one Cd lamp and 
one Hg lamp. We have found a small wavelength offset, $\Delta
\lambda$, that varies slightly as a function of the temperature of the
monochromator, $T_M$:
\begin{equation}
  \Delta\lambda = \Delta\lambda_{grating} + \beta_{grating} \times \left(T_M - T_M^0\right)
.\end{equation}
The values of $\Delta\lambda_{grating}$ and $\beta_{grating}$ are
reported in Table \ref{tab:absolute_wavelength_calibration}. This
correction was applied to all wavelength measurements.  The final
wavelength calibration error is classically derived by propagating the
uncertainties on $\Delta\lambda_{grating s}$ and $\beta_{grating}$ and on
the monochromator temperature ($\sigma_{T_M} \sim
{1}{\celsius}$). {We find that it never exceeds {1}{\AA} (for UV LEDs).}

\begin{table}[t]
\begin{center}
  \caption{Absolute wavelength calibration offsets.}
\label{tab:absolute_wavelength_calibration}
\begin{tabular}{l|cc}
\hline
\hline
Grating    &  $\Delta\lambda_{grating}$ @ $T_M^0$={25}{\celsius} &  $\beta_{grating} = d\Delta\lambda /dT$       \\
\#       &    (nm)          &  ({\celsius})\\
\hline
1        &  $+0.154 \pm 0.246$  &  $-0.02499 \pm 0.0023$  \\
2        &  $+0.020 \pm 0.059$  &  $-0.02526 \pm 0.0016$  \\
3        &  $-0.008 \pm 0.096$  &  $-0.02654 \pm 0.0022$  \\
\hline
\end{tabular}
\end{center}
\end{table}

\paragraph{Transmission}

The monochromator transmission is a rapidly varying function of
$\lambda$. It cannot be considered to be flat on the scale of one single
LED spectrum, so it has to be measured precisely.  To perform this
measurement, we placed a {800}~{\micro m} wide mask in front
of the monochromator entrance slit, at a distance of about 1 cm. The
latter was opened slightly wider than in normal operations
({1}{mm} instead of 625 \micro m), so that it contains a large, constant 
fraction of the beam.

\begin{figure}
\begin{center}
\includegraphics[width=\linewidth]{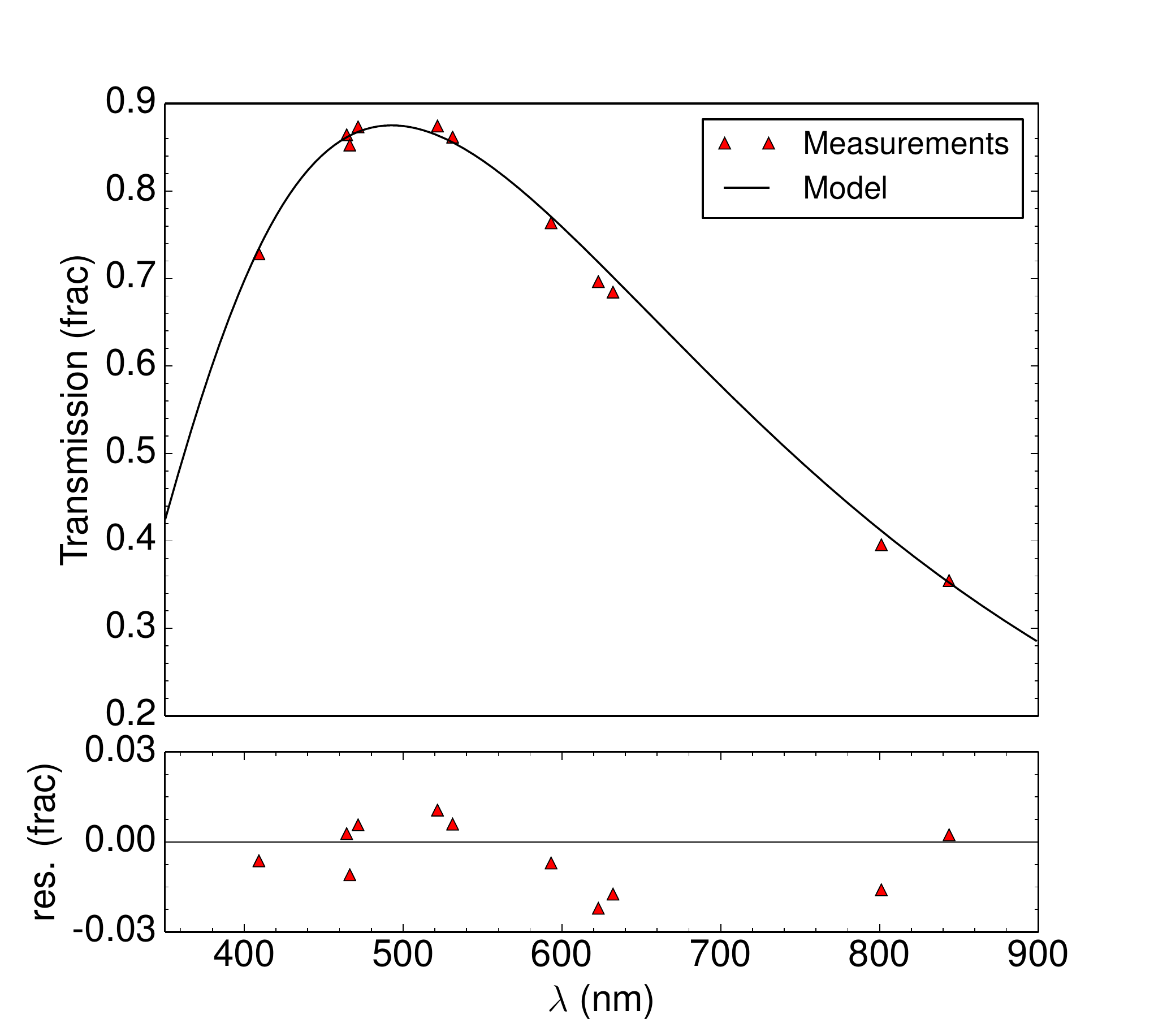}
\caption{Top panel: Model (dark line) and measurements (triangles) of monochromator grating \#2. 
  Bottom panel: fit residuals (same units as top panel). 
  \label{fig:monochromator_transmission}}
\end{center}
\end{figure}

The transmission of the monochromator at a given wavelength $\lambda$ can then
be estimated using a LED emitting in that wavelength range, by measuring the ratio between the flux measured 
with the monochromator $\phi(\lambda,LED)$ and the flux without it:
\begin{equation}
  T(\lambda) \propto \frac{\phi(\lambda,LED)}{\phi_{tot}(LED)} = 
  \frac{T(\lambda)\ \eta(\lambda)\ S_{LED}(\lambda)}{\int \eta(\lambda)\ S_{LED}(\lambda)d\lambda}
.\end{equation}
The monochromator transmission can be modelled knowing the groove
spacing (1200 g/mm). It is a simple function of the blaze angle
($\epsilon$), the Ebert angle ($\alpha$), and the focal
distance $f$: 
\begin{equation}
T(\lambda) \propto \mathrm{sinc}\left( \pi \cos(\epsilon) - \frac{\pi f}{\lambda}2\cos\left(\frac{\alpha}{2}\right)\sin(\epsilon)\sqrt{1 - \frac{\lambda^{2}}{4f^{2}\cos^{2}(\frac{\alpha}{2})}}\right)^2
\label{eqn:monochromator_transmission}
.\end{equation}
Figure \ref{fig:monochromator_transmission} shows the measurements performed over the full wavelength range of
grating \#2 ({500}{nm}). It shows excellent agreement of the model
and the observations. 
For example, for this grating, a fit of the model yields
$\theta_{blaze}= 18.1\pm 0.28 \degree$ and $\theta_{ebert} = 16.6\pm
1 \degree$, in remarkable agreement with the constructor data.

{Our transmission model does not include the
  reflectivities of the four mirrors that are inside the
  monochromator.  Inspecting the residuals of figure
  \ref{fig:monochromator_transmission}, we see a dip around
  $550-600$~nm that could be the signature of Al reflectivity.
  Incorporating an Al reflectivity function (at the fourth power
  because there are four mirrors) into the transmission model does not
  improve the fit significantly.  We find, however, that correcting
  locally for the dip has a negligible impact on the spectrum shape
  (less than 1\AA\ for the mean wavelength).  We therefore use the
  model as it is and consider that the uncertainties it carries
  essentially have no impact on the spectrum shapes. }


\section{Isotropy of the LED illumination}
\label{sec:led_illumination_isotropy}

{In this section, we report on a study conducted to look for a
  possible dependence between the LED spectra and the direction of
  emission, $\vec{u}$. We know already that the beam intensities vary by about
  1-2\% as a function of $\vec{u}$.  These
  non-uniformities are measured, as described in sections
  \ref{sec:photometric_and_spectroscopic_measurements} and
  \ref{sec:beam_maps}, and the resulting beam maps are included in the
  source emissivity model. Now, we want to verify that the {\em \emph{shape}} of
  the source spectrum also does not depend on the direction of emission. }

{In principle, we do not expect a large effect.  Indeed, the
  process that generates the LED light takes place everywhere in the
  junction.  For small angles, the photon generation and extraction
  mechanisms are the same.  However, we cannot exclude effects induced
  by the LED packaging or possibly interferences taking place within the
  junction.}

\begin{table*}[t]
  \begin{center}
    \caption{LED models tested for isotropy. }
    \label{tab:led_models_tested_for_isotropy}
    \begin{tabular}{l|cccccccc}
      \hline
      \hline
      LED model                 & Type              & Packaging         & $\left<\lambda\right>$ & $T_p\ (\Delta T)$  & $N_{\mathrm{runs}}$ & $\Delta \mathrm{angle}$ & $N_{\mathrm spectra}$ & duration \\
                                &                   &                   &    [nm]                & $\degree \mathrm{K}$   &              & [\degree]   &                 &    (days)   \\
      \hline
      APG2C1-395$^\clubsuit$     &  InGaN            &   Hemispheric     & 400.3                  &  301.5 (1.3)  &    5   & $[-7.5, 7.5]$ & 126   &  4.4 \\  
      LT W5SM-JXKX-36$^\ddagger$ &  Thin GaN         &   Flat top        & 536.9                  &  301.8 (3.9)  &   13   & $[-8.0, 8.0]$ & 465   & 25.1 \\  
      LA W5SM-GYHZ-35$^\ddagger$ &    `` (?)         &   Flat top        & 586.2                  &  303.4 (5.6)  &    6   & $[-15., 15.]$ & 298   &  5.7 \\  
      LY W5SM$^\ddagger$         &  Thinfilm InGaAlP &   Flat top        & 591.2                  &  301.3 (2.8)  &    7   & $[-6.8, 8.2]$ & 405   & 7.1 \\   
      APG2C1-810$^\clubsuit$     &  GaAlAs           &   Hemispheric     & 807.0                  &  300.9 (0.7)  &    5   & $[-7.5, 7.5]$ & 243   & 8.4 \\   
      SFH4203$^{\bowtie}$        &  GaAs             &   Flat top        & 944.3                  &  300.2 (3.6)  &    7   & $[-7.5, 7.5]$ & 1702  & 75.0 \\  
      \hline
    \end{tabular}
  \end{center}
  \begin{list}{*}{}
  \item[$\clubsuit$] Roithner Lasertechnik -- \url{http://www.roithner-laser.com/}.
  \item[$\ddagger$] Golden Dragon\textsuperscript\textregistered\ series, OSRAM Opto Semiconductors GmbH -- \url{http://www.osram-os.com}
  \item[$\bowtie$] OSRAM Opto Semiconductors GmbH -- \url{http://www.osram-os.com}
  \end{list}
\end{table*}

{This work was initiated after the DICE sources had been
  installed on site. As a consequence, it was not performed on the
  real sources, but on a small model built by gluing a selection of
  LEDs on a radiator, and connecting them to a laboratory current source.
  This system is slightly less stable in flux than the real DICE
  sources (0.1 \% stability).  However, it is stable enough for our
  purpose.}

\begin{figure*}
  \begin{center}
    \mbox{
      \subfigure[APG2C1-395 (301.5 K)]{\includegraphics[width=0.45\textwidth]{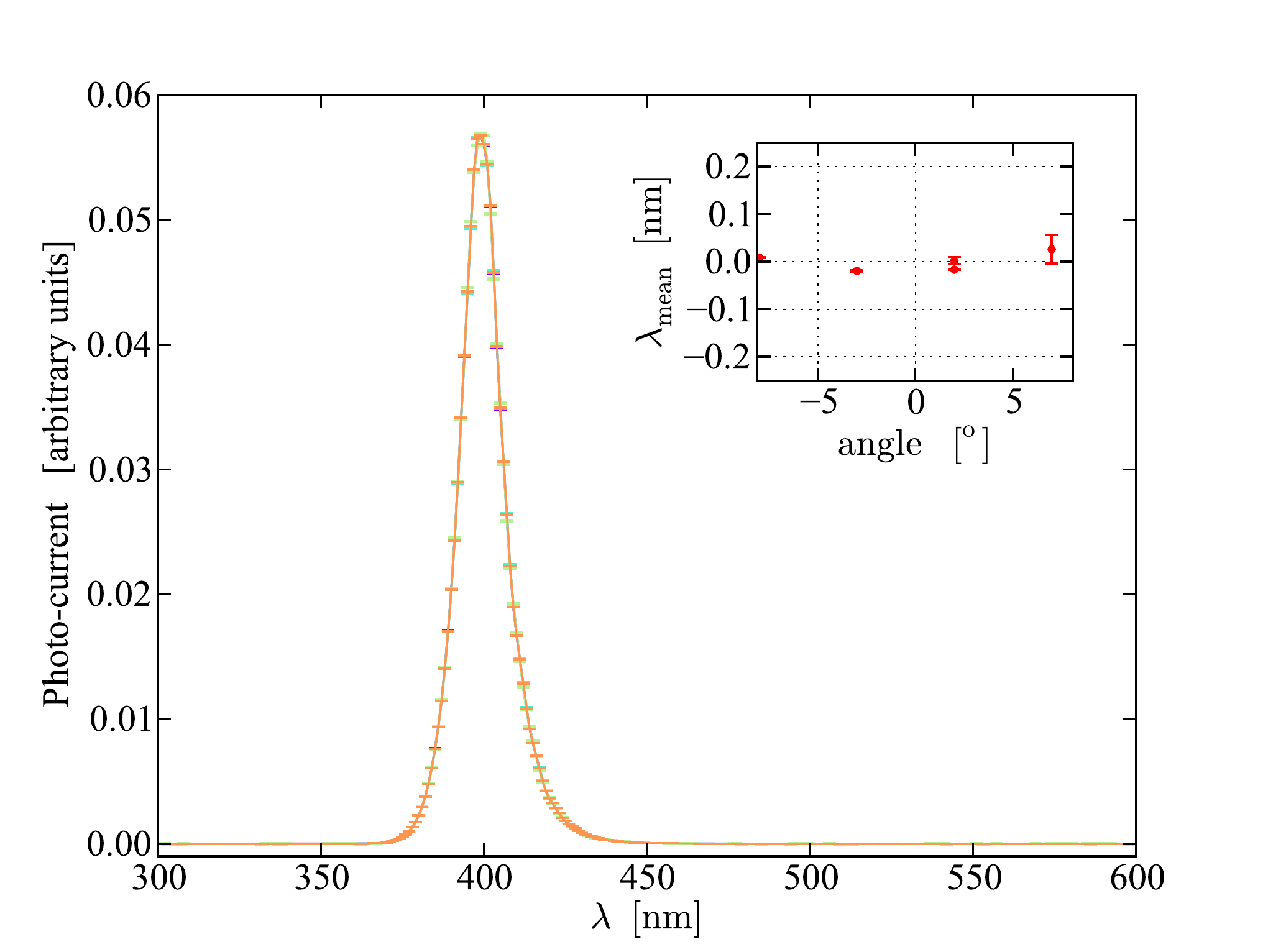}}
      \subfigure[LT W5SM-JXKX-36 (301.8 K)]{\includegraphics[width=0.45\textwidth]{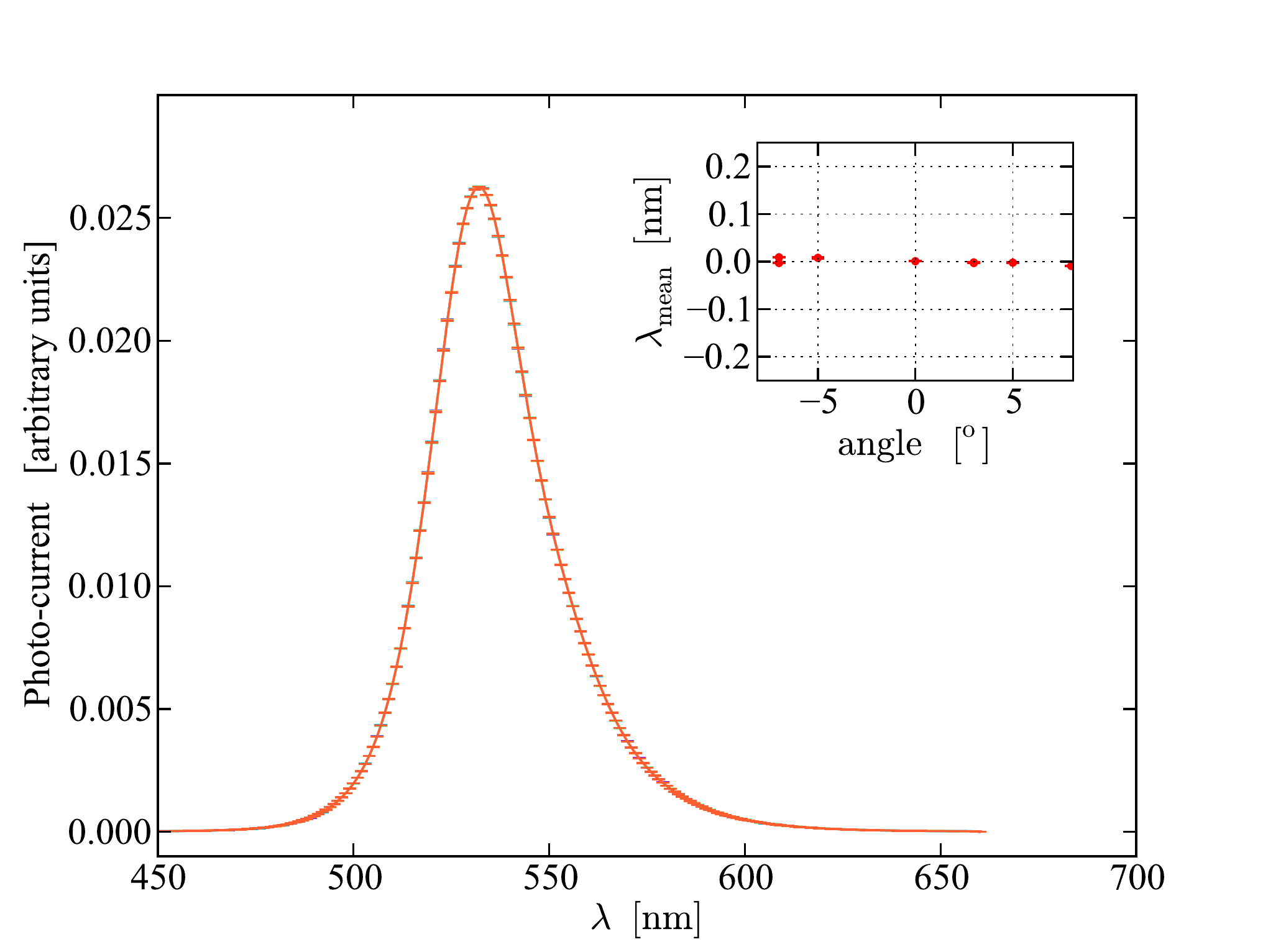}}
      }
    \mbox{
      \subfigure[LA W5SM-GYHZ-35 (303.4 K)]{\includegraphics[width=0.45\textwidth]{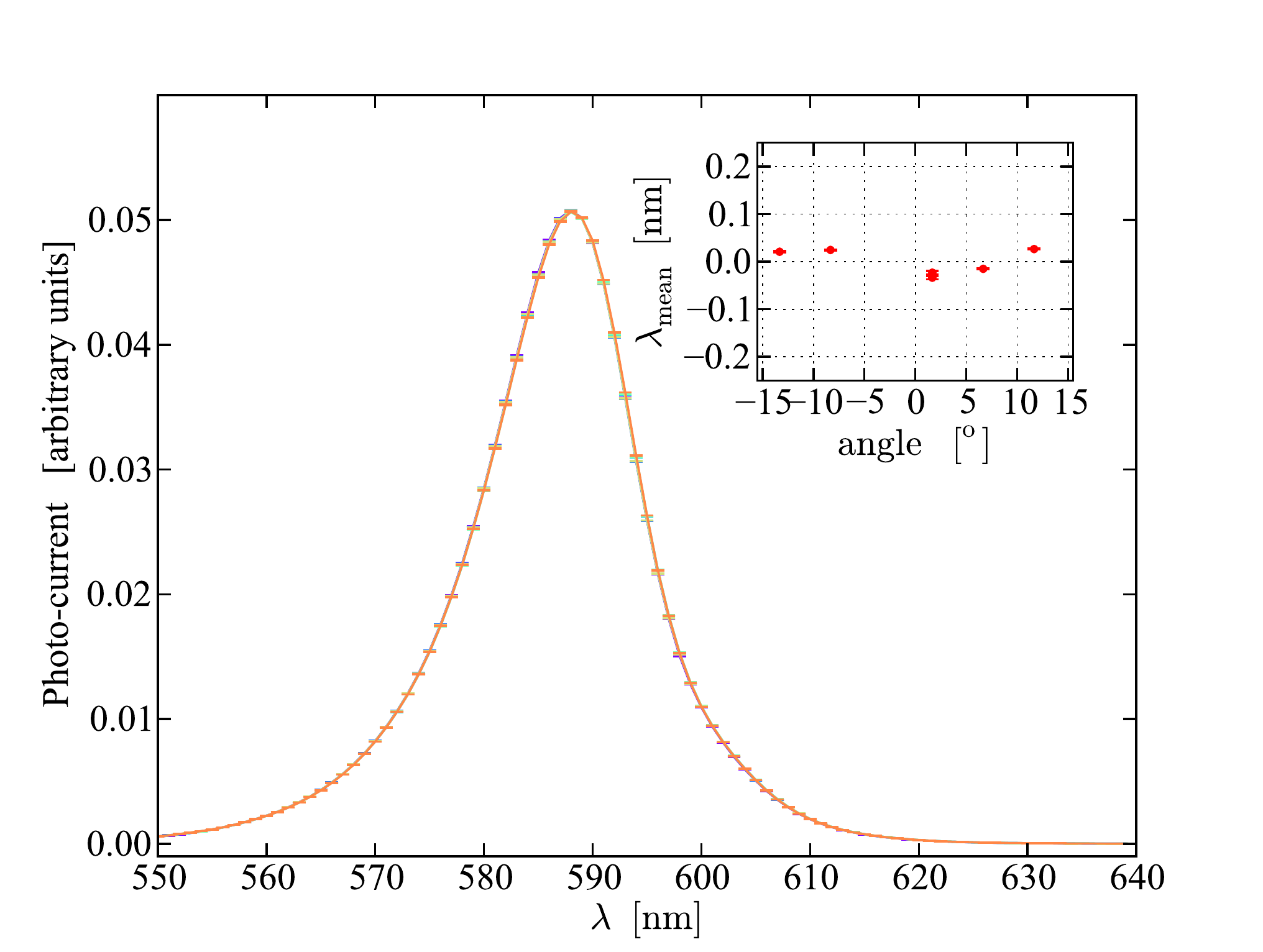}}
      \subfigure[LY W5SM (301.3 K)]{\includegraphics[width=0.45\textwidth]{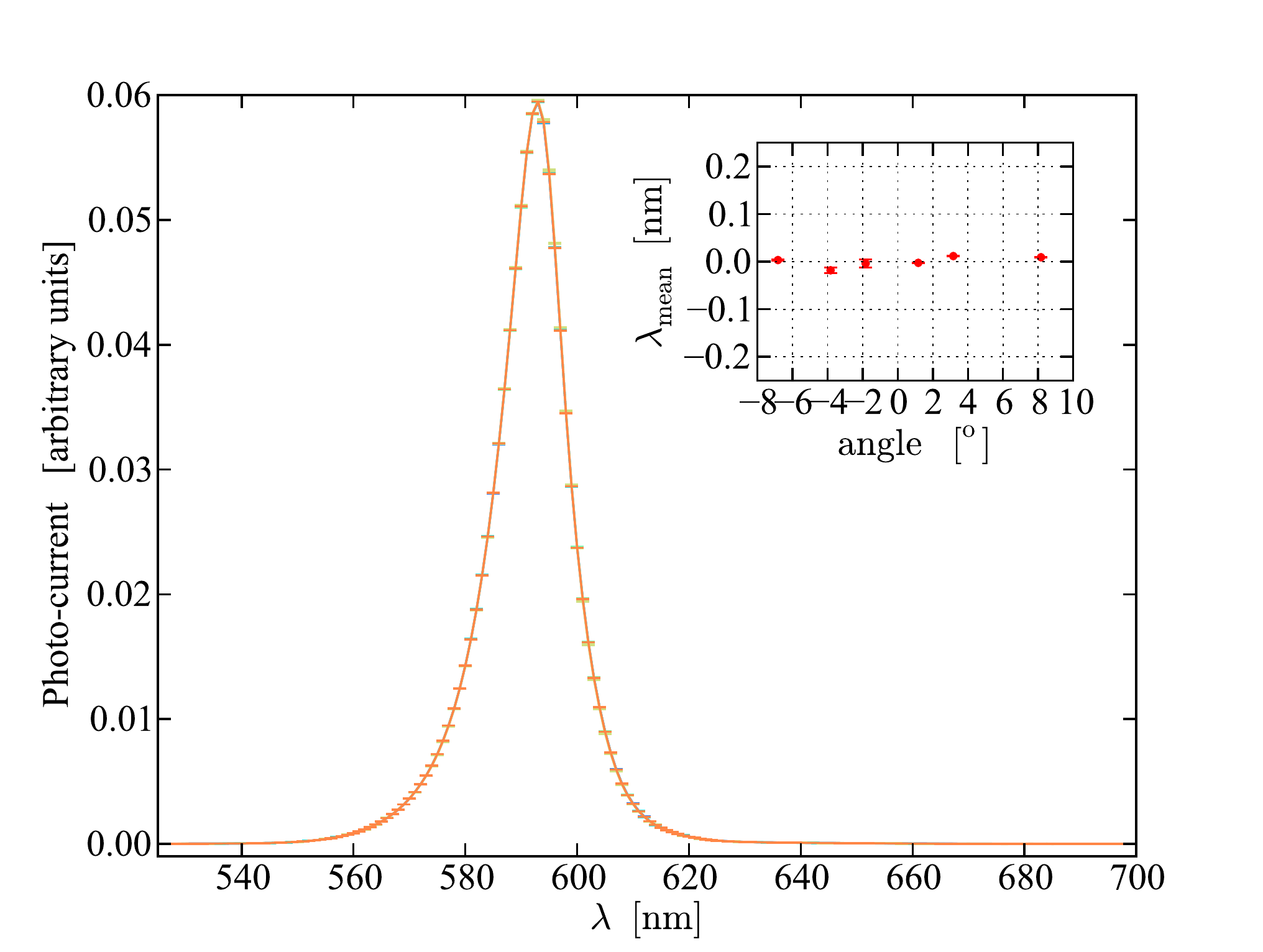}}
      }
    \mbox{
      \subfigure[APG2C1-810 (300.9 K)]{\includegraphics[width=0.45\textwidth]{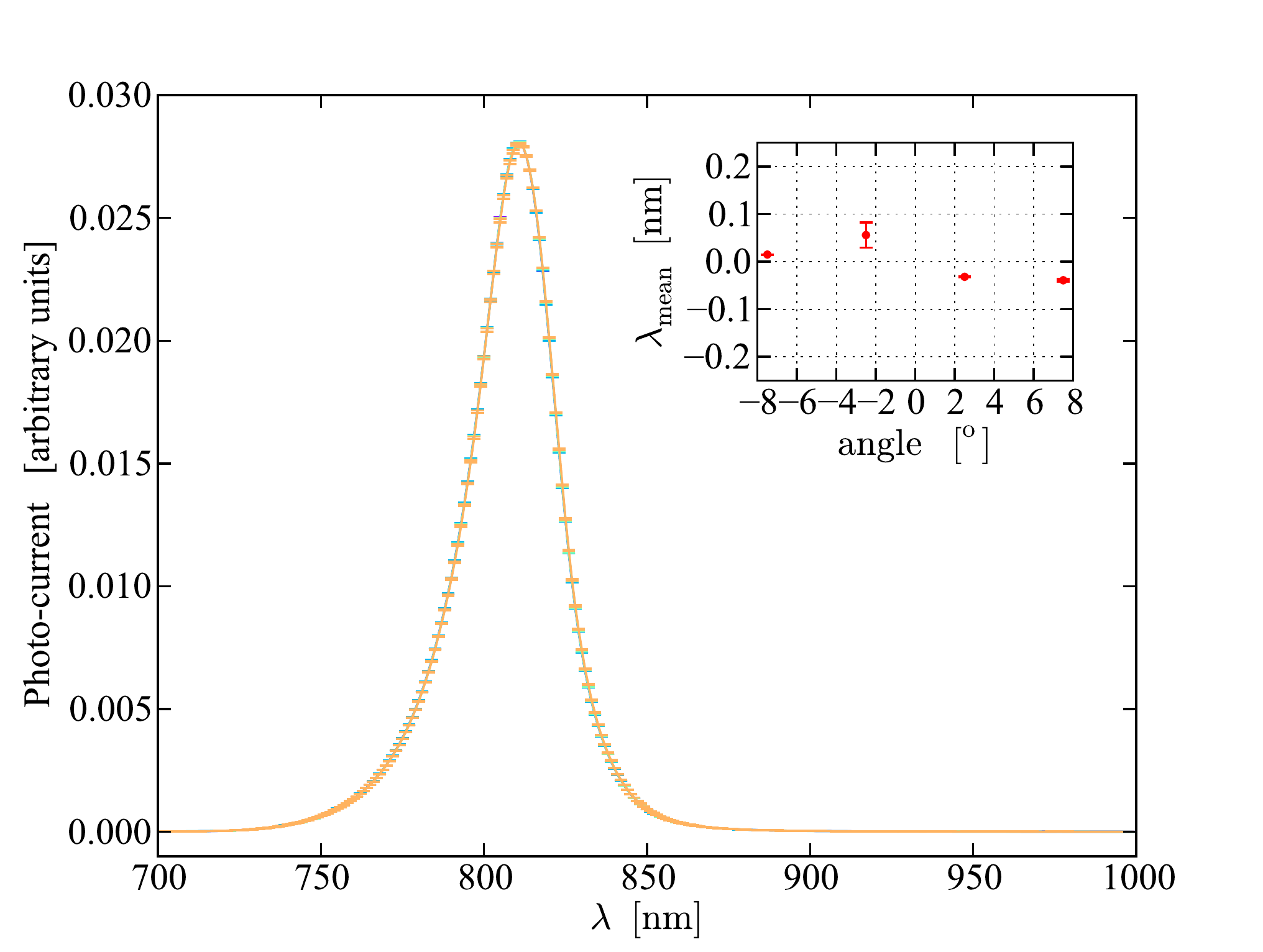}}
      \subfigure[SFH4203 (300.2 K)]{\includegraphics[width=0.45\textwidth]{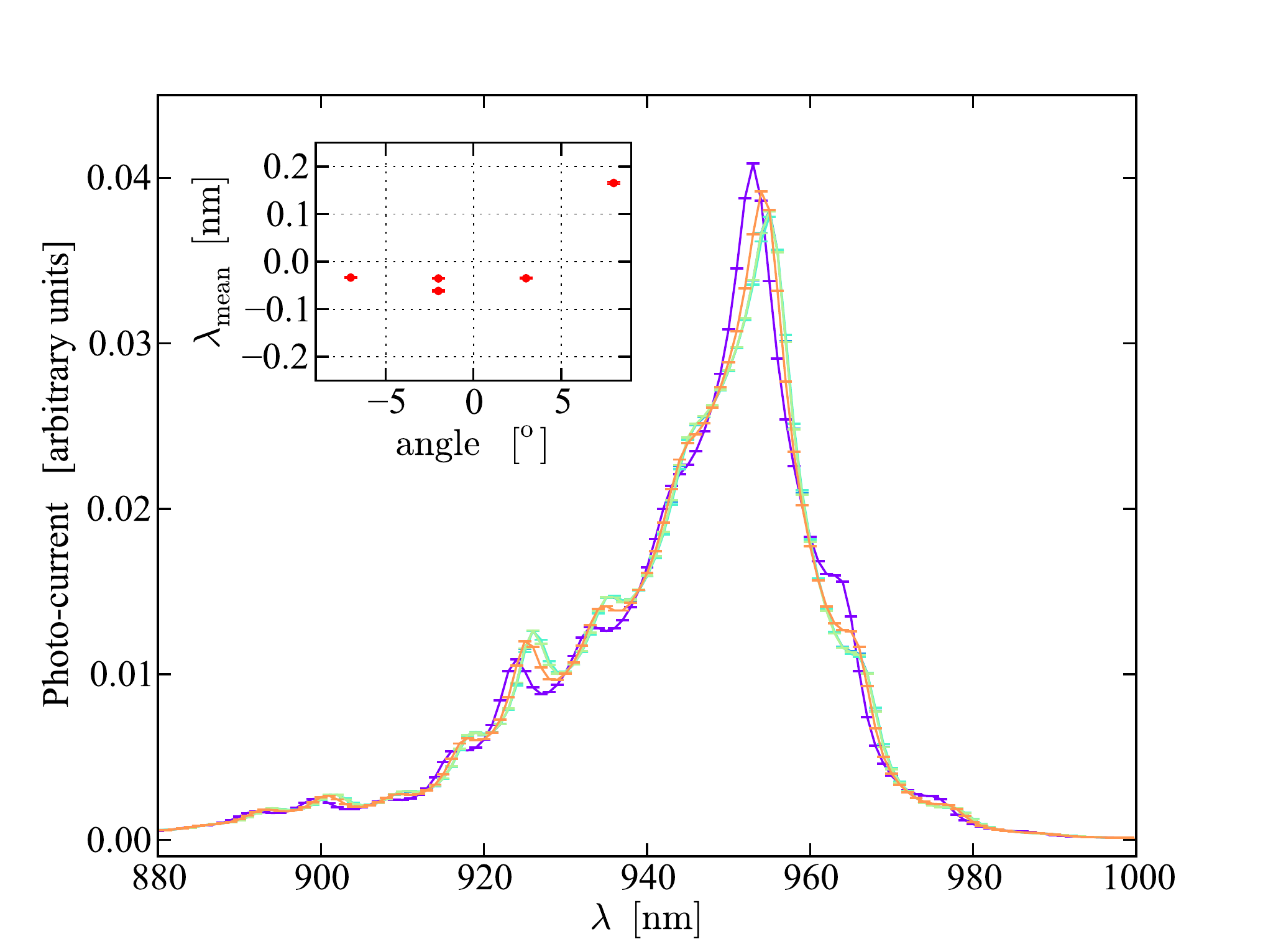}}
      }      
      
      \caption{All panels show the superposition the LED spectrum
        models, fitted on each run.  Each model therefore corresponds to a specific angle of incidence.  Except for the
        IR emitter (SFH4203), the spectra at various angles can barely
        be distinguished from each other.  The insets show the
        spectrum average wavelength $\left<\lambda\right> = \int
        \lambda S(\lambda) d\lambda / \int S(\lambda) d\lambda$
        computed on each run. This quantity varies by less than 1\AA
        (peak-to-peak) over a $\sim 15\degree$ range.  Except for SFH4203, these variations are likely to be due to
        residual temperature effects, which not well accounted for by the
        crude model used for this
        study. \label{fig:led_spectrum_isotropy}}
  \end{center}
\end{figure*}

{Our selection of LEDs is listed in Table
  \ref{tab:led_models_tested_for_isotropy}.  They are chosen so as to
  cover the full spectral range of DICE ($3800 \AA < \lambda < 9500
  \AA$).  Also, we tried to probe as many technologies and packaging
  types as possible. All these LEDs are narrow-spectrum emitters with
  typical smooth spectra (figure \ref{fig:led_spectrum_isotropy}).
  There is one exception: the
  reddest LED (SFH4203) that displays sharp features probably due to
  fringing within the substrate.  For this LED, we expect a more
  pronounced dependence of the spectrum shape on the emission
  angle.}

{The source model is mounted on a support that can rotate
  around a vertical axis, and the LED light is injected at various
  angles wwith respect to the direction of normal emission into the Digikr\"om
  DK240 monochromator that equips the bench.  A photodiode, read out
  with a Keithly picoammeter, is placed at the exit slit of the
  monochromator.}

{Information about the dataset accumulated with this setup can
  be found in Table \ref{tab:led_models_tested_for_isotropy}. For each
  LED, series of consecutive spectra are taken at four to eight
  different angles, spanning a range of about $\pm 8$ degrees, around
  the direction of normal emission.  This range is chosen to be deliberately
  larger than the 1\degree\ field of view of MegaCam in order to
  enhance our chances of detecting an effect. This dataset represents
  hundreds of spectra per LED, taken over the course of several
  weeks. Incidentally, it allowed us to check the stability of the LED
  spectra over long periods of time.  }

{We call {\em run} a series of consecutive spectra taken at a
  same angle.  Each single run is analysed independently from the
  others. The only piece of information shared from one run to another is the
  measurement of the dark current, performed during dedicated data-gathering sessions.  The analysis consists in extracting from each run
  an average spectrum template $\hat{S}(\lambda)$, modelled in practice
  as a series of 1-nm-wide bins.  If the temperature was constant with
  a precision of a tenth of a degree, its effect on the spectral
  shapes (the cooler-bluer effect) could be neglected, and the model
  could be obtained simply by averaging the flux measurements in each
  single bin.}
  
{In fact, the temperature has been found to vary by up to
  a few degrees during each single run.  As a consequence, spectra
  taken at different angles were taken at slightly different
  temperatures.  This induces spectrum-shape variations that are independent of the
  effect we are looking for.  To account for this, we fit the
  following (temperature dependent) model on each run: }
\begin{equation*}
  i(\lambda, T) = \hat{S}(\lambda) + (T + \delta T - T_p) \times \frac{\partial \hat{S}}{\partial T}(\lambda) + i_{\mathrm{dark}}
\end{equation*}
 where {$i(\lambda, T)$ is the measured photocurrent,
  $i_{\mathrm{dark}}$ the dark current, $T$ the bench temperature
  reported by the bench DAQ at the time of measurement, $T_p$ a median pivot
  temperature common to all the runs performed with a given LED and
  reported in Table \ref{tab:led_models_tested_for_isotropy},  $\delta
  T$  an offset introduced to account for the fact that $T$ lags by
  a fraction of a degree on the real LED temperature. Here,
  $\hat{S}(\lambda)$ is the average LED spectrum at the pivot
  temperature,  ${\partial \hat{S}}/{\partial T}(\lambda)$
  corresponds to the (binned) derivatives of the spectrum with respect
  to temperature, and  $i_{\mathrm{dark}}$ is measured independently.
  Therefore, the parameters that are fitted are $\hat{S}$, $\partial
  \hat{S} / \partial T$, and the $\delta T$ offsets. This model has
  slight degeneracies, which we break by adding constraints to the
  $\chi^2$ of the form $(\delta T / \sigma)^2$ with $\sigma \sim
  0.1\degree C$. }

\begin{figure*}
  \begin{center}
    \mbox{
      \subfigure[LT W5SM-JXKX-36 (301.8 K)]{\includegraphics[width=0.5\textwidth]{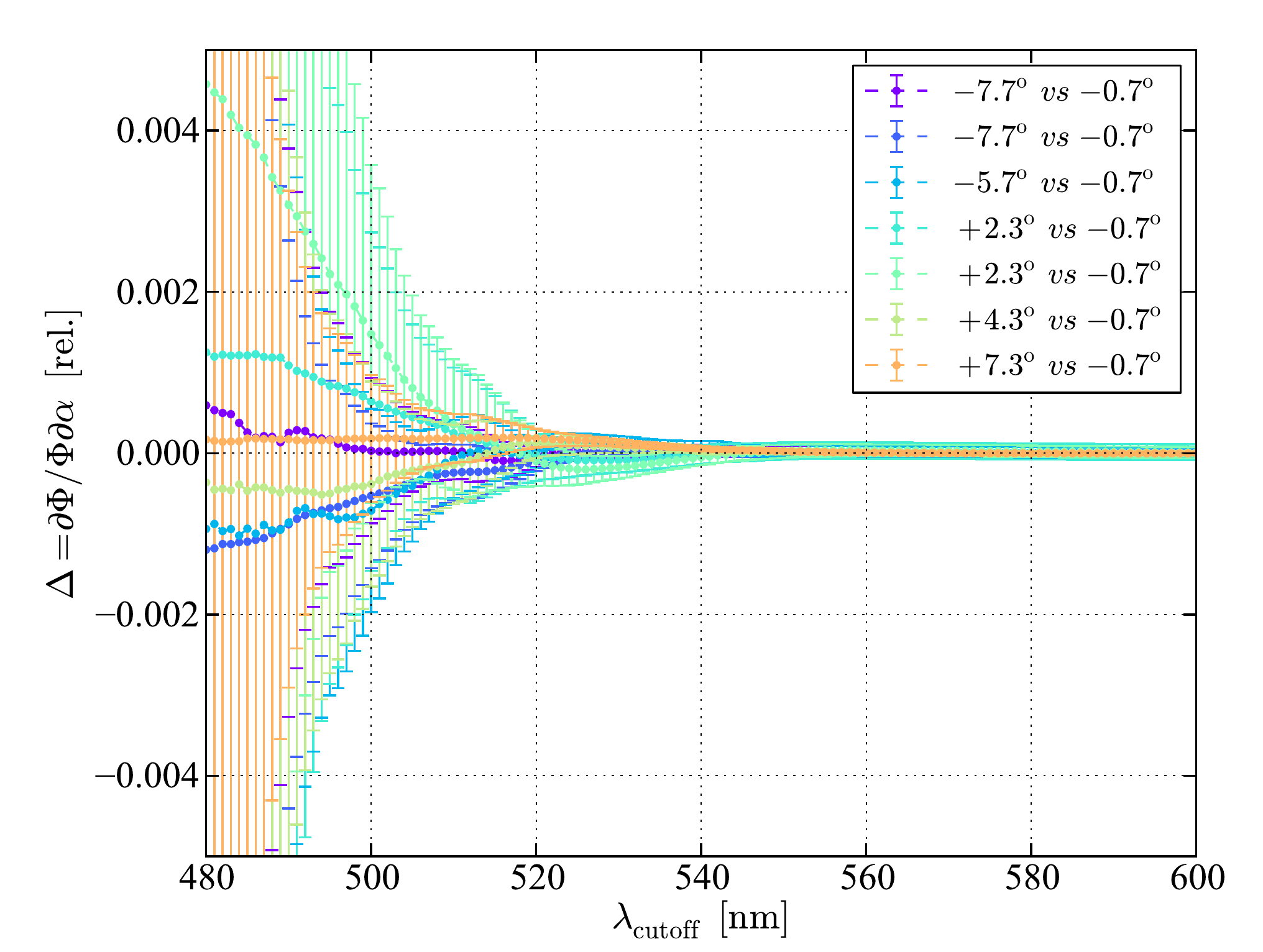}}
      \subfigure[SFH4203 (300.2 K)]{\includegraphics[width=0.5\textwidth]{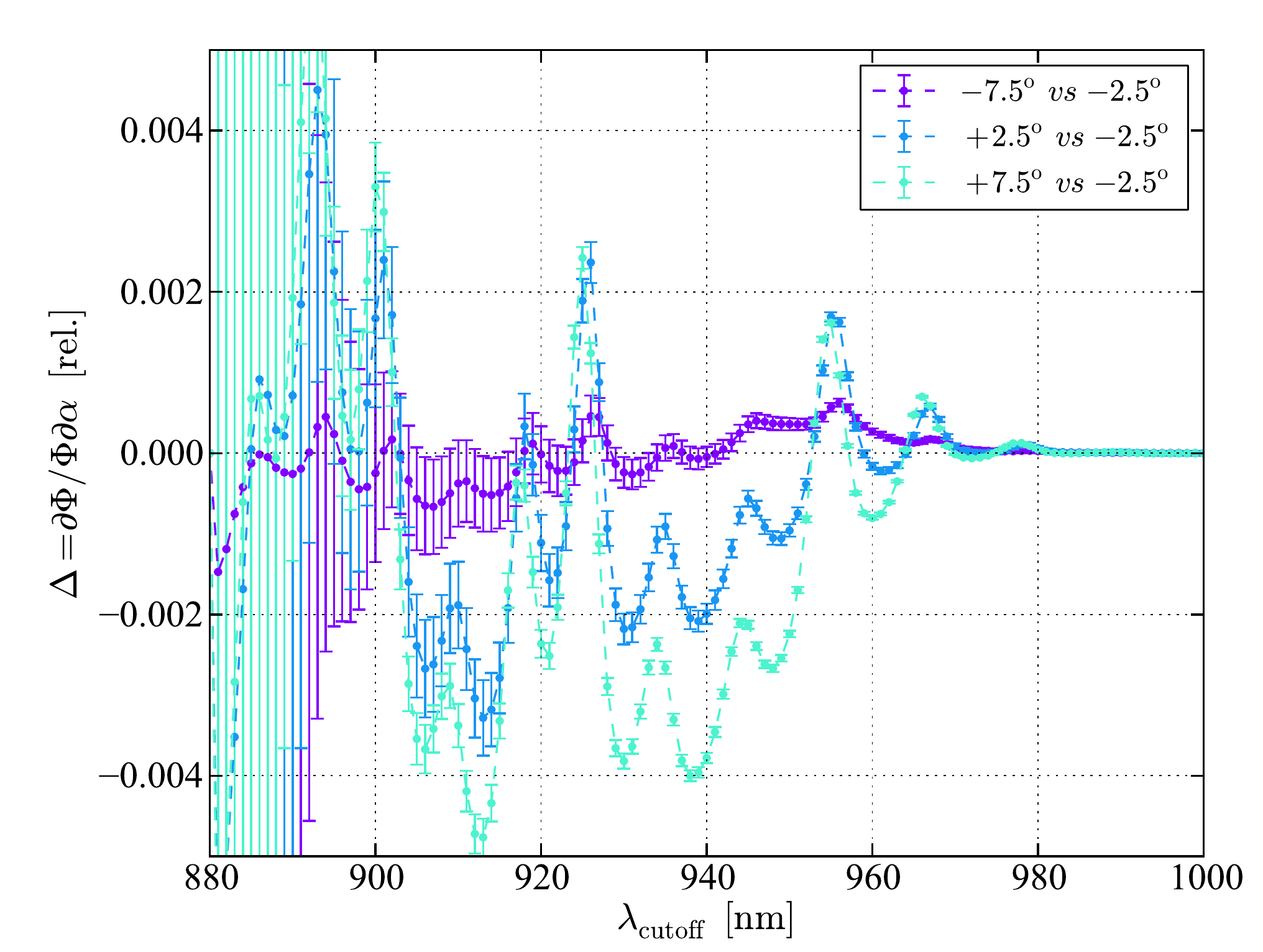}}
    }
  \end{center}
  \caption{Broadband flux sensitivity ($\Delta$) to
    emission-angle-dependent spectrum variations as a function of the
    red-cutoff of a rectangular broadband filter.  \label{fig:broadband_flux_versus_angle}}
\end{figure*}

{In figure \ref{fig:led_spectrum_isotropy}, we show the
  reconstructed LED models at their respective pivot temperatures for
  all the angles studied. For all LEDs that emit in the visible, the
  spectra taken at different angles are virtually indistinguishable.
  In the corresponding insets, we show the variations in the LED mean
  wavelengths $\left<\lambda\right>$, which are found to be stable at
  better than 1\AA\ over an angular range that is much greater than the
  acceptance of MegaCam.}

{The situation is different for the IR emitter, (SFH4203). We
  see ``spectral features'' whose positions and relative amplitudes
  depend on the angle of emission.  The question is how this
  affects the passband measurements.}

{Following the analysis sketched in section
  \ref{sec:analysis}, the bench spectra are combined with a
  (parametrised) passband model to predict broadband fluxes,
\begin{equation}
  \Phi = \int \hat{S}(\lambda) T(\lambda) d\lambda
,\end{equation}
which are then compared with the fluxes measured directly on the
calibration frames. An estimate of how much the LED broadband flux varies 
  over a 1\degree\ angle (which is precisely the size of the MegaCam focal plane) is 
\begin{equation}
\Delta = \frac{1}{\Phi}\ \frac{\partial\Phi}{\partial \alpha},\end{equation} 
where $\alpha$ is the direction of emission. }

{We use a rectangular filter to compute the broadband LED fluxes
  from the spectrum models fitted on each run. Here, $\Delta$ is computed
  numerically from the broadband fluxes at a given angle and the
  broadband flux at a reference angle.  We vary the filter cutoff
  with respect to  the LED spectrum in order to explore all the configurations.
  In figure \ref{fig:broadband_flux_versus_angle}, we show the value
  of $\Delta$ as a function of the red filter cutoff for two of our
  LEDs.  When the red filter cutoff moves towards the redder
  wavelengths, the  LED spectrum is fully encompassed by the filter
  shape, and the broadband flux becomes independent of the LED
  spectral shape.  On the other hand, when the cutoff moves towards
  bluer wavelengths, we explore the region where only a fraction of
  the LED light is integrated by the filter. This is the regime that
  constrains the filter cutoff positions.}

{For a normal LED (left), we see that $\Delta$ is of the order of $10^{-4}$. Only in the regime where a small fraction of the
  LED spectrum tail overlaps the filter do we observe small (but not
  significant) deviations around 0.1\%.  Conversely, for the IR-LED  SFH4203, we observe that the spectrum shape variations induce
  variations in the broadband fluxes of about 0.5\%. We conclude that
  except for one LED model, for which an effect was easily detected,
  no emission-angle dependent variation in the LED spectra could be
  seen over a range of angles of the order of $\pm 8\degree$. }


\section{Constraining passbands (details)}
\label{sec:passbandcal}

We now explain the estimation of the calibration parameters from series
of DICE measurements (section \ref{sec:analysis}).  The predicted flux
registered on the focal plane is given by equation
\ref{eqn:illu_model}, which can be rewritten in matrix form:
\begin{equation}
  \hat{\varphi}_{bl} = \delta\omega\ \ \vec{{\theta}}_{\hat{S}_l}^T \vec{\Sigma} 
\end{equation}
where $\delta\omega$ is the solid angle covered by the focal plane
pixel or superpixel, $\vec{\theta}_{\hat{S}_l}$ are the parameters of
the LED spectral intensity model $\hat{S}_l(\lambda)$, and
$\vec{\Sigma}$ is a vector whose components are the integrals of the
basis functions defined in section
\ref{sec:spectroscopic_calibration}, convolved with the telescope
transmission $T_b(\lambda)$: 
\begin{equation}
\vec{\Sigma}_p = \int B_p(\lambda, T)\ {T_b}(\lambda, \vec{\vartheta}_t)\ d\lambda
.\end{equation}

The calibration parameters $\vec{\vartheta}_t$ are determined by
minimizing a $\chi^2$ built from the measurements of the LED
calibration light observed through the different telescope passbands
$\phi_{bl}$, and from the corresponding predictions
$\hat{\varphi}_{bl}$. If we note $\vec{R}$ the vector of residuals, 
this $\chi^2$ writes as
\begin{equation}
  \chi^2 = \vec{R}^T (\vec{C}_{\mathrm{stat}} + \vec{C}_{\mathrm{led}}+ \vec{C}_{\mathrm{model}})^{-1} \vec{R}
,\end{equation}
where $\vec{C}_{\mathrm{stat}}$ is the covariance matrix of the flux
measurement uncertainties. It will be discussed in detail in the next
paper of the series.  Here, $\vec{C}_{\mathrm{led}}$ accounts for the LED
intrinsic variabilities and $\vec{C}_{\mathrm{model}}$ accounts for
the model uncertainties. The elements of $\vec{C}_{\mathrm{model}}$ are
propagated from the covariance matrix of the spectral intensity model
parameter, $\vec{C_\theta}$: $\mathrm{cov}(\phi_{bl}, \phi_{b'l}) =
\vec{\Sigma}_{lb}^T \vec{C_{\theta}} \vec{\Sigma}_{lb'}$.

Since $\varphi_{bl}$ is not a linear function of $\vec{\vartheta}_t$
we linearise it:
\begin{equation}
  {\varphi}_{bl} = \delta\omega\ \ \vec{\theta}_{\hat{S}_l}^T (\vec{\Sigma}_0 + \vec{Y}\ \vec{\delta\vartheta}_t)
  \label{eqn:phi_model}
\end{equation}
at each minimisation step. Here, $\vec{\Sigma}_0$ is the ``current''
value of the $\vec{\Sigma}$ vector at a given step, and ${\vec{Y}}$ is
a matrix containing its derivatives with respect to the calibration
parameters. Both are computed numerically, the full minimisation
taking a little less than five seconds on a laptop.

\paragraph{Systematics}
The systematics we have to consider come from two main sources.
First, we have the uncertainties that affect the test-bench
measurements.  Those have been discussed in section
\ref{sec:spectroscopic_calibration} and summarised as seven parameters
listed in Table \ref{tab:test_bench_systematics}.  We group them into
a vector $\vec{\eta}_b$, which comes along with a (diagonal) covariance
matrix $\vec{C}_b$.  These bench systematics have an impact on our
estimates of the LED spectral intensities, $\hat{S}_l(\lambda)$, and
therefore on our broadband flux predictions, $\varphi_{bl}$.

Second, the calibration measurements performed with the imager are
themselves affected by several systematics.  They include
uncertainties on (1) the LED temperature, (2) the positioning of the
source with respect to the telescope, (3) systematic drifts of the readout
electronics during the data-taking sequence, and (4) contamination of the
calibration frames by stray light, in particular the ghosts generated
by parasitic reflections on the optical surfaces. The most problematic
ones are those which are wavelength dependent, in particular the ghost
contamination, which is higher when illuminating the telescope with
LEDs close to the filter cutoff.  These contributions will be
discussed and quantified in detail in Paper II. We will leave them
aside for now. Our purpose is to describe the propagation
method and to evaluate the impact of the bench systematics.

All the contributions listed above are included as nuisance parameters
into the calibration fit described in section
\ref{sec:calibration_fit}, and marginalised over, using their
uncertainty estimates (encoded in matrix $\vec{C}_b$) as priors in the
$\chi^2$ -- this is the most direct and exact way to compute their
contribution to the total error budget. This means that we minimise
the following $\chi^2$:
\begin{equation}
  \chi^2 = \vec{R}^T (\vec{C}_{\mathrm{stat}} + \vec{C}_{\mathrm{led}} + \vec{C}_{\mathrm{model}})^{-1} \vec{R} + \vec{\eta}_b^T \vec{C}_b^{-1} \vec{\eta}_b 
.\end{equation}

Again, the fit is non-linear, and the model is linearised at each
step, as follows:
\begin{equation}
  \varphi_{bl} = \vec{\theta}_{\hat{S}_l}^T\ \left(\vec{\Sigma}_0 + 
  \vec{Y}\ \vec{\delta\vartheta} 
  \right) + 
  \vec{\Sigma}_0^T \vec{H}_{\hat{s}}\ \vec{\delta\eta}_b
  \label{eqn:phi_model_with_syst}
,\end{equation}
which is the equivalent of equation \ref{eqn:phi_model}, with the
systematics. The fit with systematics is very fast, and takes seconds
on a standard laptop.

\paragraph{Covariances}
\begin{figure}
  \begin{center}
    \includegraphics[width=\linewidth]{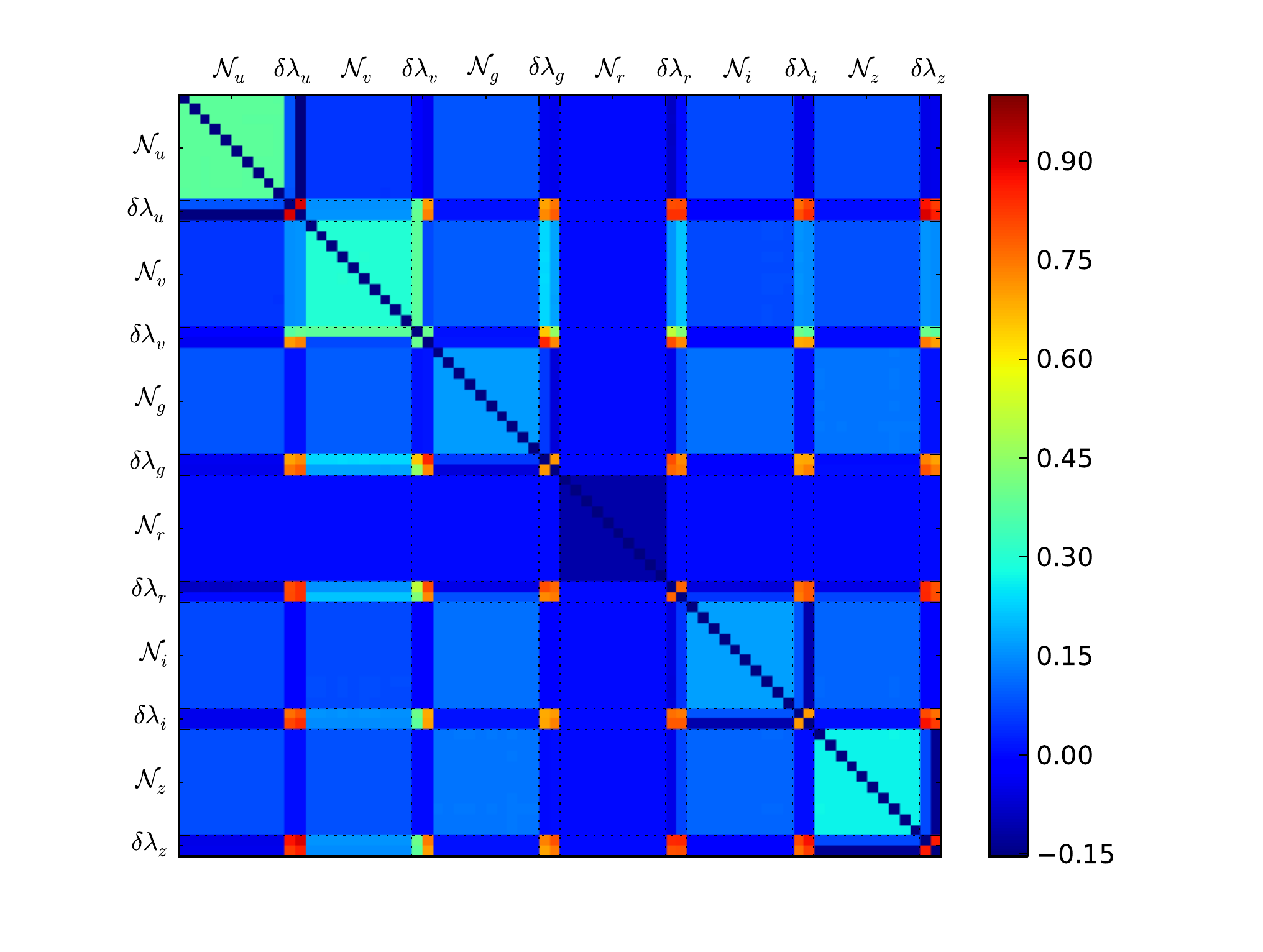}
    \caption{Correlation matrix yielded by the passband
      (syst. included). We note the strong positive correlations between
      the filter fronts.  This is because the all uncertainties on the
      filter front positions are dominated by the wavelength
      calibration of the bench monochromator.}
    \label{fig:corrmats}
  \end{center}
\end{figure}

The test bench systematics are shared by all LEDs. As a result, we expect
them to introduce sizeable off-diagonal terms in the covariance matrix
of the calibration parameters.  The calibration fit described above
yields the full (stat+syst) covariance matrix of the calibration
parameters, $\vec{\vartheta}_t$.  In figure \ref{fig:corrmats}, we
show a typical correlation matrix obtained from the fit of ten
(simulated) calibration runs. The matrix elements labelled ${\cal N}_X$
are related to the filter normalisation (relative to the r-filter
normalisation). The elements labelled $\delta\lambda_X$ are the filter
cutoff displacements. The matrix presented here corresponds to the ``best case scenario'' presented in section \ref{sec:nist_systematics} and to where 
the NIST uncertainties are all fully positively correlated (best case scenario). 
The band-to-band correlations of the filter normalisations are essentially negligible, 
while the filter's front displacements are all positively
correlated. Indeed, they all share the uncertainty on the
monochromator wavelength calibration, which is the dominant
contribution to their error budget.


\section{Cooled Large Area Photodiodes}
\label{sec:clap}

The time between two test-bench recalibrations of the illumination
system is expected to be long -- several years at least. For this
reason, it is recommended to install a calibrated photodiode on site
that can measure the light actually delivered by the device and
monitor any unexpected drift in the the calibration beam. A solution would
be to install in the dome, or even directly on the telescope, a
calibrated photodiode procured from NIST coupled with a
picoammeter. The typical irradiance of the calibration beam close to
the focal plane is about $0.5\ \mathrm{nW\ cm^{-2}}$.  A typical
photodiode that can be procured from NIST, such as the Hamamatsu S2281
with an active area of $1 \mathrm{cm^{-2}}$
and a sensitivity of $0.3\ \mathrm{A\ W^{-1}}$ at 555 nm, would
generate a current of 150 pA, which can be easily measured using a
commercial picoammeter. This solution is nevertheless expensive and
not very practical, because the photodiode and picoammeter are quite
sizeable.

The DICE team has developed miniaturised modules comprising a
Hamamatsu S3477-04 photodiode coupled with an ultra-low noise current
amplifier implemented as a custom-made ASIC. Each module is connected
with an analogue link to a small backend electronics whose main purpose
is to digitise the photodiode current and store it in a buffer. Thanks
to their small size ($\mathrm{100 mm \times 100 mm \times 15 mm}$),
these modules can be fixed directly on the telescope, close enough to
the calibration beam to intercept it. One experimental CLAP module has
been installed for tests inside MegaPrime in front of the filter
jukebox.  Besides demonstrating that this miniaturised low-cost
monitoring solution works, our goal is to investigate whether we could
gain in precision by polarising the photodiode in reverse and
operating it in photoelectric mode, as recommended by all photodiode
makers. This would have the advantage of increasing the sensitivity of
the detector, at the expense of having a stronger dark current.

The S3477-04 is slightly smaller than the S2281 ($\mathrm{5.8\ mm
  \times 5.8\ mm}$) and has a similar sensitivity. It is coupled to a
small Peltier effect, which allows one to operate it around -20
\degree C, hence reducing the dark current very significantly when the
photodiode is polarised in reverse. NIST does not provide a
calibration service for this type of detector, so we have to calibrate
it ourselves, using a NIST photodiode as a primary standard and a DICE
apparatus as an intermediate light source.


\end{document}